\documentclass[preprint2]{aastex6}
\usepackage{graphics,color}
\usepackage{subfigure}
\usepackage{natbib}
\usepackage[labelsep=none]{caption}

\newcommand\aastex{AAS\TeX}

\shorttitle{\aastex\ Changing-look AGNs}
\shortauthors{Yang et al.}

\begin{document}

\title{Discovery of 21 New Changing-look AGNs in Northern Sky}

\email{qianyang.astro@pku.edu.cn}

\author{Qian Yang\altaffilmark{1,2}, Xue-Bing Wu\altaffilmark{1,2}, Xiaohui Fan\altaffilmark{3,2}, Linhua Jiang\altaffilmark{2}, Ian McGreer\altaffilmark{3}, Jinyi Shangguan\altaffilmark{1,2}, Su Yao\altaffilmark{2}, Bingquan Wang\altaffilmark{1}, Ravi Joshi\altaffilmark{2}, Richard Green\altaffilmark{3}, Feige Wang\altaffilmark{1,2}, Xiaotong Feng\altaffilmark{1,2}, Yuming Fu\altaffilmark{1,2}, Jinyi Yang\altaffilmark{1,2}, Yuanqi Liu\altaffilmark{1}}

\altaffiltext{1}{Department of Astronomy, School of Physics, Peking University, Beijing 100871, China}
\altaffiltext{2}{Kavli Institute for Astronomy and Astrophysics, Peking University, Beijing 100871, China}
\altaffiltext{3}{Steward Observatory, University of Arizona, 933 North Cherry Avenue, Tucson, AZ 85721, USA}

\begin{abstract}
The rare case of changing-look (CL) AGNs, with the appearance or disappearance of broad Balmer emission lines within a few years, challenges our understanding of the AGN unified model. We present a sample of 21 new CL AGNs at $0.08<z<0.58$, which doubles the number of such objects known to date. These new CL AGNs were discovered by several ways, from (1) repeat spectra in the SDSS, (2) repeat spectra in the Large Sky Area Multi-Object Fiber Spectroscopic Telescope (LAMOST) and SDSS, and (3) photometric variability and new spectroscopic observations. We use the photometric data from surveys, including the SDSS imaging survey, the Pan-STARRS1, the DESI Legacy imaging survey, the Wide-field Infrared Survey Explorer (WISE), the Catalina Real-time Transient Survey, and the Palomar Transient Factory. The estimated upper limits of transition timescale of the CL AGNs in this sample spans from 0.9 to 13 years in the rest frame. The continuum flux in the optical and mid-infrared becomes brighter when the CL AGNs turn on, or vice versa. Variations of more than 0.2 mag in $W1$ band were detected in 15 CL AGNs during the transition. The optical and mid-infrared variability is not consistent with the scenario of variable obscuration in 10 CL AGNs at more than $3\sigma$ confidence level. We confirm a bluer-when-brighter trend in the optical. However, the mid-infrared WISE colors $W1-W2$ become redder when the objects become brighter in the $W1$ band, possibly due to a stronger hot dust contribution in the $W2$ band when the AGN activity becomes stronger. The physical mechanism of type transition is important for understanding the evolution of AGNs.
\end{abstract}


\section{Introduction} \label{sec:introduction}
Active Galactic Nucleus (AGNs) are classified into Type 1 and Type 2 AGNs based on emission line features \citep[e.g.,][]{Seyfert1943, Khachikian1971}. Type 1 AGNs show broad (1,000-20,000 km s$^{-1}$) and narrow (300-1,000 km s$^{-1}$) emission lines, while only narrow emission lines are present in Type 2 AGNs \citep[and references therein]{Netzer2015}. This dichotomy is explained in the unified model \citep{Antonucci1993, Urry1995} as a viewing angle effect due to the obscuration of the broad-line region (BLR). Intermediate AGN types exist. Types 1.8 (1.9) is classified by a broad H$\alpha$ and a weak (absent) broad H$\beta$ line \citep{Osterbrock1981}. Type 1.5 is an intermediate type between Type 1 and Type 2 with an apparent narrow H$\beta$ profile superimposed on broad H$\beta$ components \citep{Osterbrock1976, Osterbrock1977}. The intermediate type objects are explained as Type 2 objects observed in scattered light \citep{Antonucci1985}, or partial obscuration by optically thin dust \citep{Stern2012}. An alternative scenario is that the different classes are evolutionary \citep{Penston1984, Korista2004, Wang2007, Elitzur2014}. \citet{Penston1984} proposed that the Type 2 class is possibly Type 1 class in which the continuum source is temporarily off.

Some AGNs were observed to change between different spectral types, so-called changing-look (CL) AGNs. The term ``changing-look" was originally used in X-ray, in which objects were found to change from Compton-thick to Compton-thin, or vice versa \citep[e.g.,][]{Matt2003, Bianchi2005, Ballo2008, Piconcelli2007, Risaliti2009, Marchese2012, Ricci2016}. Lately, this term is widen to describe objects with optical spectral type transitions \citep[e.g.][]{Denney2014, LaMassa2015, MacLeod2016, Ruan2016, Runnoe2016, McElroy2016, Gezari2017}. The physical mechanisms of the changes are still under debate. The main plausible mechanisms are (1) variable obscuration due to the movement of obscuring material, in a scenario that the dusty toroidal structure obscuring the BLR has a patchy distribution \citep[e.g.,][]{Nenkova2008a, Nenkova2008b, Elitzur2012} or an accelerating outflow \citep[e.g.,][]{Shapovalova2010}; (2) variable accretion rate, in an evolutionary paradigm that AGN follows an evolutionary sequence from Type 1 to intermediate type and later to Type 2, or vice versa \citep[e.g.,][]{Penston1984, Elitzur2014}; or (3) a tidal disruption event (TDE) of a star disrupted by the supermassive black hole (SMBH), which may also result in a change of classification \citep{Eracleous1995, Merloni2015, Blanchard2017}. The nature of the type transition is important for the understanding of the evolution of AGNs.

The spectral type changes, with the appearance or disappearance of broad Balmer emission lines, have been detected in only a small number of AGNs. Long timescale observations of a handful of AGNs show that some AGNs have changed from Type 1 to Type 2, and back and forth. Mrk 1018 had changed from a Type 1.9 to Type 1 Seyfert, and changed back to Type 1.9 after 30 years \citep{Cohen1986, McElroy2016}. Observations in more than 40 years reveals that Mrk 590 had changed from Type 1.5 to Type 1 and changed back to Type 1.9-2 \citep{Denney2014}. NGC 2617, which was a Seyfert 1.8 galaxy \citep{Moran1996}, changed to a Seyfert 1 \citep{Shappee2014}. These cases are better explained as changes in its luminosity than obscuration \citep{McElroy2016, Denney2014, Shappee2014}. The broad emission lines of NGC 4151, which was originally Type 1.5 \citep{Osterbrock1977}, had once disappeared \citep{Antonucci1983, Lyutyj1984, Penston1984} and then returned \citep[e.g.,][]{Shapovalova2010}. The variation in NGC 4151 is explained as probably caused by an accelerating outflow originating very close to the black hole \citep[BH,][]{Shapovalova2010}. Such back-and-forth type changes put forward questions on what mechanisms generate these changes.

In recent years, the discoveries of CL quasars show that such transitional phenomenon happens in higher redshift and can occur in more luminous and massive systems. Until now, there are fewer than 20 known CL quasars. \citet{LaMassa2015} reported the first CL quasar, J0159+0033, changed from Type 1 to Type 1.9. They demonstrated that variable absorption does not explain the observed timescales and a large scale obscuration material is needed for this scenario. \citet{Merloni2015} argued that J0159+0033 could be a luminous flare produced by a TDE. \citet{Runnoe2016} reported a CL quasar, J1011+5442, ``turning-off" within a rest frame time of approximately 500 days. They argued that the transition timescale is inconsistent with an abrupt change in the reddening towards the central engine, and the decaying light curve with a prolonged bright state preceding the decay is not consistent with a decaying TDE. If the type changes were caused by the obscuration of the quasar, high linear polarization would be expected \citep{Hutsemekers2017, Marin2017}. \citet{Hutsemekers2017} measured the polarization of J1011+5442, and found null polarization, suggesting that the type transition was not due to variable obscuration. A rapid ``turn-on" of a quasar, J1554+3629, was detected by iPTF in a timescale shorter than one year \citep{Gezari2017}. They implied that a factor of 10 brightening in UV and X-ray continuum flux is more likely caused by an intrinsic change in the accretion rate. Some systematic archival searches for objects in the Sloan Digital Sky Survey (SDSS) with repeat spectroscopy found more CL quasars \citep{Ruan2016, MacLeod2016}. They implied that changes in the accretion rate can better explain the transition timescale and emission line properties than variable dust obscuration. \citet{Sheng2017} argued that the large variability amplitude of CL AGNs in mid-infrared supports the scenario of changes in the accretion rate other than that of varying obscuration. In a recent TDE (PS16dtm), broad Balmer emission lines appeared accompanied by the significant increase in the continuum flux, as well as strong Fe II and He II emissions \citep{Blanchard2017}.

The frequency and timescale of such transient in the universe are interesting issues. \citet{Martini2003} suggested that the number of turn-off quasars in a large number of quasar can be used to measure or set a lower limit to the episodic lifetime of quasars. In the search by \citet{MacLeod2016}, out of more than 1000 quasars with $g$-band variability larger than one mag, only ten objects show variable broad emission lines \citep{MacLeod2016}. The majority of highly variable quasars did not exhibit emerging or disappearing broad emission lines. \citet{Rumbaugh2018} identified $\sim$ 1000 extreme variability quasars (EVQs) with a maximum $g$-band magnitude change of more than one mag with the SDSS and 3-Year Dark Energy Survey \citep[DES,][]{Flaugher2005} imaging. They claimed that these EVQs are good candidates for CL AGNs.

CL AGNs provide perfect cases to study the connection between AGNs and their host galaxies. There is a tight correlation between BH mass, $M_{\rm BH}$, and the velocity dispersion, $\sigma_*$, of the bulge component in nearby galaxies \citep{Kormendy2013}. \citet{Gezari2017} reported that the $M_{\rm BH}\ (2_{-1.5}^{+4}\times10^8M_{\odot})$ of J1554+3629 ($z = 0.237$), estimated from the quasar spectrum after ``turning-on" with emerging broad Balmer emission lines, is in good agreement with its $M_{\rm BH}\ (1_{-0.7}^{+2}\times10^8M_{\odot})$ inferred from $\sigma_*$. CL AGNs provide exceptional cases to study the central BHs and their host galaxies at higher redshift. On the other hand, the ``turning-off" CL quasars provide perfect cases to study the host galaxies of quasars in detail, avoiding contamination from the luminous central engines.

The CL AGNs, with the appearance or disappearance of broad Balmer lines in a time scale of years, challenge our understanding of the AGN unification model. Motivated by the issues about the physical mechanism behind the transition, and the questions about the frequency and timescale of the type transition, we conduct a survey for CL AGNs. The investigations consist of (1) repeat spectroscopy in the SDSS spectral archive, (2) repeat spectroscopy in the SDSS and the Large Sky Area Multi-Object Fiber Spectroscopic Telescope (LAMOST) spectral archive, and (3) searching for CL candidates based on photometric variability.

This paper is organized as follows. Section \ref{sec:data} describes the spectroscopic and imaging data. Section \ref{sec:method} outlines the selection methods and spectroscopic observations. In Section \ref{sec:results}, we present the new CL AGNs and their variability in optical and mid-infrared. In Section \ref{sec:discussion}, we discuss the CL AGN color variability and the timescale of type transition. We summarize the paper in Section \ref{sec:summary}. In this work we adopt a standard $\Lambda$CDM cosmology with $\Omega_{\Lambda}=0.7$, $\Omega_m=0.3$, and $H_0=70$ km s$^{-1}$ Mpc$^{-1}$. Through this paper, all magnitudes are in AB magnitudes.

\section{Data} \label{sec:data}
\subsection{Spectroscopic Data}
\subsubsection{SDSS Spectroscopy} \label{subsec:sdss_spec}
There are 4,851,200 spectra in the SDSS Fourteenth Data Release \citep[DR14;][]{Abolfathi2017} taken by the Sloan Foundation 2.5m telescope \citep{Gunn2006} at Apache Point Observatory. The spectra are from the SDSS-I/II with a wavelength coverage from 3800 to 9100 ${\rm \AA}$, and the Baryon Oscillation Spectroscopic Survey \citep[BOSS;][]{Dawson2013} spectrograph of the SDSS-III \citep{Eisenstein2011} with a wavelength coverage from 3600 to 10400 ${\rm \AA}$ \citep{Smee2013}. The spectral resolution is 1500 at 3800 ${\rm \AA}$ and 2500 at 9000 ${\rm \AA}$. The SDSS spectroscopic pipelines classify the objects as galaxies (``GALAXY"), stars (``STAR"), or quasars (``QSO"), through the comparison of individual spectrum with galaxy, QSO, and stellar templates \citep{Bolton2012, Hutchinson2016}. The spectral quality is described by a confidence flag ``zWarning", which is 0 or 16 for good data without identified problems \citep[e.g.,][]{Stoughton2002}. We carry out searches of CL AGNs from SDSS galaxies or quasars.

\subsubsection{LAMOST Spectroscopy} \label{subsec:lamost}
LAMOST is a 4 meter reflecting Schmidt telescope equipped with 4000 fibers with a 5$^{\circ}$ field of view \citep{Cui2012, Zhao2012}. The wavelength coverage of LAMOST ranges from 3700 ${\rm \AA}$ to 9000 ${\rm \AA}$ with two arms \citep{Du2016}, a blue arm (3700$-$5900 ${\rm \AA}$) and a red arm (5700$-$9000 ${\rm \AA}$). The overall spectral resolution of LAMOST is approximately 1800. The data is reduced with LAMOST pipelines \citep{Luo2012}. In this paper, we utilize the LAMOST spectra from data release one to five \citep{Luo2015, He2016}.

\subsection{Imaging Data}
CL AGNs showed continuum flux changes in the optical and mid-infrared. Apart from objects with repeat spectroscopy, we carry out additional searches of CL AGNs based on imaging data. We briefly introduce the imaging data as follows.

The SDSS imaging survey scanned the sky in five filters, $ugriz$ \citep{Fukugita1996}, covering 11,663 deg$^2$ in SDSS-I/II from 2000 to 2007 \citep{Abazajian2009} and additional 3,000 deg$^2$ in SDSS-III in 2008.

The Pan-STARRS1 \citep[PS1;][]{Chambers2016} survey used a 1.8-meter telescope with a 1.4 Gigapixel camera to image the sky in five broadband filters ($grizy$). The observations cover three-quarters of the sky several times per filter. We use the PS1 magnitudes in the stack catalog, from co-added images made from the multiple exposures. Using a quasar composite spectrum from \citet{Vanden2001} convolved with the PS1 and SDSS $g$-band filter curves \citep[Equation 2 in][]{Wu2004}, the magnitude difference, $\delta(g_{\rm PS1} - g_{\rm SDSS})$, is between $-$0.065 mag and 0.008 mag at redshift $z<2$.

The DESI Legacy imaging survey (DELS; Dey et al. 2017 in preparation) commenced imaging surveys, including the DECam Legacy Survey (DECaLS) $g$, $r$, and $z$-bands, the $g$, $r$-bands Beijing-Arizona Sky Survey \citep[BASS;][]{Zou2017}, and the Mayall $z$-band Legacy Survey (MzLS). The imaging survey commenced from 2014 and will complete in 2018, covering 14000 deg$^2$ sky. The magnitude difference, $\delta(g_{\rm DELS} - g_{\rm SDSS})$, is between $-$0.053 mag and 0.005 mag at $z<2$, using the composite spectrum convolved with the DELS and SDSS $g$-band filter curves.

The Wide-field Infrared Survey Explorer \citep[WISE;][]{Wright2010} mapped the all-sky from January to July in 2010 in four bands centered at wavelengths of 3.4, 4.6, 12, and 22 $\mu$m ($W1$, $W2$, $W3$, and $W4$). The secondary cryogen survey and Near-Earth Object Wide-field Infrared Survey Explorer \citep[NEOWISE;][]{Mainzer2011} Post-Cryogenic Mission mapped the sky from Aug 2010 to Feb 2011. The NEOWISE Reactivation Mission \citep[NEOWISE-R;][]{Mainzer2014} surveys the sky in $W1$ and $W2$ bands from 2013. We use WISE multi-epoch photometry from individual single-exposure images in $W1$ and $W2$ bands.

The Catalina Real-time Transient Survey \citep[CRTS;][]{Drake2009} uses data from the Catalina Sky Survey which repeatedly covers 26,000 deg$^2$ on the sky. The CRTS photometric data are unfiltered and calibrated to $V$-band magnitude. We apply a constant offset to the CRTS magnitudes to match the simultaneous $g$-band PS1 magnitude.

The Palomar Transient Factory \citep[PTF;][]{Law2009} is a wide-field survey covering approximately 30,000 deg$^2$ in $g$, $r$ bands from 2009 to 2012.

\section{Target Selection and Observation} \label{sec:method}

\subsection{SDSS Repeat Spectroscopy}
We carried out a systematic investigation in the SDSS DR14 spectral archive, in which 87\% spectra are in good quality ($zWarning=0\ {\rm or}\ 16$). Cross-matching the spectra with a radius of 2$\arcsec$ results in 350,609 objects repeatedly observed ($\geq2$ epochs). There were 175,575 repeatedly observed objects classified as ``GALAXY" by the SDSS pipeline in at least one epoch spectrum. Among them, 2,023 objects were classified as ``GALAXY" in one epoch of spectrum and ``QSO" in another epoch of spectrum. We visually checked all spectra of these objects to search for CL AGNs. 

As the definitions of AGN spectral types depend on the strength of broad emission lines, we qualitatively define a detection of a broad emission line if the signal-to-noise ratio, S/N, of the emission line is higher than 5; and non-detection if the S/N is lower than 1. Specifically, a weak detection is when the S/N is between 1 and 3, and an intermediate detection is when the S/N is between 3 and 5. Visual check is not able to accurately obtain the quantitative changes of emission lines, but can qualitatively distinguish the cases with dramatic changes. More specifically, the visual inspection is easier to distinguish $>5\sigma$ and $<1\sigma$ detections than to distinguish $3-5\sigma$ and 1-3$\sigma$ detections. Therefore, the visual inspection process is inclined to select AGNs changed between Type 1 and Type 2 (or Type 1.9), with appearing or disappearing broad H$\alpha$ (or H$\beta$) emission line. Nevertheless, the visual check selects AGNs changed between Type 1 and Type 1.8 with a lower completeness. Objects with distinct companions (within 2$\arcsec$) in the SDSS image were also excluded in the visual check process. From the visual inspection, we selected 9 CL AGNs. Four of the selected CL AGNs were reported previously, including the CL quasar J0159+0033 in \citet{LaMassa2015}, J0126$-$0839 and J2336+0017 in \citet{Ruan2016}, and J1011+5442 in \citet{Runnoe2016}. We recovered all the changing-look quasars in these three works. The new CL AGNs found in the SDSS archive are J1104+6343, J1118+3203, J1150+3632, J1358+4934, and J1533+0110. We describe the objects rejected by visual inspection in the appendix. In Section \ref{sec:fitting}, we describe fitting of H$\alpha$ and H$\beta$ emission lines of these objects to quantitatively understand the changes of CL AGNs selected from the visual inspection. The ratio of CL AGNs, with appearance or disappearance of broad Balmer emission lines, is roughly 0.006\%, out of objects that were identified as galaxies and repeatedly observed. This ratio may be affected by some issues, for example the time intervals of the repeat spectra, spectroscopic survey selection bias, spectral quality, and possible selection bias of visual inspection. The selection steps are summarized in Table \ref{tab:spectra_sdss}.

\subsection{LAMOST and SDSS Repeat Spectroscopy}
We cross-matched the LAMOST spectral archive with the SDSS spectra archive. The selection steps are summarized in Table \ref{tab:spectra_lamost}. There were 155,220 objects, that were classified as ``GALAXY" (75\%) or ``QSO" (25\%) with good spectral quality in the SDSS, were observed in the LAMOST. Comparing the SDSS spectra with the LAMOST spectra, we fitted the Balmer emission lines in spectra from the SDSS and the LAMOST with Gaussian profiles. We selected the objects with changes in the emission line flux density larger than $2\times10^{-18}$ ${\rm erg}\ {\rm s}^{-1}\ {\rm cm}^{-2}\ {\rm \AA}^{-1}$, which is a conservative criterion avoiding missing CL AGNs at this step. There were 8,181 objects with potential emission-line variation. We visually inspected all these objects, and identified 10 CL AGNs, including 8 turn-on CL AGNs (J0831+3646, J0909+4747, J0937+2602, J1115+0544, J1132+0357, J1447+2833, J1545 +2511, and J1552+2737) and 2 turn-off AGNs (J0849+2747 and J1152+3209). We obtained flux calibration to match the narrow emission line flux of the SDSS spectra, assuming a constant narrow emission line flux within a few tens years. We describe the objects visually rejected in the appendix. In this survey, about 0.007\% of the galaxies from the LAMOST and SDSS cross-matched sample were proved to be CL AGNs with appearance of broad Balmer emission lines.

\subsection{Photometric Variability}
Most of the reported CL AGNs were discovered by repeat spectroscopy. Meanwhile, the optical and mid-infrared flux of CL AGNs varies following the type transition. We conducted additional CL AGN searches based on photometric variability. We searched CL AGN candidates from objects which were spectroscopically identified as ``GALAXY" but brightened in later photometric data, and from objects which were spectroscopically identified as ``QSO" but significantly dimmed.

\begin{deluxetable*}{llr}[htbp]
\tablecaption{\ CL AGNs selection from SDSS repeat spectra \label{tab:spectra_sdss}}
\tablewidth{0pt}
\tablehead{
\colhead{Note} &
\colhead{Selection} &
\colhead{Number}
}
\startdata
Spectra in SDSS DR14 & All & 4,851,200 spectra\\
Spectra with good quality & $zWarning = 0\ {\rm or}\ 16$ & 4,196,290 spectra \\
Objects with repeat spectra & 2$\arcsec$ coordinates cross-match & 350,609 objects\\
Galaxies with repeat spectra & classified as ``GALAXY" at one epoch & 175,575 objects \\
Classification changed between QSO and Galaxy & classified as ``QSO" at another epoch & 2,023 objects\\
Visual check & appearing or disappering broad H$\beta$ & 9 (4 known) CL AGNs\\
\enddata
\end{deluxetable*}

\begin{deluxetable*}{llr}[htbp]
\tablecaption{\ CL AGNs selection from SDSS and LAMOST repeat spectra \label{tab:spectra_lamost}}
\tablewidth{0pt}
\setlength{\tabcolsep}{9pt}
\tablehead{
\colhead{Note} &
\colhead{Selection} &
\colhead{Number}
}
\startdata
Spectra with good quality & $zWarning = 0\ {\rm or}\ 16$ & 4,196,290 spectra \\
SDSS QSO/Galaxy & classified as ``GALAXY" or ``QSO" & 3,223,478 spectra \\
Repeatedly observed by LAMOST & 2$\arcsec$ cross-match with LAMOST & 155,220 objects\\
Possible variable Balmer lines & program automatically check emission-line variation & 8,181 objects\\
Visual check & appearing or disappering broad H$\beta$ & 10 CL AGNs \\
\enddata
\end{deluxetable*}

\begin{deluxetable*}{llr}[htbp]
\tablecaption{\ Turn-on CL AGN candidate selection based on imaging data \label{tab:photometric_galaxy}} 
\tablewidth{0pt}
\setlength{\tabcolsep}{12pt}
\tablehead{
\colhead{Note} &
\colhead{Selection} &
\colhead{Number}
}
\startdata
Spectra in SDSS DR14 & All & 4,851,200 spectra\\
WISE single epoch detected & 2$\arcsec$ cross-match with the WISE single epoch data & 4,196,290 spectra \\
Galaxies with good spectra & class $=$ ``GALAXY" and ($zWarning = 0\ {\rm or}\ 16$) & 2,494,319 spectra \\
WISE brighten and redder & $\Delta W1<-0.2$ and $\Delta(W1-W2)>0.1$ & 28,395 objects \\
Optical brighten & $\Delta g<0$ and $g<19$ & 2,147 objects \\
Redshift & $z>0.1$ & 660 objects\\
Visual check light curves & obvious brighten trend (CRTS/PTF) & 59 objects \\
Observed & $\cdots$ & 17 objects \\
Confirmed & $\cdots$ & 6 (1 known) CL AGNs \\
\enddata
\end{deluxetable*}

\begin{deluxetable*}{llr}[htbp]
\tablecaption{\ Turn-off CL AGN candidate selection based on imaging data \label{tab:photometric_qso}} 
\tablewidth{0pt}
\setlength{\tabcolsep}{12pt}
\tablehead{
\colhead{Note} &
\colhead{Selection} &
\colhead{Number}
}
\startdata
SDSS QSO in DR7 \& DR12 & All & 346,464 objects \\
WISE single epoch detected & 2$\arcsec$ cross-match with the WISE single epoch photometry & 326,124 objects \\
WISE dim and bluer & $\Delta W1>0.2$ and $\Delta(W1-W2)<-0.1$ & 6,847 objects \\
Optical dim & $\Delta g>1$ & 232 objects \\
Observed & $\cdots$ & 1 objects \\
Confirmed & $\cdots$ & 1 CL AGN \\
\enddata
\end{deluxetable*}

\subsubsection{Turn-on CL AGNs Selected from Imaging Data}
We carried out a search for turn-on CL AGNs from SDSS galaxies, which became brighter later. The selection procedures are shown in Table \ref{tab:photometric_galaxy}. Motivated by the rapid transition happened in J1011+5442 and J1554+3629, we first considered recent imaging data. The WISE multi-epoch data is available in the recent years, from 2010 to 2017. We started from galaxies detected by WISE multi-epoch data, and there were 28,395 galaxies became brighter ($\Delta W1 < -0.2$ mag) and redder ($\Delta(W1-W2)>0.1$ mag) in mid-infrared. More detail discussion about the color criteria in mid-infrared will be provided in Section \ref{sec:color}. Among them, 2,147 galaxies also became brighter in optical $g$-band and had $g<19$ mag. We checked the CRTS and PTF light curves of these galaxies at $z>0.1$, and picked out 59 turn-on CL AGN candidates with obvious trend of increasing flux in the optical and mid-infrared. This selection method recovered J0831+3646 and J1554+3629. Other turn-on CL AGNs from repeat spectroscopy were missed in this process due to smaller redshift or smaller variability. More systematic searches can be extended to lower redshift and fainter objects with smaller variability.

Optical long-slit spectroscopic observations for some CL AGN candidates were carried out using the Xinglong 2.16 m telescope (XLT) in China and Palomar P200/DBSP spectrograph. The Xinglong 2.16 m telescope is located at the NAOC observatory. It is equipped with the Beijing Faint Object Spectrograph and Camera (BFOSC). We observed 17 CL AGN candidates in April 2017, including a CL AGN J1554+3629 discovered by \citet{Gezari2017}, using the BFOSC and Grism 4 (G4) with dispersion of 198 ${\rm \AA}$/mm and a wavelength coverage from 3850 to 8300 ${\rm \AA}$. The grism yields a resolution of $R \sim$265 or 340 at 5000 ${\rm \AA}$ using a 2$\farcs$3 or 1$\farcs$8 slit \citep{Fan2016}. The spectra were obtained using a
1$\farcs$8 slit when seeing$<2\arcsec$ or a 2$\farcs$3 slit when $2\arcsec<$seeing$<3\arcsec$. J1259+5515 was observed by DBSP after it was observed by XLT. The DBSP spectrum was obtained using P200/DBSP with grating G600 at blue side ($R\sim$1200 at 5000 ${\rm \AA}$) and G316 at red side ($R\sim$642 at 5000 ${\rm \AA}$) with a 1.5$\arcsec$ slit under seeing $\sim1.5\arcsec$. The spectra were reduced using standard IRAF\footnote{IRAF is distributed by the National Optical Astronomy Observatory, which is operated by the Association of Universities for Research in Astronomy (AURA) under cooperative agreement with the National Science Foundation.} routines \citep{Tody1986, Tody1993}. The flux of J1259+5515 obtained with DBSP and XLT in two days are in good agreement. Five new CL AGNs (J1003+3525, J1110-0003, J1259+5515, J1319+6753, and J1550+4139), with emerging broad H$\beta$, were confirmed by the XLT and DBSP spectra. The fiber diameters of SDSS, BOSS, and LAMOST are 3$\arcsec$, 2$\arcsec$, and 3.3$\arcsec$, respectively. The fiber diameters are large enough to include the light from the nuclear. The slit widths of long-slit spectra are slightly smaller than the fiber diameters, leading to less host galaxy light contribution to the spectra. The photometric variability of these CL AGNs rules out the possibility that the different spectra are merely caused by differences between fiber and slit spectra.

\begin{deluxetable*}{ccclcclc}[htbp]
\tablecaption{\ CL AGNs \label{tab:cl}}
\tablewidth{0pt}
\tablehead{
\colhead{Name} & \colhead{R.A.} &
\colhead{Decl.} & \colhead{Redshift} &
\colhead{Transition} &
\colhead{Epoch} &
\colhead{Instrument2} &
\colhead{Reference}
}
\startdata
J0831+3646 & 08:31:32.25 & +36:46:17.2 & 0.19501 & Turn-on & [52312, 57367] & LAMOST & This work \\
J0849+2747 & 08:49:57.78 & +27:47:28.9 & 0.29854 & Turn-off & [53350, 56628] & LAMOST & This work \\
J0909+4747 & 09:09:32.02 & +47:47:30.6 & 0.11694 & Turn-on & [52620, 57745] & LAMOST & This work \\
J0937+2602 & 09:37:30.32 & +26:02:32.1 & 0.16219 & Turn-on & [54524, 57369] & LAMOST & This work \\
J1003+3525 & 10:03:23.47 & +35:25:03.8 & 0.11886 & Turn-on & [53389, 57867] & XLT & This work \\
J1104+0118 & 11:04:55.17 & +01:18:56.6 & 0.57514 & Turn-off & [52374, 57867] & XLT & This work \\
J1104+6343 & 11:04:23.21 & +63:43:05.3 & 0.16427 & Turn-off & [52370, 54498] & SDSS & This work \\
J1110$-$0003 & 11:10:25.44 & $-$00:03:34.0 & 0.21922 & Turn-on & [51984, 57864] & XLT & This work \\
J1115+0544 & 11:15:36.57 & +05:44:49.7 & 0.08995 & Turn-on & [52326, 57393] & LAMOST & This work \\
J1118+3203 & 11:18:29.64 & +32:03:59.9 & 0.3651 & Turn-off & [53431, 56367] & BOSS & This work \\
J1132+0357 & 11:32:29.14 & +03:57:29.0 & 0.09089 & Turn-on & [52642, 57392] & LAMOST & This work \\
J1150+3632 & 11:50:39.32 & +36:32:58.4 & 0.34004 & Turn-off & [53436, 57422] & BOSS & This work \\
J1152+3209 & 11:52:27.48 & +32:09:59.4 & 0.37432 & Turn-off & [53446, 57844] & LAMOST & This work \\
J1259+5515 & 12:59:16.74 & +55:15:07.2 & 0.19865 & Turn-on & [52707, 57863] & XLT/DBSP & This work \\
J1319+6753 & 13:19:30.75 & +67:53:55.4 & 0.16643 & Turn-on & [51988, 57867] & XLT & This work \\
J1358+4934 & 13:58:55.82 & +49:34:14.1 & 0.11592 & Turn-on & [53438, 54553] & SDSS & This work \\
J1447+2833 & 14:47:54.23 & +28:33:24.1 & 0.16344 & Turn-on & [53764, 57071] & LAMOST & This work \\
J1533+0110 & 15:33:55.99 & +01:10:29.7 & 0.14268 & Turn-on & [51989, 54561] & SDSS & This work \\
J1545+2511 & 15:45:29.64 & +25:11:27.9 & 0.11696 & Turn-on & [53846, 57891] & LAMOST & This work \\
J1550+4139 & 15:50:17.24 & +41:39:02.2 & 0.22014 & Turn-on & [52468, 57864] & XLT & This work \\
J1552+2737 & 15:52:58.30 & +27:37:28.4 & 0.08648 & Turn-on & [53498, 56722] & LAMOST & This work \\
\tableline
J0126$-$0839 & 01:26:48.08 & $-$08:39:48.0 & 0.19791 & Turn-off & [52163, 54465] & SDSS & \citet{Ruan2016} \\
J0159+0033 & 01:59:57.64 & +00:33:10.5 & 0.31204 & Turn-off & [51871, 55201] & BOSS & \citet{LaMassa2015} \\
J1011+5442 & 10:11:52.98 & +54:42:06.4 & 0.24639 & Turn-off & [52652, 57073] & BOSS & \citet{Runnoe2016} \\
J1554+3629 & 15:54:40.26 & +36:29:51.9 & 0.23683 & Turn-on  & [53172, 57862] & XLT & \citet{Gezari2017} \\
J2336+0017 & 23:36:02.98 & +00:17:28.7 & 0.24283 & Turn-off & [52199, 55449] & BOSS & \citet{Ruan2016} \\
\enddata
\tablecomments{Epoch describes the MJD of the two epoch spectra. Instrument2 describes the spectrograph or telescope of the recent spectrum. XLT represents the Xinglong 2.16m telescope at NAOC. }
\end{deluxetable*}

\begin{figure*}[htbp]
 \centering
 \hspace{0cm}
 \subfigure{
 \hspace{-1.0cm}
  \includegraphics[width=3.9in]{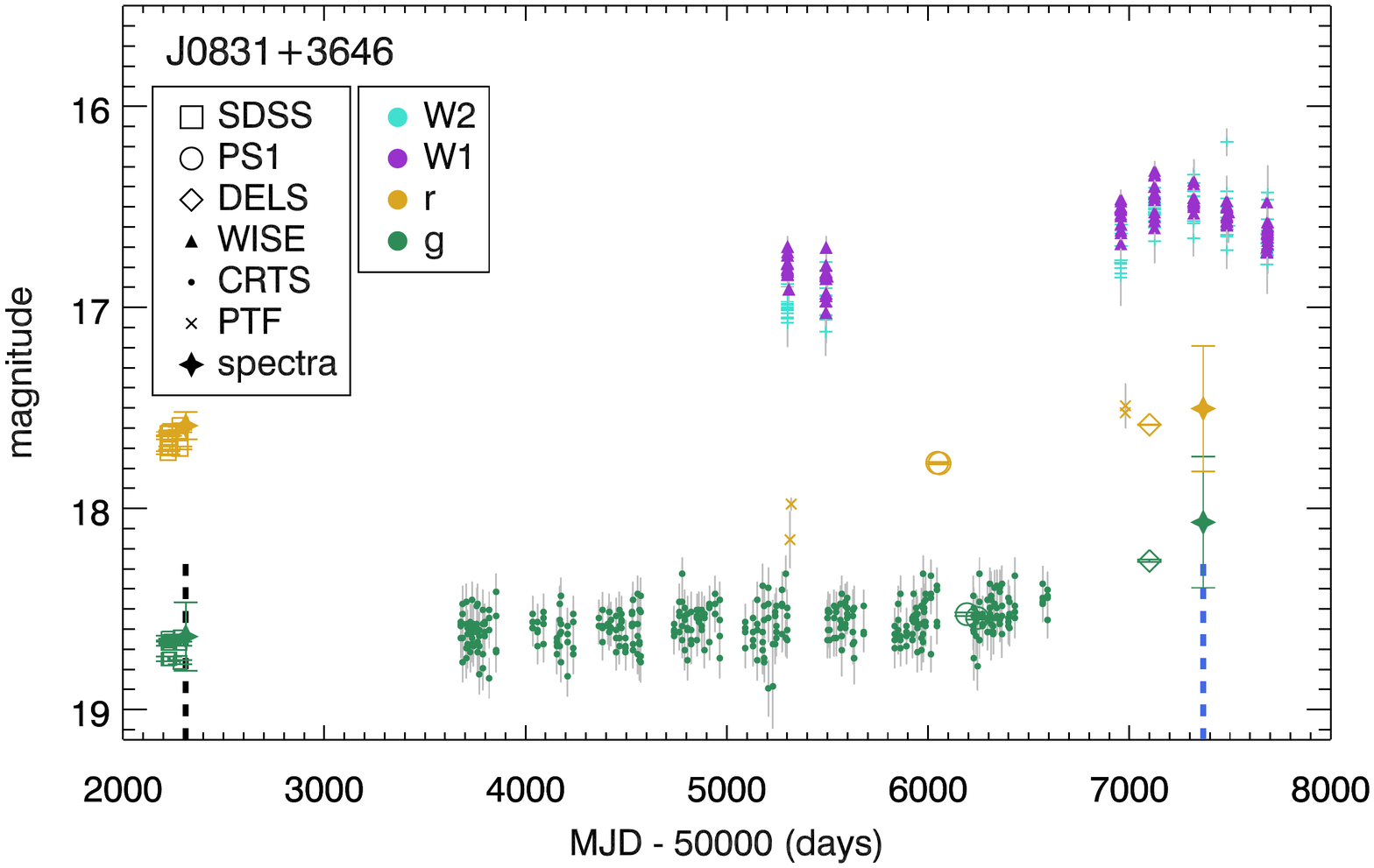}}
 \hspace{-1.4cm}
 \subfigure{
  \includegraphics[width=3.9in]{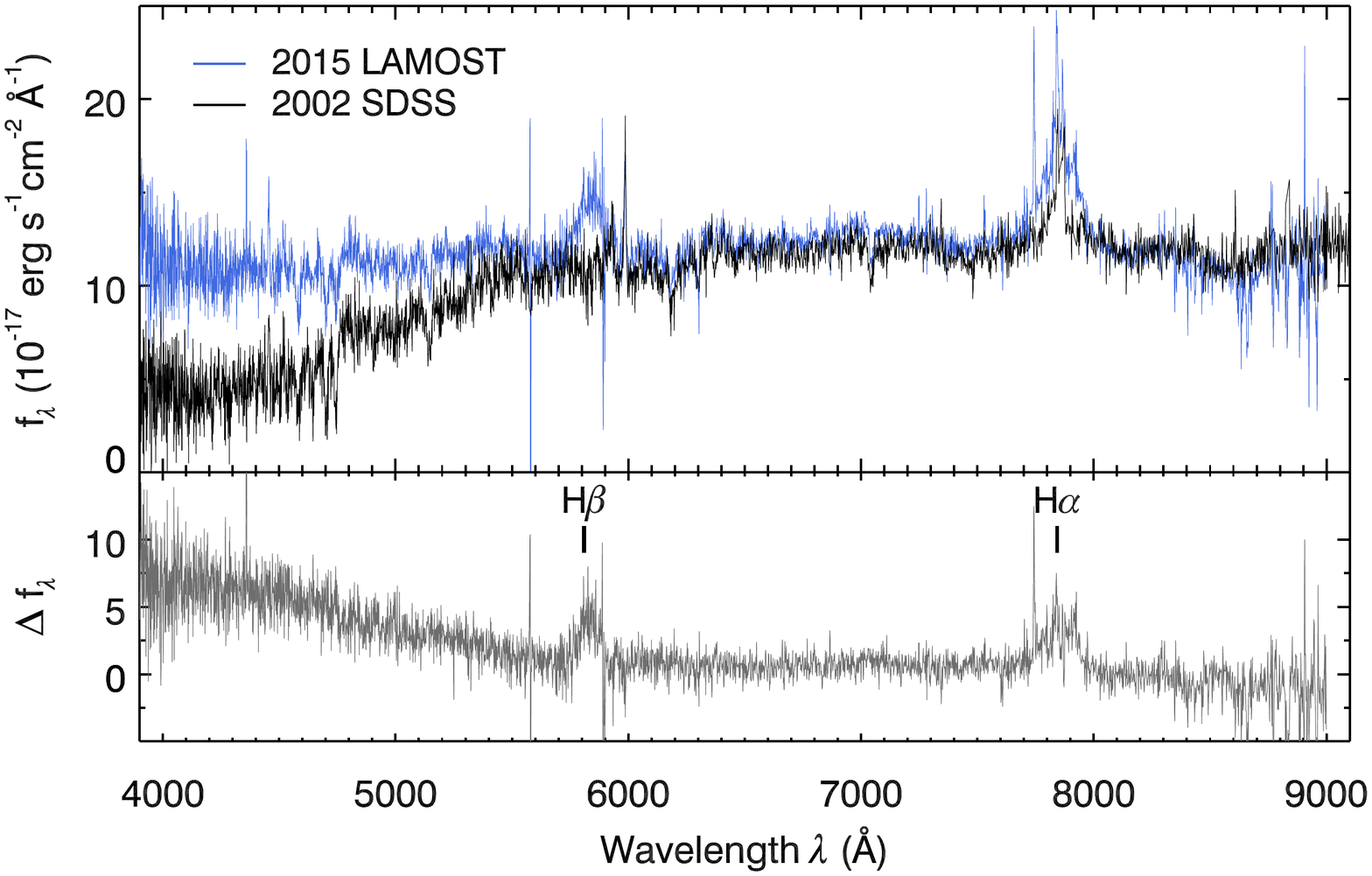}}\\
 \vspace{-1cm}
  \caption{\label{fig:cl_example} An example of the light curves (left panel), and spectra (right panel) of a new CL AGN J0831+3646. In the left panel, all photometric data are expressed in magnitude, and the x-axis is MJD$-50000$. Photometric data in distinct bands are shown in different colors, including the $g$ (green), $r$ (yellow), $W1$ (magenta), and $W2$ (cyan) bands photometry. Different shapes represent data from different surveys, including SDSS (open square), PS1 (open circle), DELS (open diamond), WISE $W1$ (solid triangle), WISE $W2$ (cross), CRTS (solid dots), and PTF (x shape). The vertical dashed lines shows the epochs of the spectra. The right panel shows the spectral flux density, $f_{\lambda}$, of CL AGNs in units of $10^{-17}$ erg s$^{-1}$ cm$^{-2}$ ${\rm \AA}^{-1}$. The early epoch SDSS spectrum is plotted in black, and the new epoch spectrum is colored as blue if the AGN turned on, or red if the AGN turned off. The lower panel shows the spectra difference, $\Delta f_{\lambda}$, between the bright epoch and faint epoch spectra. The vertical bars mark the locations of appeared or disappeared broad emission lines. The light curves and spectra of other 20 CL AGNs are in the appendix.}
\end{figure*}

\subsection{Turn-off CL AGNs Selected from Imaging Data}
We conducted a survey of turn-off CL AGNs from SDSS quasar catalogs (shown in Table \ref{tab:photometric_qso}). There are 346,464 quasars \citep{Yang2017} in the SDSS Data Release 7 Quasar catalog \citep{Schneider2010} and the Data Release 12 Quasar catalog \citep{Paris2017}. Crossing-match with a radius of 2$\arcsec$, 94\% quasars were detected in the WISE multi-epoch data. There were 6,847 quasars dimmed ($\Delta W1>0.2$ mag) and became bluer ($\Delta(W1-W2)<-0.1$ mag) in mid-infrared. 

During a timescale of months to years, the continuum variability of quasars is typically 0.2 mag in optical \citep[e.g.,][]{Vanden2004, Wilhite2005, MacLeod2012}. Among the quasars with mid-infrared variability, 232 quasars also dimmed significantly in optical (more than 1 mag in $g$ band). We recovered two turn-off CL AGNs, including J0849+2747 and J1011+5442. In April 2017, We observed one candidate (with the XLT using a 2$\farcs$3 slit), J1104+0118, with $g$-band and $W1$-band magnitudes dimmed by more than 1.8 mag and 0.6 mag, respectively. The continuum flux of J1104+0118 obviously became much redder than that in the SDSS spectrum, and the broad Mg II emission line was not detected in the new spectrum. Higher quality spectra are needed for further analysis in detail. A more complete survey of CL AGNs will need more spectroscopic follow-up.

\section{Results} \label{sec:results}
\subsection{New CL AGNs} \label{sec:fitting}
In total, there are 21 new CL AGNs found in our surveys (summarized in Table \ref{tab:cl}). Among the new CL AGNs, five were found by repeat spectra from the SDSS, ten were discovered based on repeat spectra from the SDSS and the LAMOST, and six were selected from photometric variability in optical and mid-infrared imaging data, and confirmed by new spectroscopy. The five known CL AGNs, we recovered, are also listed in Table \ref{tab:cl}. Figure \ref{fig:cl_example} shows an example of the light curves (left panels), and spectra (right panels) of a new CL AGN, J0831+3646. The light curves and spectra of other 20 new CL AGNs are listed in the appendix. In Figure \ref{fig:cl_example}, the spectra separating for 13 years show that there was no H$\beta$ emission in 2002 in the SDSS (black), while the broad H$\beta$, as well as broad H$\alpha$, emission lines emerged in the spectra taken in 2015 in the LAMOST (blue). The residual spectrum (gray spectrum at the bottom of the right panel), between the bright epoch and faint epoch spectra, distinctly shows the emerging of broad H$\beta$ and H$\alpha$. The left panel in Figure \ref{fig:cl_example} shows the photometric data from several surveys, including SDSS, PS1, DELS, WISE, CRTS, and PTF. The imaging data taken in separated epochs help identify the variability of objects.

We fit the spectra of all the CL AGNs with a quasar spectral fitting procedure \citep[QSfit,][]{Calderone2017}, considering the contribution from the AGN continuum, the Balmer continuum, host galaxy, blended iron lines, broad emission lines, and narrow emission lines. The spectroscopic measurements are summarized in Table \ref{tab:spec}. The later epoch spectrum is calibrated to the earlier SDSS spectrum assuming a constant narrow emission line during a few years. Only in 4 out of these 21 objects, the S/N of the faint state spectrum is lower than that of the bright state spectrum. The weaker Balmer emission line intensity at the faint state is not a result of a worse observational condition.

According to the definition of AGN spectral types, we mark the spectral type as Type 1 when there are ($>5\sigma$) broad H$\beta$ components detected; and Type 2 when the S/N of broad H$\alpha$ is lower than 5. Specifically, when ($>5\sigma$) H$\alpha$ detected, we mark the spectal type as Type 1.9 (1.8 or 1.5) when there is no, $<1\sigma$, (weak, 1-3$\sigma$, or intermediate, 3-5$\sigma$) broad H$\beta$ detected.  As the Balmer emission lines disappeared in some spectra, we free the parameter limit of the amplitude of the line fitting, which was assumed to be positive. The fitting results are shown in Table \ref{tab:spec}. The negative luminosity means that there is absorption instead of emission lines. As shown in Table \ref{tab:spec}, six CL AGNs transited between Type 1 and Type 2 classes, ten CL AGNs transited between Type 1 and Type 1.9 classes. Besides, the fitting procedure recognizes $>5\sigma$ broad H$\beta$ emission lines in both bright and faint state spectra of J1118+3203 and J1152+3209, and $<1\sigma$ broad H$\beta$ lines in both bright and faint state spectra of J1552+2737. Therefore, these three AGNs fail to fit the quantitative definition above. While the dramatic H$\beta$ luminosity changes show that H$\beta$ faded in J1118+3203 and J1152+3209 (marked as ``off$^*$" in Table \ref{tab:spec}) and enhanced in J1152+3209 (marked as ``on$^*$" in Table \ref{tab:spec}). Among them, 15 AGNs turned on, with broad Balmer emission lines emerging (or enhancing), as well as increased broadband flux. The broad Balmer lines of 6 CL AGNs disappeared (or faded), accompanied by dimming in the continuum. Among the 21 CL AGNs, 17 CL AGNs showed broad H$\alpha$, H$\beta$, H$\gamma$, or even H$\delta$ appeared or disappeared. Only broad H$\alpha$ and H$\beta$ changes were detected in J0831+3646, J1319+6753, and J1552+2737. While obvious emerging of broad H$\alpha$ and H$\beta$ emission lines, and increased flux in the optical and mid-infrared confirmed their changes.

\begin{deluxetable*}{clrrrrrrrrrrrlll}
\rotate
\tablecaption{\ Spectral type transition of CL AGNs \label{tab:spec}}
\setlength{\tabcolsep}{3pt}
\tablehead{
\colhead{Name} &
\colhead{Redshift} &
\colhead{$\lambda L_{5100}$} &
\colhead{S/N$_1$} &
\colhead{S/N$_2$} &
\colhead{$L_{\rm H\beta, 1}$} & 
\colhead{$L_{\rm H\beta, 2}$} &
\colhead{$L_{\rm H\alpha, 1}$} & 
\colhead{$L_{\rm H\alpha, 2}$} &
\colhead{$\rm S/N_{H\beta_1}$} &
\colhead{$\rm S/N_{H\beta_2}$} &
\colhead{$\rm S/N_{H\alpha_1}$} &
\colhead{$\rm S/N_{H\alpha_2}$} &
\colhead{E1} &
\colhead{E2} &
\colhead{Type}\\
\colhead{} &
\colhead{} &
\colhead{($10^{42}$ erg s$^{-1}$)} &
\colhead{} &
\colhead{} &
\colhead{($10^{41}$ erg s$^{-1}$)} &
\colhead{($10^{41}$ erg s$^{-1}$)} &
\colhead{($10^{41}$ erg s$^{-1}$)} &
\colhead{($10^{41}$ erg s$^{-1}$)} &
\colhead{} &
\colhead{} &
\colhead{} &
\colhead{} &
\colhead{}
}
\startdata
J0831+3646 & 0.19501 & $18.23\pm0.24$ & 12.3 & 10.9 & $-4.05\pm0.33$ & $2.85\pm0.39$ & $3.60\pm1.00$ & $10.19\pm0.42$ & -12.1 & 7.3 & 3.6 & 24.2 & 2 & 1 & on\\
J0849+2747 & 0.29854 & $63.42\pm1.48$ & 9.3 & 2.7 & $8.56\pm0.39$ & $-2.81\pm2.03$ & $22.76\pm1.95$ & $2.49\pm0.97$ & 22.2 & -1.4 & 11.7 & 2.6 & 1 & 2 & off\\
J0909+4747 & 0.11694 & $6.97\pm0.20$ & 11.0 & 8.5 & $-1.88\pm0.13$ & $1.32\pm0.18$ & $1.48\pm0.13$ & $7.39\pm0.13$ & -14.9 & 7.5 & 11.8 & 57.1 & 1.9 & 1 & on\\
J0937+2602 & 0.16219 & $6.85\pm0.32$ & 20.9 & 7.1 & $-1.71\pm0.13$ & $1.75\pm0.20$ & $3.66\pm0.24$ & $11.93\pm2.04$ & -12.8 & 8.6 & 15.4 & 5.9 & 1.9 & 1 & on\\
J1003+3525 & 0.11886 & $45.77\pm1.64$ & 15.6 & 6.9 & $-1.40\pm0.12$ & $10.30\pm1.02$ & $2.34\pm0.57$ & $12.58\pm0.63$ & -11.7 & 10.1 & 4.1 & 19.9 & 2 & 1 & on\\
J1104+0118 & 0.57514 & $302.41\pm6.30$ & 7.0 & 0.7 & $79.89\pm5.19$ & $-43.38\pm13.61$ & $\cdots$ & $\cdots$ & 15.4 & -3.2 & $\cdots$ & $\cdots$ & 1 & 2 & off\\
J1104+6343 & 0.16427 & $4.46\pm0.49$ & 6.0 & 6.1 & $1.14\pm0.14$ & $0.33\pm0.11$ & $4.90\pm0.16$ & $0.79\pm0.13$ & 7.9 & 3.0 & 31.6 & 6.2 & 1 & 1.8 & off\\
J1110$-$0003 & 0.21922 & $48.61\pm1.34$ & 8.0 & 6.1 & $0.27\pm0.10$ & $6.70\pm0.92$ & $1.78\pm0.71$ & $11.98\pm1.01$ & 2.8 & 7.3 & 2.5 & 11.9 & 2 & 1 & on\\
J1115+0544 & 0.08995 & $17.27\pm0.27$ & 18.9 & 8.7 & $-1.23\pm0.07$ & $2.46\pm0.18$ & $0.05\pm0.07$ & $7.92\pm1.11$ & -18.4 & 14.0 & 0.7 & 7.2 & 2 & 1 & on\\
J1118+3203 & 0.3651 & $56.48\pm2.29$ & 4.9 & 4.9 & $11.76\pm1.11$ & $2.90\pm0.51$ & $\cdots$ & $10.31\pm0.57$ & 10.6 & 5.7 & $\cdots$ & 18.0 & 1 & 1 & off$^*$ \\
J1132+0357 & 0.09089 & $16.26\pm0.38$ & 17.3 & 7.4 & $-3.23\pm0.08$ & $4.74\pm0.40$ & $0.78\pm0.05$ & $2.40\pm0.17$ & -40.2 & 11.8 & 15.9 & 14.1 & 1.9 & 1 & on\\
J1150+3632 & 0.34004 & $39.01\pm2.28$ & 4.9 & 5.2 & $5.02\pm0.53$ & $-3.29\pm0.49$ & $22.24\pm1.25$ & $3.18\pm0.40$ & 9.4 & -6.7 & 17.8 & 7.9 & 1 & 1.9 & off\\
J1152+3209 & 0.37432 & $138.05\pm2.68$ & 11.9 & 3.7 & $38.31\pm1.16$ & $10.53\pm1.29$ & $\cdots$ & $\cdots$ & 33.0 & 8.2 & $\cdots$ & $\cdots$ & 1 & 1 & off$^*$ \\
J1259+5515 & 0.19865 & $17.95\pm0.91$ & 7.4 & 3.5 & $0.02\pm0.24$ & $5.53\pm0.82$ & $4.30\pm0.38$ & $9.21\pm3.88$ & 0.1 & 6.8 & 11.4 & 2.4 & 1.9 & 1 & on\\
J1319+6753 & 0.16643 & $30.90\pm1.46$ & 12.1 & 7.3 & $0.36\pm0.33$ & $7.96\pm1.16$ & $2.37\pm0.27$ & $5.90\pm4.43$ & 1.1 & 6.8 & 8.6 & 1.3 & 1.9 & 1 & on\\
J1358+4934 & 0.11592 & $5.50\pm0.19$ & 7.8 & 13.1 & $0.36\pm0.09$ & $0.96\pm0.10$ & $0.74\pm0.06$ & $1.98\pm0.10$ & 3.8 & 9.2 & 13.4 & 19.0 & 1.9 & 1 & on\\
J1447+2833 & 0.16344 & $66.68\pm0.97$ & 7.8 & 13.1 & $-1.16\pm0.29$ & $2.49\pm0.24$ & $2.27\pm0.19$ & $11.63\pm0.25$ & -3.9 & 10.6 & 12.2 & 47.2 & 1.9 & 1 & on\\
J1533+0110 & 0.14268 & $<1.00$ & 14.2 & 13.4 & $-3.98\pm0.17$ & $0.54\pm0.12$ & $1.00\pm0.15$ & $3.97\pm0.29$ & -23.6 & 4.4 & 6.9 & 13.7 & 1.9 & 1.5 & on\\
J1545+2511 & 0.11696 & $6.81\pm0.06$ & 19.6 & 17.1 & $-2.16\pm0.15$ & $0.71\pm0.11$ & $1.69\pm0.12$ & $5.90\pm0.14$ & -14.4 & 6.7 & 14.4 & 41.3 & 1.9 & 1 & on\\
J1550+4139 & 0.22014 & $28.26\pm1.30$ & 11.6 & 8.1 & $-4.41\pm0.47$ & $15.71\pm1.48$ & $2.47\pm0.40$ & $14.01\pm0.89$ & -9.5 & 10.6 & 6.2 & 15.8 & 1.9 & 1 & on\\
J1552+2737 & 0.08648 & $<1.00$ & 12.3 & 7.8 & $-3.67\pm0.05$ & $-1.34\pm0.10$ & $0.38\pm0.04$ & $2.70\pm0.05$ & -69.2 & -13.9 & 9.8 & 59.1 & 1.9 & 1.9 & on$^*$ \\
\enddata
\tablecomments{$\lambda L_{5100}$ is the continuum luminosity at 5100 ${\rm \AA}$ in the bright epoch spectrum. J1533+0110 and J1552+2737 are too red to fit a power-law continuum, with a upper limit of $1.00\times10^{42}$ erg s$^{-1}$. S/N$_1$ and S/N$_2$ are the median S/N pixel$^{-1}$ of the former and recent epoch spectra, respectively. $L_{\rm H\beta, 1}$. ($L_{\rm H\alpha, 1}$) and $L_{\rm H\beta, 2}$ ($L_{\rm H\alpha, 2}$) are the luminosities of broad H$\beta$ (H$\alpha$) component in the former and recent epoch spectra. The negative luminosity means that there is absorption instead of emission lines. There is no H$\alpha$ data when H$\alpha$ moves out of the range of the spectrum. $\rm S/N_{H\beta_1}$ ($\rm S/N_{H\alpha_1}$) and $\rm S/N_{H\beta_2}$ ($\rm S/N_{H\alpha_2}$) are the S/N of broad H$\beta$ (H$\alpha$) component in the former and recent epoch spectra. E1 and E2 describe the spectral types of the former and recent epoch spectra. Type describes the transition type.}
\end{deluxetable*}
\clearpage

\begin{deluxetable*}{clrrrrrrc}
\tablecaption{\ Variability of Changing-look AGNs \label{tab:photometry}}
\tablewidth{0pt}
\tablehead{
\colhead{Name} &
\colhead{Survey} &
\colhead{$\Delta g_{\rm phot}$} &
\colhead{$\Delta (g-r)_{\rm phot}$} &
\colhead{$\Delta g_{\rm spec}$} &
\colhead{$\Delta(g-r)_{\rm spec}$} &
\colhead{$\Delta W1$} &
\colhead{$\Delta(W1-W2)$} &
\colhead{$\delta_{\rm CRTS}$}\\
\colhead{} & \colhead{} &
\colhead{(mag)} & \colhead{(mag)} & \colhead{(mag)} & \colhead{(mag)} & \colhead{(mag)} & \colhead{(mag)} & \colhead{(mag)}
}
\startdata
J0831+3646 & DELS & $-0.41\pm0.01$ & $-0.35\pm0.02$ & $-0.57\pm0.37$ & $-0.48\pm0.49$ & $-0.35\pm0.10$ & $0.13\pm0.14$ & 1.15\\
J0849+2747 & DELS & $0.31\pm0.01$ & $0.29\pm0.02$ & $1.00\pm0.65$ & $0.69\pm0.80$ & $0.28\pm0.13$ & $-0.24\pm0.21$ & 1.17\\
J0909+4747 & PS1 & $0.06\pm0.01$ & $-0.03\pm0.02$ & $-0.71\pm0.29$ & $-0.39\pm0.39$ & $-0.42\pm0.10$ & $0.10\pm0.18$ & 0.90\\
J0937+2602 & DELS & $-$0.17$\pm$0.01 & $-$0.06$\pm$0.01 & $-$0.29$\pm$0.32 & $-$0.24$\pm$0.44 & $-$0.41$\pm$0.10 & 0.07$\pm$0.15 & 0.92\\
J1003+3525 & DELS & $-$0.52$\pm$0.01 & $-$0.27$\pm$0.01 & $-$1.17$\pm$0.21 & $-$0.54$\pm$0.29 & $-$0.58$\pm$0.16 & 0.16$\pm$0.26 & 0.92\\
J1104+0118 & DELS & 1.89$\pm$0.03 & 1.48$\pm$0.04 & 6.14$\pm$111.71 & 4.98$\pm$111.72 & 0.68$\pm$0.11 & $-$0.56$\pm$0.26 & 1.68\\
J1104+6343 & PS1 & 0.55$\pm$0.02 & 0.44$\pm$0.03 & 0.08$\pm$0.46 & 0.11$\pm$0.53 & 0.31$\pm$0.24 & 0.03$\pm$0.36 & 0.93\\
J1110$-$0003 & DELS & $-$0.42$\pm$0.02 & $-$0.33$\pm$0.02 & $-$0.87$\pm$0.40 & $-$0.43$\pm$0.53 & $-$0.65$\pm$0.12 & 0.11$\pm$0.27 & 0.95\\
J1115+0544 & DELS & $-$0.21$\pm$0.01 & $-$0.11$\pm$0.01 & $-$0.82$\pm$0.22 & $-$0.44$\pm$0.30 & $-$1.01$\pm$0.10 & 0.59$\pm$0.16 & 0.97\\
J1118+3203 & PS1 & 0.67$\pm$0.04 & 0.49$\pm$0.05 & 0.80$\pm$0.33 & 0.30$\pm$0.43 & 0.15$\pm$0.41 & 0.20$\pm$0.58 & 1.50\\
J1132+0357 & PS1 & $-$0.11$\pm$0.01 & $-$0.15$\pm$0.01 & $-$1.16$\pm$0.19 & $-$0.55$\pm$0.26 & $-$0.56$\pm$0.05 & 0.19$\pm$0.12 & 0.91\\
J1150+3632 & DELS & 1.59$\pm$0.02 & 0.92$\pm$0.03 & 0.95$\pm$0.41 & 0.50$\pm$0.47 & 0.30$\pm$0.24 & 0.18$\pm$0.37 & 1.38 \\
J1152+3209 & PS1 & 1.36$\pm$0.03 & 0.99$\pm$0.03 & 1.39$\pm$0.60 & 0.84$\pm$0.76 & 0.19$\pm$0.08 & $-$0.27$\pm$0.13 & $\cdots$\\
J1259+5515 & DELS & $-$0.06$\pm$0.02 & $-$0.05$\pm$0.02 & $-$0.77$\pm$0.93 & $-$0.40$\pm$0.99 & $-$0.54$\pm$0.10 & 0.26$\pm$0.26 & 1.06\\
J1319+6753 & PS1 & $-$0.28$\pm$0.01 & $-$0.26$\pm$0.01 & $-$0.46$\pm$0.40 & $-$0.17$\pm$0.47 & $-$0.23$\pm$0.07 & 0.10$\pm$0.12 & 0.70\\
J1358+4934 & DELS & 0.41$\pm$0.01 & 0.20$\pm$0.02 & $-$0.46$\pm$0.20 & $-$0.15$\pm$0.24 & 0.00$\pm$0.16 & 0.00$\pm$0.23 & 0.80\\
J1447+2833 & PS1 & $-$0.15$\pm$0.01 & $-$0.18$\pm$0.01 & $-$0.72$\pm$0.20 & $-$0.46$\pm$0.29 & $-$0.45$\pm$0.05 & 0.06$\pm$0.08 & 0.85\\
J1533+0110 & DELS & $-$0.14$\pm$0.01 & $-$0.05$\pm$0.01 & $-$0.28$\pm$0.18 & $-$0.21$\pm$0.20 & $-$0.12$\pm$0.06 & 0.05$\pm$0.15 & 1.13\\
J1545+2511 & PS1 & $-$0.16$\pm$0.01 & $-$0.14$\pm$0.01 & $-$0.16$\pm$0.48 & 0.03$\pm$0.52 & $-$0.10$\pm$0.04 & 0.14$\pm$0.07 & 0.84\\
J1550+4139 & DELS & $-$0.25$\pm$0.01 & $-$0.13$\pm$0.02 & $-$0.55$\pm$0.49 & $-$0.24$\pm$0.57 & $-$0.34$\pm$0.06 & 0.12$\pm$0.15 & 1.14\\
J1552+2737 & PS1 & 0.03$\pm$0.01 & $-$0.26$\pm$0.01 & $-$0.45$\pm$0.36 & $-$0.37$\pm$0.47 & $-$0.25$\pm$0.08 & 0.12$\pm$0.17 & 0.89\\
\tableline
J0126$-$0839 & PS1 & 0.43$\pm$0.01 & 0.15$\pm$0.02 & 0.29$\pm$0.18 & 0.25$\pm$0.21 & 0.00$\pm$0.09 & 0.00$\pm$0.27 & 0.91 \\
J0159+0033 & DELS & 0.28$\pm$0.02 & 0.22$\pm$0.02 & 1.15$\pm$0.24 & 0.60$\pm$0.28 & 0.12$\pm$0.21 & $-$0.01$\pm$0.41 & 0.95\\
J1011+5442 & PS1 & 1.04$\pm$0.02 & 0.53$\pm$0.02 & 1.91$\pm$0.12 & 0.46$\pm$0.15 & 1.30$\pm$0.21 & $-$0.50$\pm$0.36 & 0.86\\
J1554+3629 & DELS & $-0.56\pm0.01$ & $-0.42\pm0.02$ & $-1.37\pm0.39$ & $-0.67\pm0.49$ & $-0.74\pm0.08$ & $0.16\pm0.17$ & 1.22\\
J2336+0017 & DELS & 0.41$\pm$0.02 & 0.30$\pm$0.02 & 0.41$\pm$0.40 & 0.22$\pm$0.45 & 0.14$\pm$0.28 & $-$0.17$\pm$0.42 & 1.06 \\
\enddata
\tablecomments{$\Delta g_{\rm phot}$ and $\Delta (g-r)_{\rm phot}$ are the $g$-band variability and $g-r$ color variability from imaging data, while $\Delta g_{\rm spec}$ and $\Delta(g-r)_{\rm spec}$ are from spectrophotometry.}
\end{deluxetable*}

\begin{figure}[htbp]
\hspace*{-0.5cm}
\epsscale{1.1}
\plotone{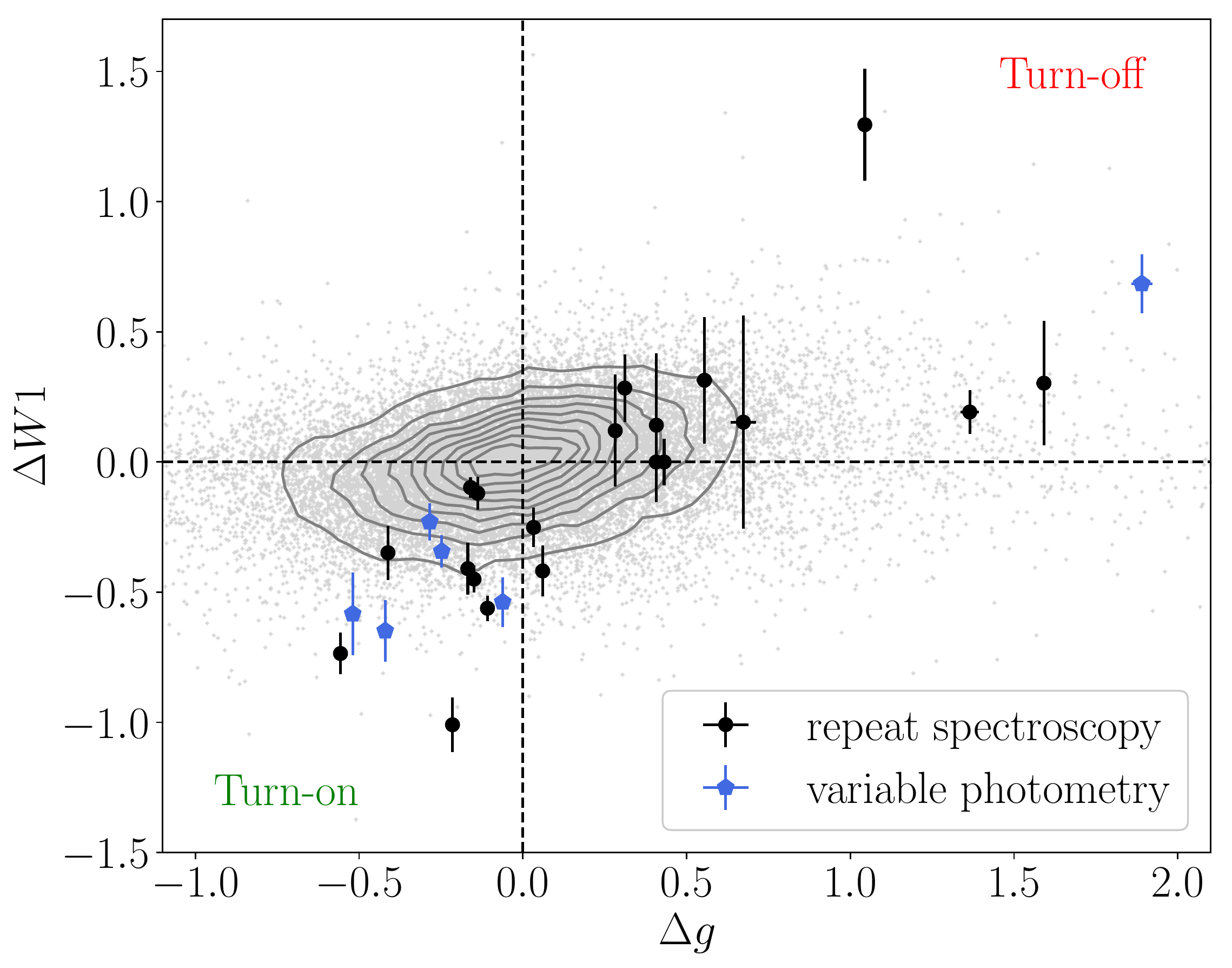}
\vspace{-0.0cm}
\caption{\label{fig:variability} The variability of CL AGNs in optical $g$-band and mid-infrared $W1$-band. The broadband flux in optical and mid-infrared have the same trend. The CL AGNs marked as black dots are from repeat spectroscopy, and the blue pentagon shows CL AGNs selected from variable photometry. The left bottom region, where both $g$ and $W1$ band became brighten, is a region for turn-on CL AGN selection; the top right area, where both $g$ and $W1$ dimmed, is useful for turn-off CL AGN selection. The gray dots and the contours shows quasars with similar redshift range, $z<0.7$.}
\end{figure}

\subsection{CL AGNs Optical and Mid-infared Variability}
The continuum flux varies along with the CL AGN type transition. Table \ref{tab:photometry} summaries the photometric variability of CL AGNs. We obtain the $g$-band variability, $\Delta g$, from the magnitude offset between PS1 $g$-band magnitude, or the DELS $g$-band magnitude when available, and the SDSS $g$-band magnitude. We use the first epoch SDSS photometry if there is data at more than one epoch. The uncertainties of $\Delta g$ are calculated from propagation of SDSS $g$-band magnitude uncertainty and PS1 $g$-band magnitude uncertainty, or the DELS $g$-band magnitude uncertainty when available. Similarly, the $g-r$ color variation, $\Delta (g-r)$, is calculated by the color offset between PS1 $g-r$ color, or the DELS $g-r$ color when available, and the SDSS $g-r$ color. The $g$-band variability ranges from $-$1.89 to 0.52 mag in this sample. As in some CL AGN cases, the PS1 or DELS images were taken before the type transition. We also calculate the spectrophotometry with the spectra convolved with the SDSS filters. The spectrophotometry is calibrated by setting the first epoch SDSS spectrophotometry equal to the SDSS photometry at the closest epoch. The variations of spectrophotometry in $g$-band and $g-r$ color are also listed in Table \ref{tab:photometry}. In the mid-infrared, there is a series of exposures within one day with a long interval of half a year in WISE multi-epoch data. We calculate the 3$\sigma$-clipped mean magnitude of WISE in $W1$ and $W2$ every half a year. The uncertainties of $W1$ and $W2$ are obtained from the 3$\sigma$-clipped standard deviation of $W1$ and $W2$ magnitude every half a year. There are less than ten epochs of WISE mean magnitude for a single object. For the turn-on objects, the WISE variability is calculated from the offset between the brightest epoch photometry and the first epoch photometry. For turn-off objects, the WISE variability is calculated from the offset between the last epoch photometry and the brightest epoch photometry. The magnitude and color variability is calculated between the later epoch and the former epoch. Therefore, the magnitude variability is negative if an object becomes brighter, and the color variability is negative if the object becomes bluer. Figure \ref{fig:variability} shows the optical broadband flux changes along with mid-infrared flux. In our sample, the mid-infrared variability $\Delta W1$ ranges from $-$1.01 to 0.68, and 15 of them varied for more than 0.2 mag ($|\Delta W1|>0.2$ mag). The mid-infrared flux is not significantly affected by dust extinction \citep{Weingartner2001}. In the scenario of variable obscuration, the variation in $W1$ band due to dust extinction yields a factor of $\sim 21$ variability in $g$-band magnitude, according to the extinction curve in optical and mid-infrared even considering micrometer-sized grains \citep{Wang2015}. A variability of 0.2 mag in $W1$ band expects approximately 4.2 mag variability in $g$ band. The optical variability, from photometric and spectrophotometric data, is not consistent with the scenario of variable obscuration in 10 CL AGNs in our sample at more than $3\sigma$ confidence level (J0831+3646, J0937+2602, J1003+3525, J1104+0118, J1110$-$0003, J1115+0544, J1132+0357, J1259+5515, J1447+2833, and J1550+4139) and in 8 CL AGNs between $1\sigma$ and $3\sigma$ confidence level (J0849+2747, J1104+6343, J1150+3632, J1152+3209, J1319+6753, J1533+0110, J1545+2511, and J1552+2737). The accuracy of spectrophotometry is lower than that of photometry. Following-up high accuracy photometric data after the type transition can better constrain the mechanism of type transition. Using mid-infrared variability to select CL AGNs, we were inclined to select CL AGNs with intrinsic changes instead of varying obscuration.

\section{Discussion} \label{sec:discussion}

\subsection{CL AGNs Color Variability} \label{sec:color}
\begin{figure*}[htbp]
  \centering
  \hspace{0cm}
  \subfigure{
  \includegraphics[width=3.2in]{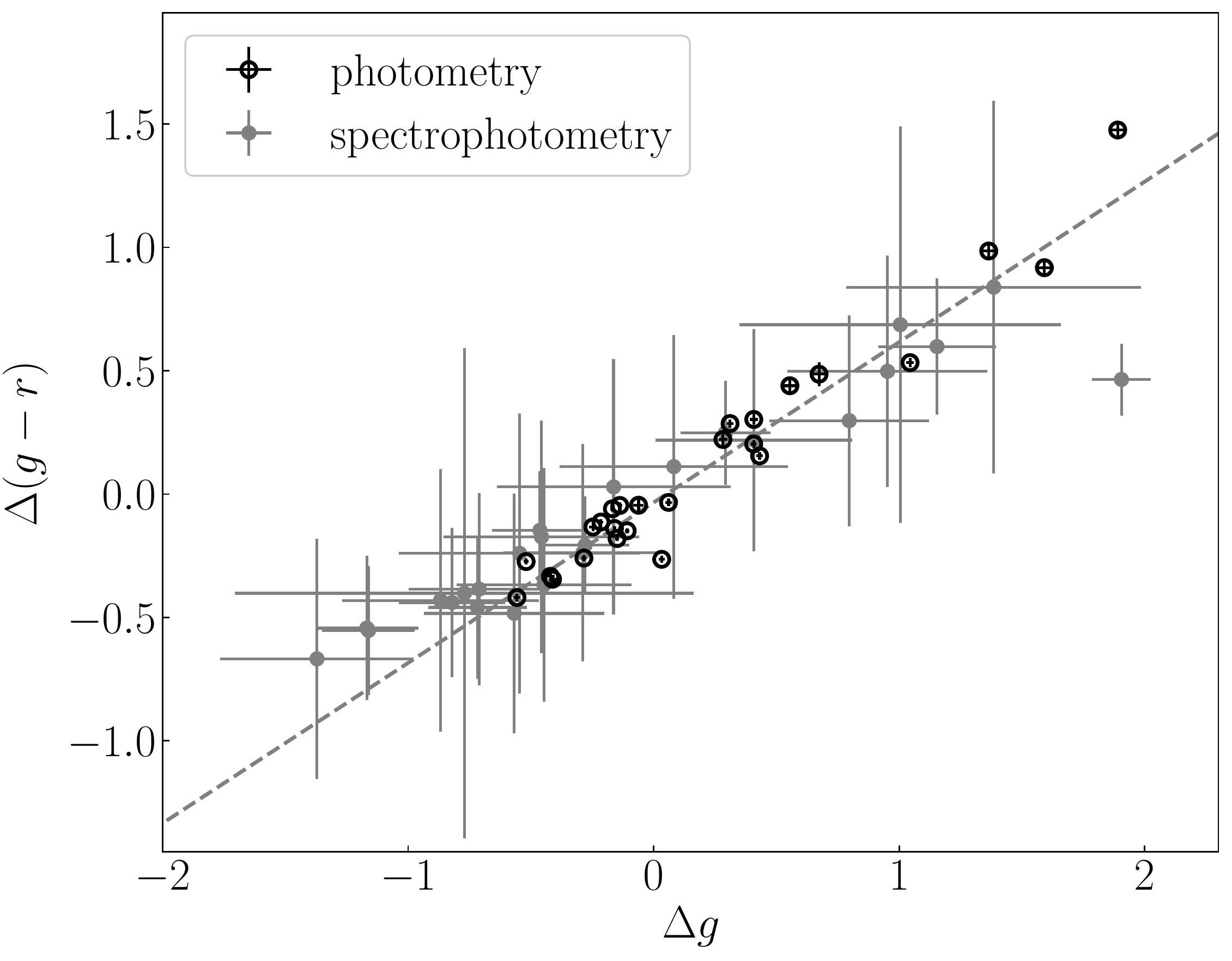}}
 \hspace{-0cm}
 \subfigure{
  \includegraphics[width=3.2in]{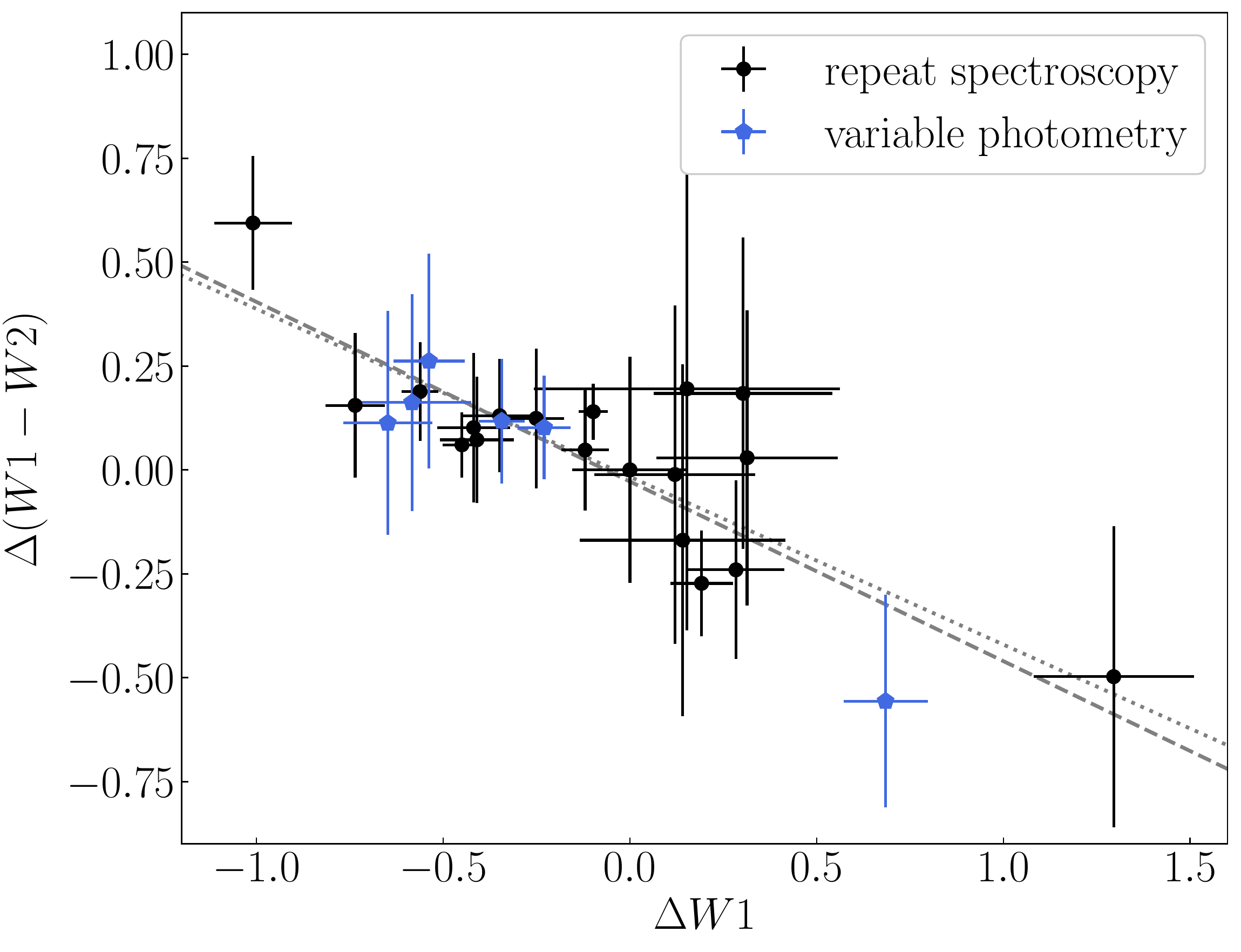}}\\
 \vspace{-0cm}
\caption{\label{fig:color} The color variability versus magnitude variability of CL AGNs in optical (left panel) and mid-infrared (right panel). A bluer-when-brighter chromatism is confirmed in the optical. However, the mid-infrared $W1-W2$ color is redder when brighter. The opposite color change trend in the mid-infrared is possibly due to a stronger contribution from the AGN dust torus when the AGN turns on.}
\end{figure*}

The optical and mid-infrared colors vary following flux changes. Figure \ref{fig:color} shows the color variability versus the magnitude variability in optical (left panel) and mid-infrared (right panel). The bluer-when-brighter trend is a well known trend in the optical \citep[e.g.,][]{Wilhite2005, Schmidt2012, Zuo2012, Ruan2014}. In the mid-infrared, there is a trend that $W1-W2$ is redder when brighter. A similar $W1-W2$ color dependence on AGN luminosity is reported in the Swift/Burst Alert Telescope  AGN sample \citep{Ichikawa2017}.

\begin{figure*}[htbp]
 \centering
 \subfigure{
  \includegraphics[width=3.2in]{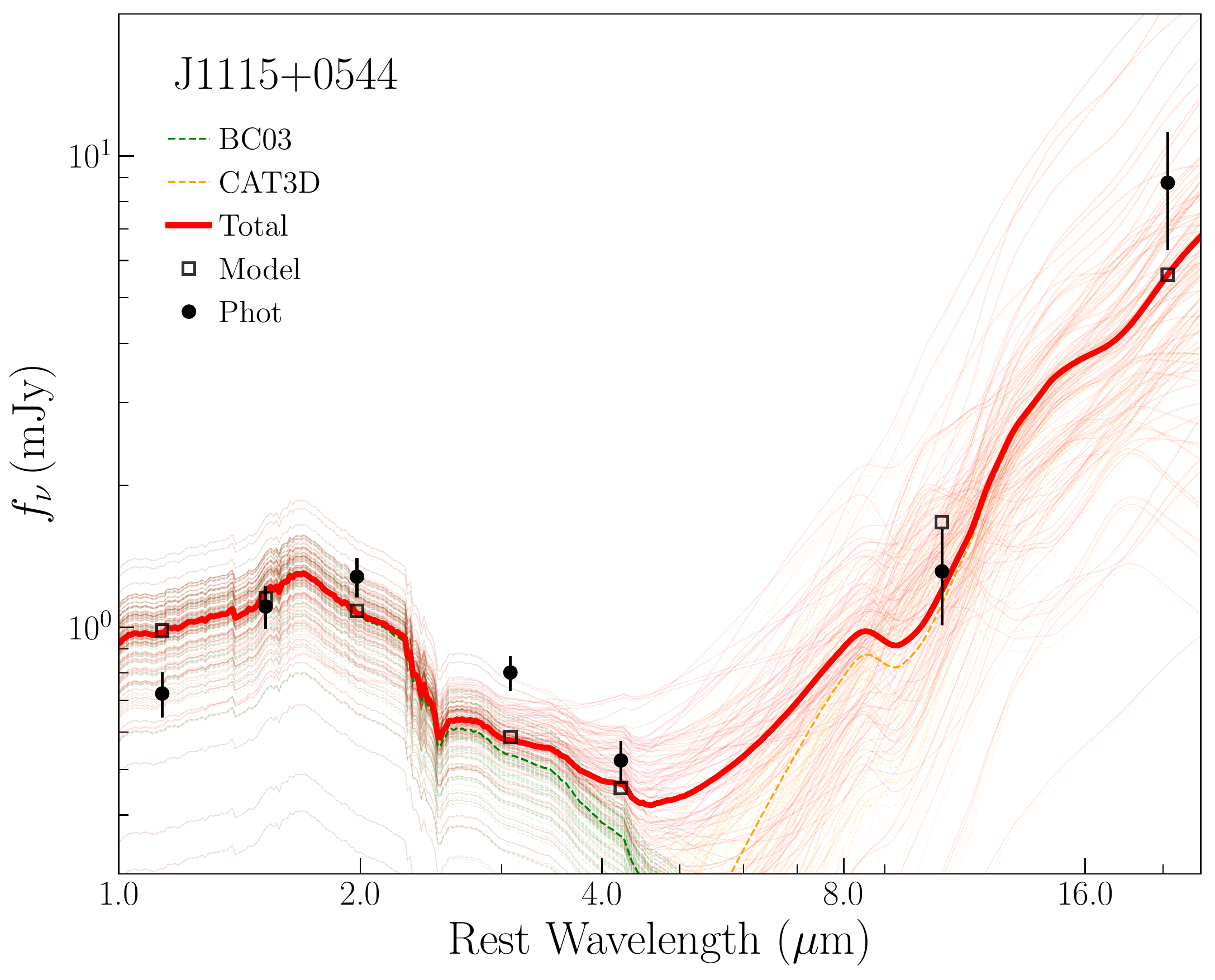}}
 \subfigure{
  \includegraphics[width=3.2in]{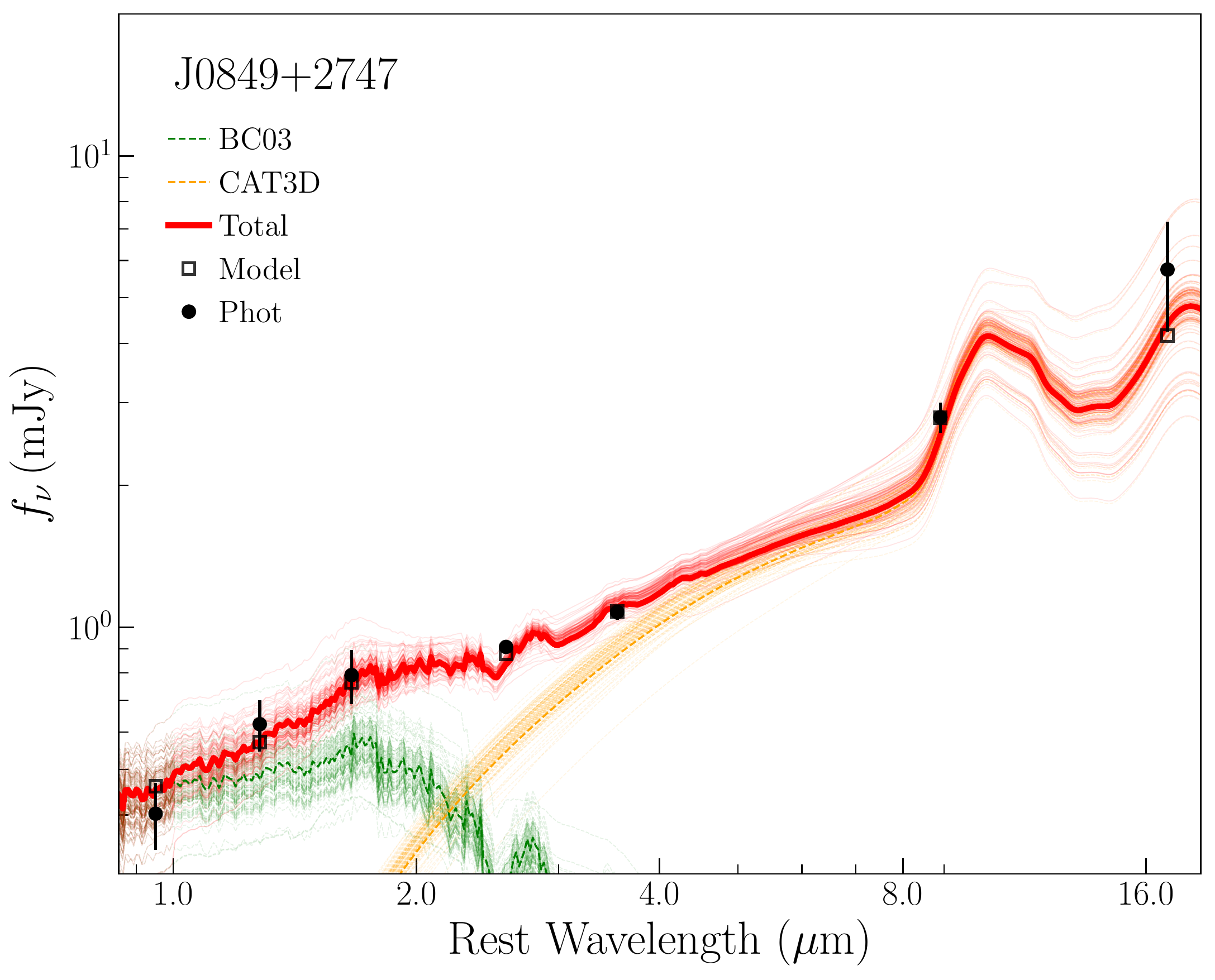}}\\
  \caption{\label{fig:sed} The infrared SED of two CL AGNs, J1115+0544 (left panel) and J0849+2747 (right panel). The green, red, and brown dashed lines are the BC03, CAT3D, and the combined models, while the light dotted lines are 100 random samples of the final model and each component from the MCMC method to demonstrate the uncertainty of the fitting. The black circles with errorbars are the observed data (2MASS $J$, $H$, $K_s$, and WISE $W1$, $W2$, $W3$, $W4$ data from short to long wavelength). The empty squares are the modeled photometric data at the observed bands.}
\end{figure*}

\begin{deluxetable*}{clccrrrr}[htbp]
\tablecaption{\ Timescale of Changing-look AGN Transition \label{tab:time}}
\tablewidth{0pt}
\tablehead{
\colhead{Name} &
\colhead{Redshift} &
\colhead{Epoch1} &
\colhead{Epoch2} &
\colhead{$\Delta t_{\rm spec}$} &
\colhead{Epoch1$^*$} &
\colhead{Epoch2$^*$} &
\colhead{$\Delta t$} \\
\colhead{} & \colhead{} &
\colhead{} & \colhead{} &
\colhead{(yr)} & \colhead{} & \colhead{} & \colhead{(yr)}
}
\startdata
J0831+3646 & 0.19501 & 52312 & 57367 & 11.6 & PS1 56244 & $\cdots$ & 2.6 \\
J0849+2747 & 0.29854 & 53350 & 56628 & 6.9  &  $\cdots$ & $\cdots$  &  6.9 \\
J0909+4747 & 0.11694 & 52620 & 57745 & 12.6 &  $\cdots$ & $\cdots$  & 12.6  \\
J0937+2602 & 0.16219 & 54524 & 57369 & 6.7  & PS1 55845 & $\cdots$  & 3.6 \\
J1003+3525 & 0.11886 & 53389 & 57867 & 11.0 & PS1 55726 & $\cdots$  & 5.2 \\
J1104+0118 & 0.57514 & 52374 & 57867 & 9.6  & $\cdots$  & PS1 55902 & 6.1 \\
J1104+6343 & 0.16427 & 52370 & 54498 & 5.0 &  $\cdots$ & $\cdots$  &  5.0 \\
J1110$-$0003 & 0.21922 & 51984 & 57864 & 13.2 &   & WISE 57003 & 11.3 \\
J1115+0544 & 0.08995 & 52326 & 57393 & 12.7 & WISE 57002 & WISE 57367 & 0.9 \\
J1118+3203 & 0.3651  & 53431 & 56367 & 5.9 & $\cdots$  & $\cdots$  & 5.9  \\
J1132+0357 & 0.09089 & 52642 & 57392 & 11.9 & CRTS 56665 & $\cdots$  & 1.8 \\
J1150+3632 & 0.34004 & 53436 & 57422 & 8.1 & $\cdots$  & $\cdots$  & 8.1  \\
J1152+3209 & 0.37432 & 53446 & 57844 & 8.8 & $\cdots$  & PS1 56063 & 5.2 \\
J1259+5515 & 0.19865 & 52707 & 57863 & 11.8 & WISE 55534 & WISE 56806 & 2.9 \\
J1319+6753 & 0.16643 & 51988 & 57867 & 13.8 & $\cdots$  & WISE 56982 & 11.7 \\
J1358+4934 & 0.11592 & 53438 & 54553 & 2.7 & $\cdots$  & $\cdots$  &  2.7 \\
J1447+2833 & 0.16344 & 53764 & 57071 & 7.8 &  $\cdots$ &  $\cdots$ &  7.8 \\
J1533+0110 & 0.14268 & 51989 & 54561 & 6.2 & $\cdots$  &  $\cdots$ &  6.2 \\
J1545+2511 & 0.11696 & 53846 & 57891 & 9.9 &  $\cdots$ & $\cdots$  &  9.9 \\
J1550+4139 & 0.22014 & 52468 & 57864 & 12.1 & PS1 56233  & $\cdots$  & 3.7 \\
J1552+2737 & 0.08648 & 53498 & 56722 & 8.1 & PS1 55987  & $\cdots$  & 1.9 \\
\enddata
\tablecomments{$\Delta g_{\rm phot}$ and $\Delta (g-r)_{\rm phot}$ are the $g$-band variability and $g-r$ color variability from imaging data, while $\Delta g_{\rm spec}$ and $\Delta(g-r)_{\rm spec}$ are from spectrophotometry.}
\end{deluxetable*}

The global SED variation helps to understand the color changes in the infrared. We construct the infrared SED using the WISE data and the near-infrared data from the Two Micron All Sky Survey \citep[2MASS;][]{Skrutskie2006}, which scanned the entire sky from 1997 to 2001 in three bands $J$ (1.25 $\mu$m), $H$ (1.65 $\mu$m) and $K_s$ (2.17 $\mu$m). As there is no multi-epoch near-infrared and WISE $W3$ and $W4$ data, we show examples of two different objects in Figure \ref{fig:sed}. The two objects are J1115+0544 (left panel) at faint state before it turned on and J0849+2747 (right panel) before it turned off.

We fit the SED with the stellar and dust torus emission components. The simple stellar synthesis model by \citet[][BC03]{Bruzual2003} with age 5 Gyr and solar metallicity is used for the stellar emission. The dust torus component is described by a new version of the radiative transfer model, CAT3D \citep{Honig2017}. Due to the limitation of the data, we adopt the templates without the wind component. We perform the SED fitting with a new Markov chain Monte Carlo (MCMC) method \citep{Shangguan2018} combining the stellar and dust torus models. 

In the SED before the AGN turned on, the $W1$ and $W2$ bands generally follow the stellar emission. In the SED after the AGN turned on, the $W1$ and $W2$ bands, especially $W2$, was strongly effected by the dust torus radiation. Therefore, it is possible that the $W1-W2$ shows as blue without strong AGN contribution, and turns to be red after the AGN turning on. Thus the mid-infrared $W1-W2$ color of CL AGNs turns to be redder when brighter. 

In the optical, the color-magnitude variation relation (dashed line in left panel in Figure \ref{fig:color}) of CL AGNs (listed in Table \ref{tab:cl}) is fitted by a least square fitting algorithm as\\
\begin{equation} \label{eq:1}
 \Delta(g-r) = (0.650\pm0.042) \Delta g - (0.034\pm0.017).
\end{equation}

The CL AGNs selected from repeat spectroscopy are not biased by WISE variability selection criteria. For CL AGNs selected from repeat spectroscopy, the mid-infrared color-magnitude relation (dotted line in left panel in Figure \ref{fig:color}) is fitted as\\
\begin{equation} \label{eq:2}
 \Delta(W1-W2) = (-0.404\pm0.078)\Delta W1 - (0.016\pm0.033).
\end{equation}
Thus a selection criterion based on mid-infrared variability as $|\Delta (W1-W2)|>0.1$ when $|\Delta W1|>0.2$ is reasonable. With all the CL AGNs (in Table \ref{tab:cl}), the color-magnitude relation in the mid-infrared (dashed line in the left panel in Figure \ref{fig:color}) is fitted as\\
\begin{equation} \label{eq:3}
 \Delta(W1-W2) = (-0.432\pm0.066)\Delta W1 - (0.028\pm0.028).
\end{equation}
The CL AGNs, selected from WISE variability, obey the color-magnitude relation well. The mid-infrared variability and color variability are good tracers for CL AGNs with intrinsic changes.

\subsection{The Timescale of the Type Transition}
The current data coverage is not good enough to measure the timescale of transition. Therefore, we only roughly estimate the upper limit of type transition as follows. The upper limit of the transition timescale is obtained by the time interval between the former and recent spectroscopic epoch ($\Delta t_{\rm spec}$ in the rest frame in Table \ref{tab:time}). We further prescribe a limit to the timescale with the light curve changes approaching the transition. For example, although the two spectra of J1115+0544 were separated by 14 years, there was no variability detected in SDSS, PS1, and CRTS imaging data. The rapid flux brightening was detected by WISE in 2015. The $W1$ of J1115+0544 increased for more than 1 mag in a short interval, less than one year in the rest frame. The light curves of CL AGNs keep quiescent at faint state, namely before the transition of turn-on CL AGNs or after the transition of turn-off CL AGNs. The CL AGNs usually vary in their AGN phase. In some cases (for example J0831+3646, J1110$-$0003, J1115+0544, J1259+5515, and J1319+6753), the WISE flux increased when the AGN turned on, and decreased with a smaller amplitude later on, possibly due to the variability of AGN accretion rate. On the other hand, it is also possible that the flux increase is due to the TDEs instead of AGN activity. In this scenario, the later-on light curve should continually decrease following the typical light curve of TDEs, which is proportional to $t^{-5/3}$ \citep{Rees1988, Lodato2009, Guillochon2013}. Following-up photometric data is needed to distinguish the scenarios of AGN accretion rate variability and TDEs. The epoch approaching the type transition is obtained by a recent quiescent image epoch at faint state, or the brightest image epoch in their AGN phase. In Table \ref{tab:time}, Epoch1$^*$ and Epoch2$^*$ show the imaging data epoch (and imaging survey) approaching the transition before and after the type transition. $\Delta t$, obtained from photometric variation approaching the type transition when available, is a better upper limit of the transition timescale in the rest frame than $\Delta t_{\rm spec}$. Therefore, the upper limit of the type transition timescales ranges from 0.9 to 12.6 years in the rest frame.

The short transition timescale is not consistent with the scenario of variable obscuration \citep[e.g.,][]{LaMassa2015, Gezari2017, Sheng2017}. In a scenario of variable accretion rate, the CL timescale in this sample is much shorter than the inflow timescale  of gas in the inner parts of the accretion disk, as discussed in previous works \citep{LaMassa2015, MacLeod2016, Runnoe2016, Gezari2017}. So modeling CL AGNs via changes in accretion rates is far from a settled matter. A detail analysis on the mechanisms of the type transition will be present in a subsequent paper.

\section{Sumary} \label{sec:summary}
We present surveys of CL AGNs in the SDSS spectra archive, the LAMOST spectra archive, and observation for some CL AGNs candidates selected from photometric data. In total, we discover 21 new CL AGNs at $0.08<z<0.58$. Among the new CL AGNs, five were found by repeat spectra from the SDSS, ten were discovered from repeat spectra in the SDSS and the LAMOST, and six were selected from photometric variability and confirmed by new spectroscopy. From our surveys, approximately 0.006\% (0.007\%) galaxies with repeat spectroscopy in the SDSS (SDSS and LAMOST) are CL AGNs, with obvious broad Balmer emission lines changes.

The physical mechanism of type transition is important for understanding the evolution of AGNs. The mid-infrared flux changes with the optical continuum flux. Variations of more than 0.2 mag in the mid-infrared were detected in 15 CL AGNs during the transition, and such variability expects $\sim$ 4.2 mag variability in $g$ band. In the scenario of variable obscuration, the variation in $W1$ band due to dust extinction yields a factor of $\sim 21$ variability in $g$ band. The optical variability is not consistent with the scenario of variable obscuration in 10 CL AGNs in our sample at more than $3\sigma$ confidence level and in 8 CL AGNs between $1\sigma$ and $3\sigma$ confidence level. Following-up high accuracy photometric data after the type transition can better constrain the mechanism of type transition.

The optical and mid-infrared colors change with flux variation. A bluer-when-brighter chromatism is confirmed in optical bands. However, the mid-infrared $W1-W2$ color is redder when brighter. The opposite color change trend in the mid-infrared is possibly caused by a stronger contribution from the AGN dust torus when the AGNs turn on. The mid-infrared variability and colors are good tracers for CL AGNs with intrinsic variability.

The upper limit of type transition timescales ranges from 0.9 to 12.6 years in the rest frame. The mid-infrared emission of J1115+0544 increased for more than 1 mag in a short interval, less than one year in the rest frame. The timescale of type transition will help distinguish the mechanism of the changes. Following-up photometric data is needed to distinguish the scenarios of AGN accretion rate variability and TDEs based on the decreasing light curve of TDEs ($\sim t^{-5/3}$).

The photometric variability of CL AGNs provide ways to select CL AGNs from large area surveys. In the future, the Large Synoptic Survey Telescope \citep[LSST;][]{Ivezic2008}, with multi-epoch and multi-band data, will provide powerful data for CL AGN selection and monitoring.

There are some future works in our subsequent papers. We will analyze the spectra and imaging data in more detail, and discuss their probable transition mechanism. The rare CL AGNs provide exceptional cases for M-$\sigma_*$ relation studies at higher redshift with faint-state spectra and AGN-phase spectra. We plan to estimate their black hole masses from the M-$\sigma_*$ relation, and compare with their viral black hole masses obtained from the single epoch spectra. The images of CL AGNs at faint state are useful for studies of AGN host galaxies avoiding contamination from the luminous central engines. We will do a statistical study on the host galaxies properties of CL AGNs.\\

The work is supported by the the National Key R\&D Program of China (2016YFA0400703), National Key Basic Research Program of China 2014CB845700, the NSFC grant No.11373008 and No.11533001. We thank Luis Ho, Yue Shen, Arjun Dey, Nicholas Ross, Aaron Meisner, and Ning Jiang, Yanxia Xie for very helpful discussions. We thank Shu Wang for providing the extinction data.

We acknowledge the use of SDSS data. Funding for SDSS-III has been provided by the Alfred P. Sloan Foundation, the Participating Institutions, the National Science Foundation, and the U.S. Department of Energy Office of Science. The SDSS-III website is \url{http://www.sdss3.org/}. SDSS-III is managed by the Astrophysical Research Consortium for the Participating Institutions of the SDSS-III Collaboration including the University of Arizona, the Brazilian Participation Group, Brookhaven National Laboratory, Carnegie Mellon University, University of Florida, the French Participation Group, the German Participation Group, Harvard University, the Instituto de Astrofisica de Canarias, the Michigan State/Notre Dame/JINA Participation Group, Johns Hopkins University, Lawrence Berkeley National Laboratory, Max Planck Institute for Astrophysics, Max Planck Institute for Extraterrestrial Physics, New Mexico State University, New York University, Ohio State University, Pennsylvania State University, University of Portsmouth, Princeton University, the Spanish Participation Group, University of Tokyo, University of Utah, Vanderbilt University, University of Virginia, University of Washington, and Yale University.

We acknowledge the use of LAMOST data. The Large Sky Area Multi-Object Fiber Spectroscopic Telescope (LAMOST, also named Guoshoujing Telescope) is a National Major Scientific Project built by the Chinese Academy of Sciences. Funding for the project has been provided by the National Development and Reform Commission. LAMOST is operated and managed by the National Astronomical Observatories, Chinese Academy of Sciences.

This research has made use of PS1, DELS, WISE, CRTS, and PTF imaging data. The PS1 has been made possible through contributions by the Institute for Astronomy, the University of Hawaii, the Pan-STARRS Project Office, the Max-Planck Society and its participating institutes, the Max Planck Institute for Astronomy, Heidelberg and the Max Planck Institute for Extraterrestrial Physics, Garching, The Johns Hopkins University, Durham University, the University of Edinburgh, Queen's University Belfast, the Harvard-Smithsonian Center for Astrophysics, the Las Cumbres Observatory Global Telescope Network Incorporated, the National Central University of Taiwan, the Space Telescope Science Institute, the National Aeronautics and Space Administration under Grant No. NNX08AR22G issued through the Planetary Science Division of the NASA Science Mission Directorate, the National Science Foundation under Grant No. AST-1238877, the University of Maryland, and Eotvos Lorand University (ELTE). We acknowledge the use of DESI Legacy imaging survey, and the website is \url{http://legacysurvey.org}. This research has made use of the NASA/ IPAC Infrared Science Archive, which is operated by the Jet Propulsion Laboratory, California Institute of Technology, under contract with the National Aeronautics and Space Administration. This publication makes use of data products from the \emph{Wide-field Infrared Survey Explorer}, which is a joint project of the University of California, Los Angeles, and the Jet Propulsion Laboratory/California Institute of Technology, funded by the National Aeronautics and Space Administration. The CSS survey is funded by the National Aeronautics and Space Administration under Grant No. NNG05GF22G issued through the Science Mission Directorate Near-Earth Objects Observations Programme. The CRTS survey is supported by the US National Science Foundation under grants AST-0909182 and AST-1313422. We acknowledge the use of PTF data, and the website is \url{https://www.ptf.caltech.edu}.

We acknowledge the use of the Xinglong 2.16 m telescope and the Palomar Hale 5m telescope. This work has made use of the TOPCAT \citep{Taylor2005}. We thank the Chinese Virtual Observatory, and the website is \url{http://www.china-vo.org}.

\facilities{Sloan, PS1, IRSA, WISE, PTF, Beijing:2.16m (BFOSC), Palomar P200/Caltech}

\appendix
\renewcommand{\thefigure}{A.\arabic{figure}}

The light curves and spectra of other 20 new CL AGNs, in addition to one shown in Figure \ref{fig:cl_example}, are provided in Figure \ref{fig:cl_others}.

\begin{figure*}
 \centering
 \hspace{0cm}
  \subfigure{
   \hspace{-1.0cm}
  \includegraphics[width=3.9in]{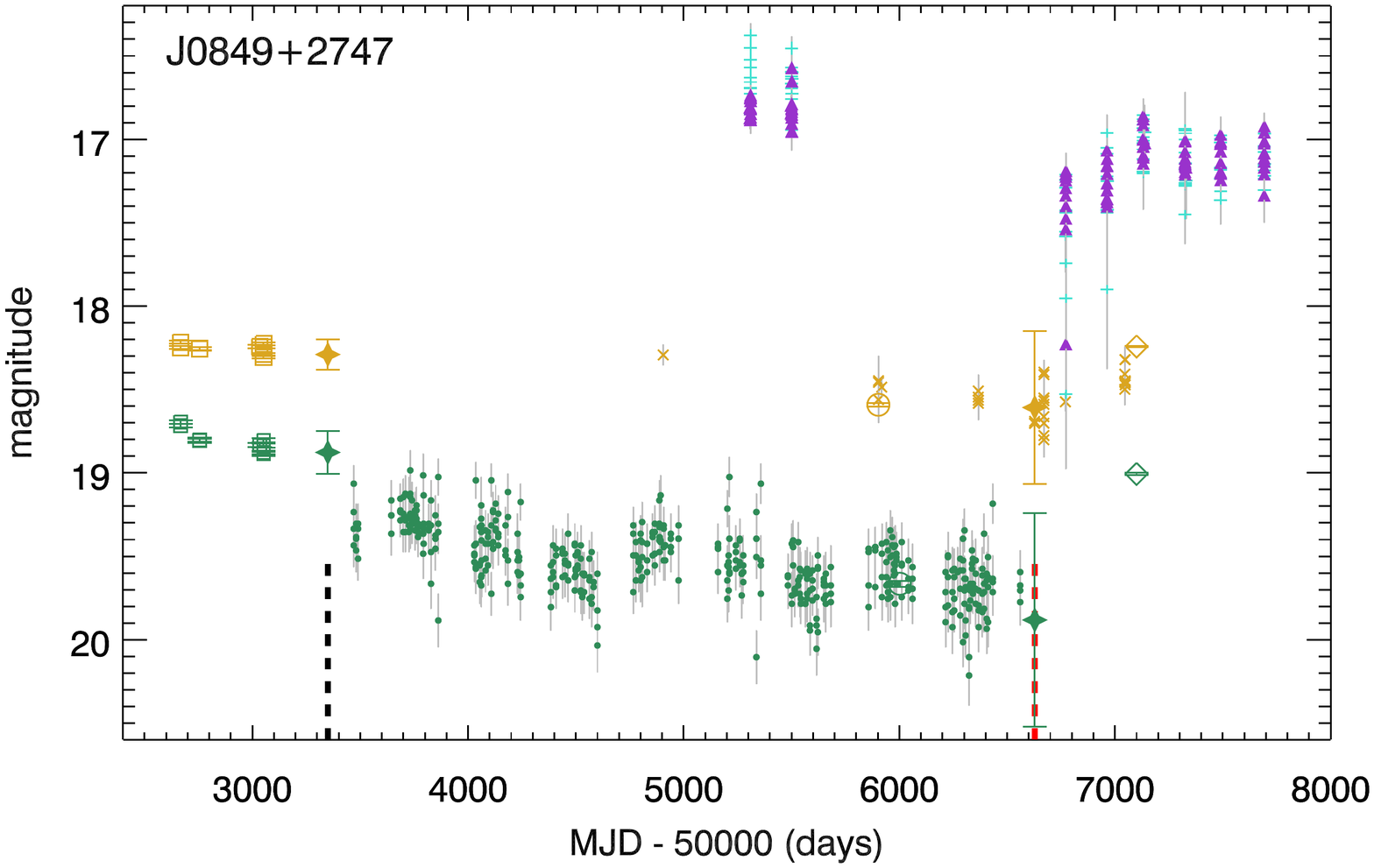}}
 \hspace{-1.4cm}
 \subfigure{
  \includegraphics[width=3.9in]{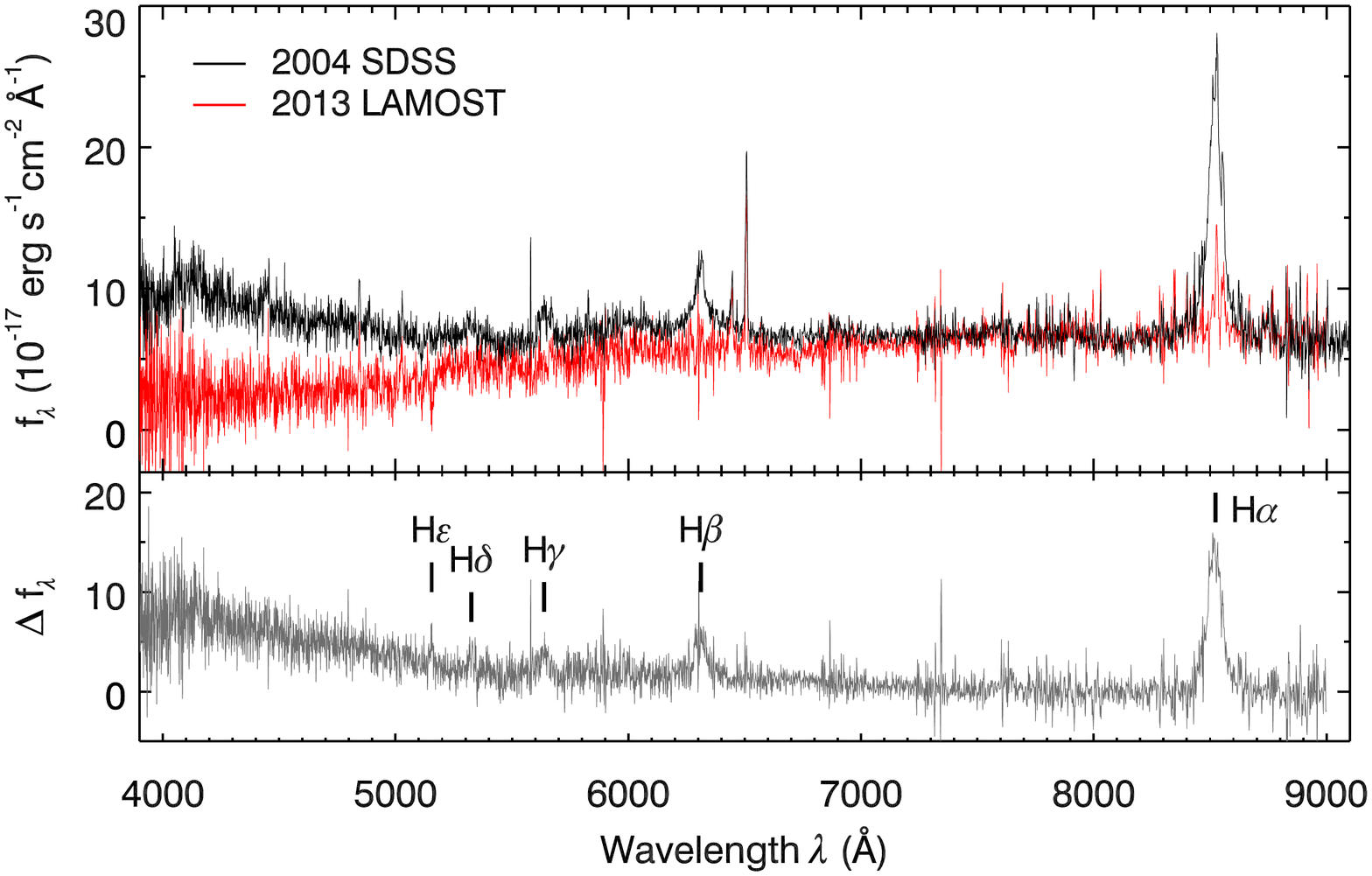}}\\
 \vspace{-2.4cm}

 \hspace{0cm}
 \subfigure{
  \hspace{-1.0cm}
  \includegraphics[width=3.9in]{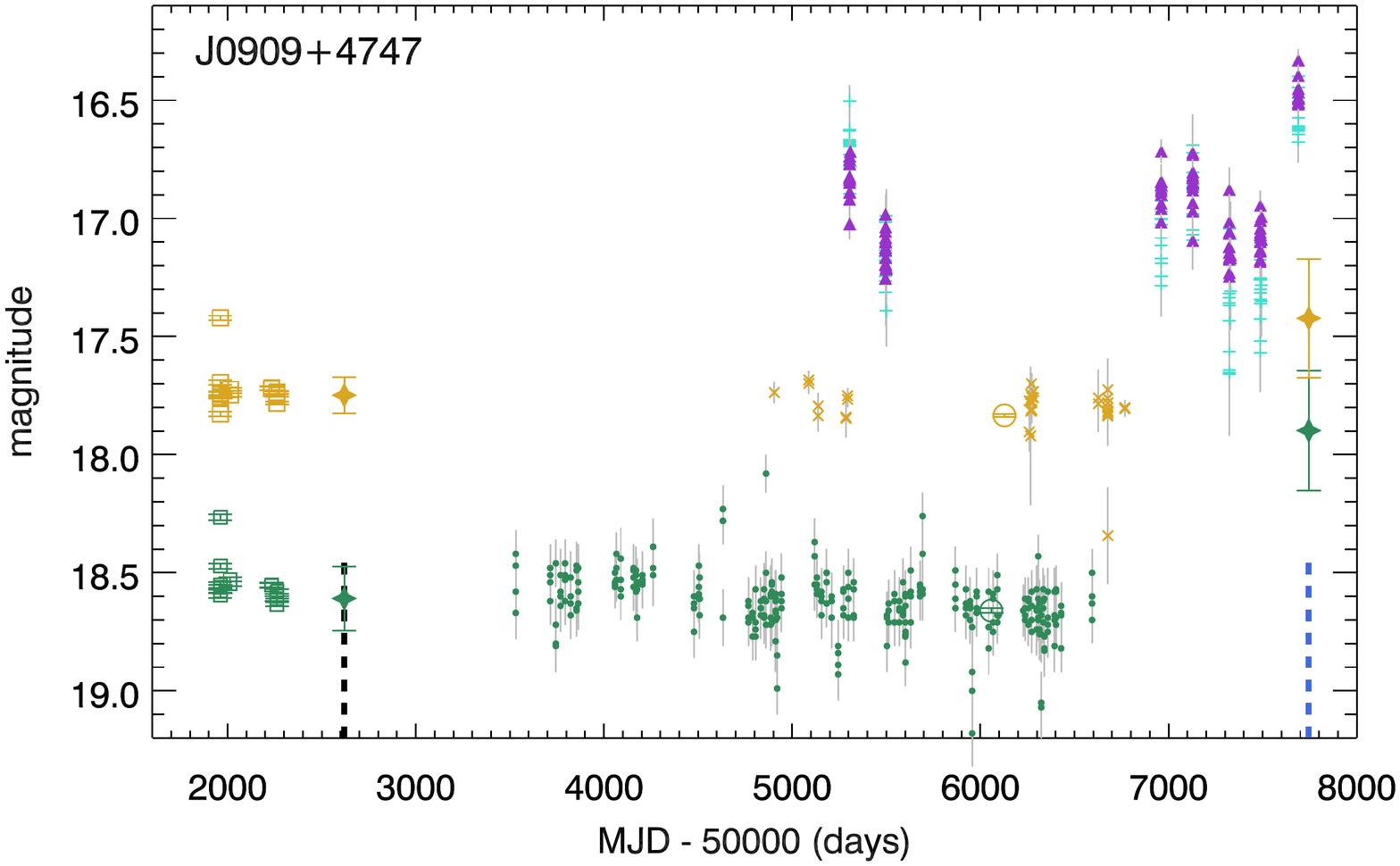}}
 \hspace{-1.4cm}
 \subfigure{
  \includegraphics[width=3.9in]{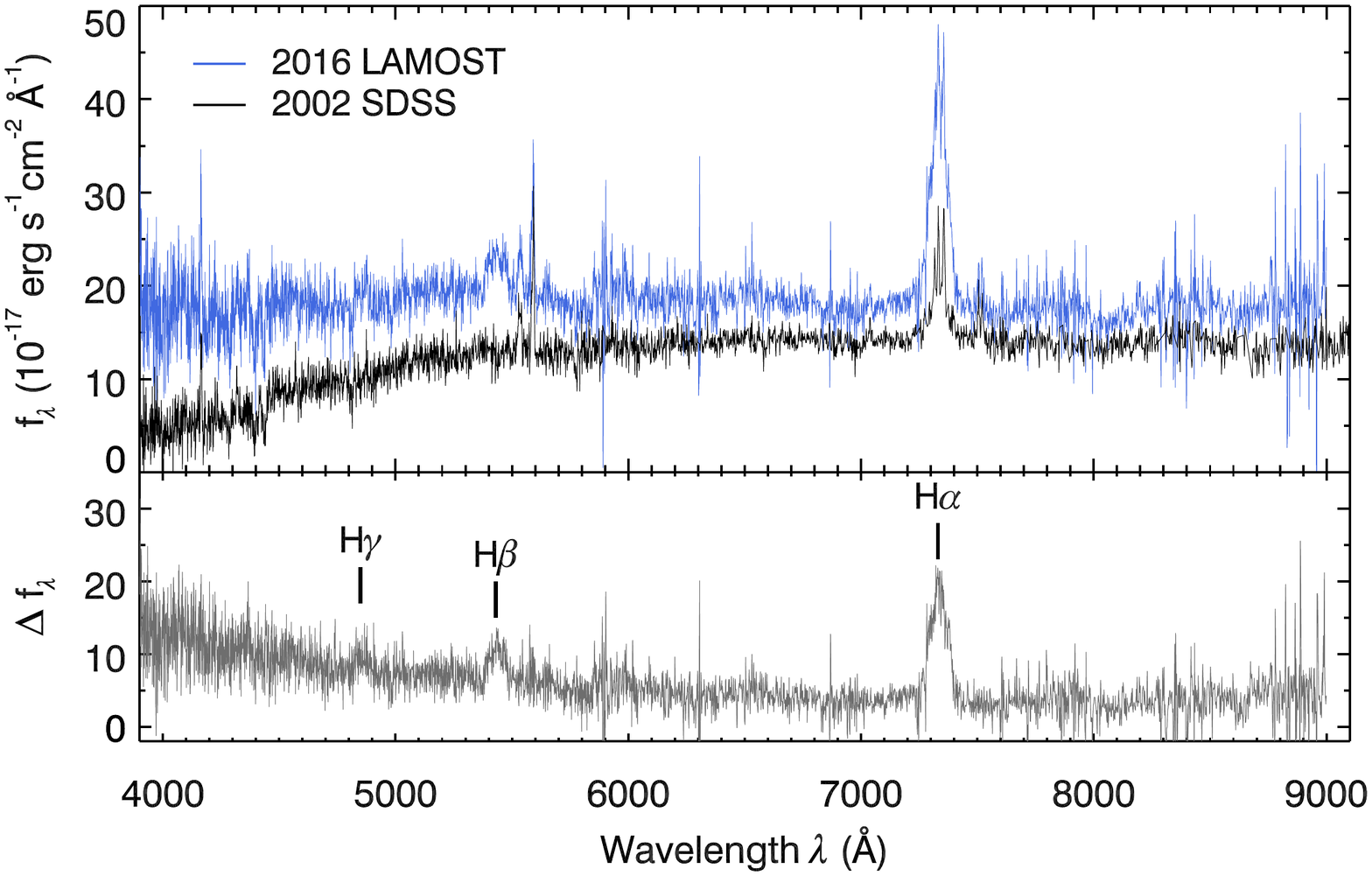}}\\
  \vspace{-2.4cm}

 \hspace{0cm}
 \subfigure{
  \hspace{-1.0cm}
  \includegraphics[width=3.9in]{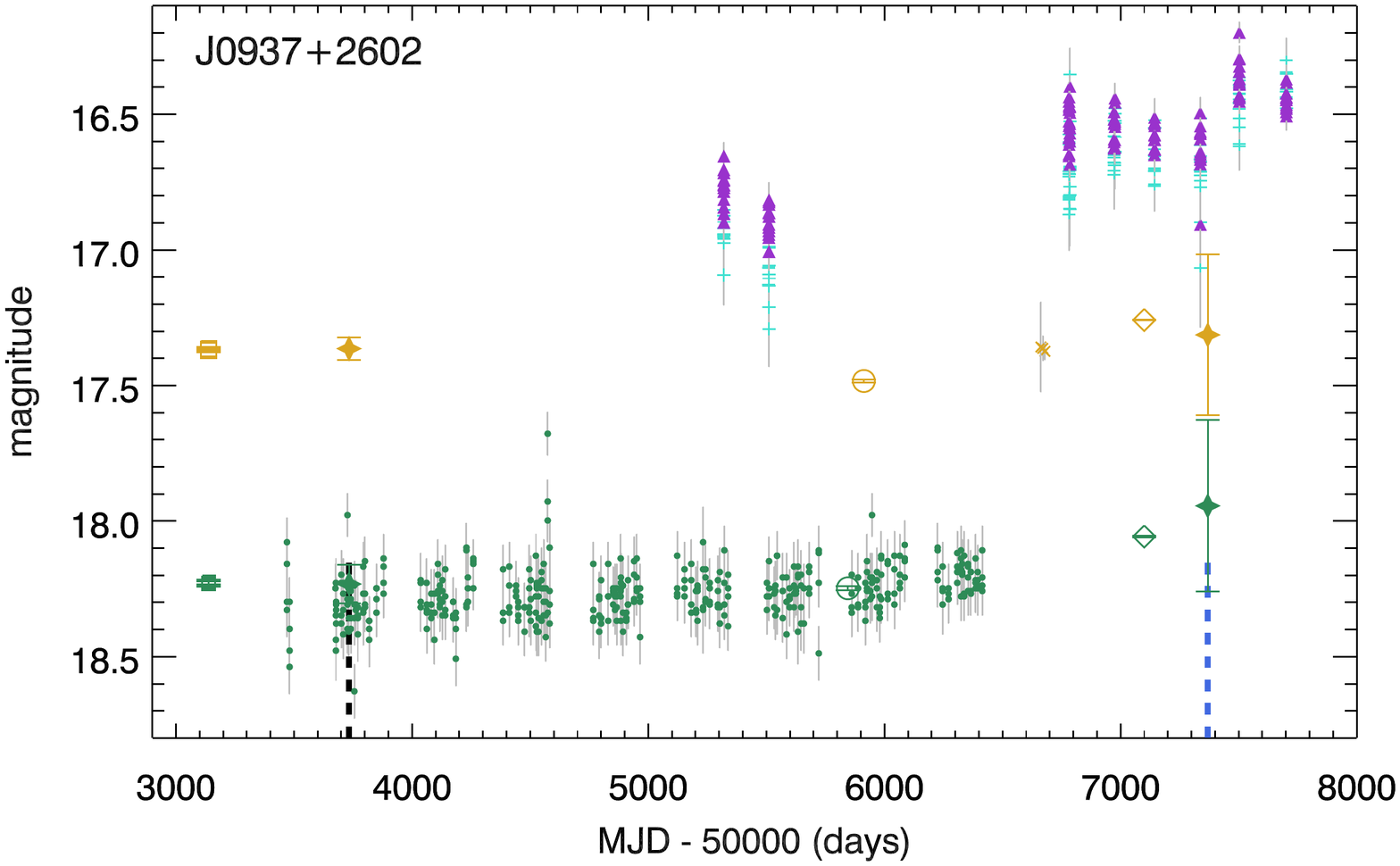}}
 \hspace{-1.4cm}
 \subfigure{
  \includegraphics[width=3.9in]{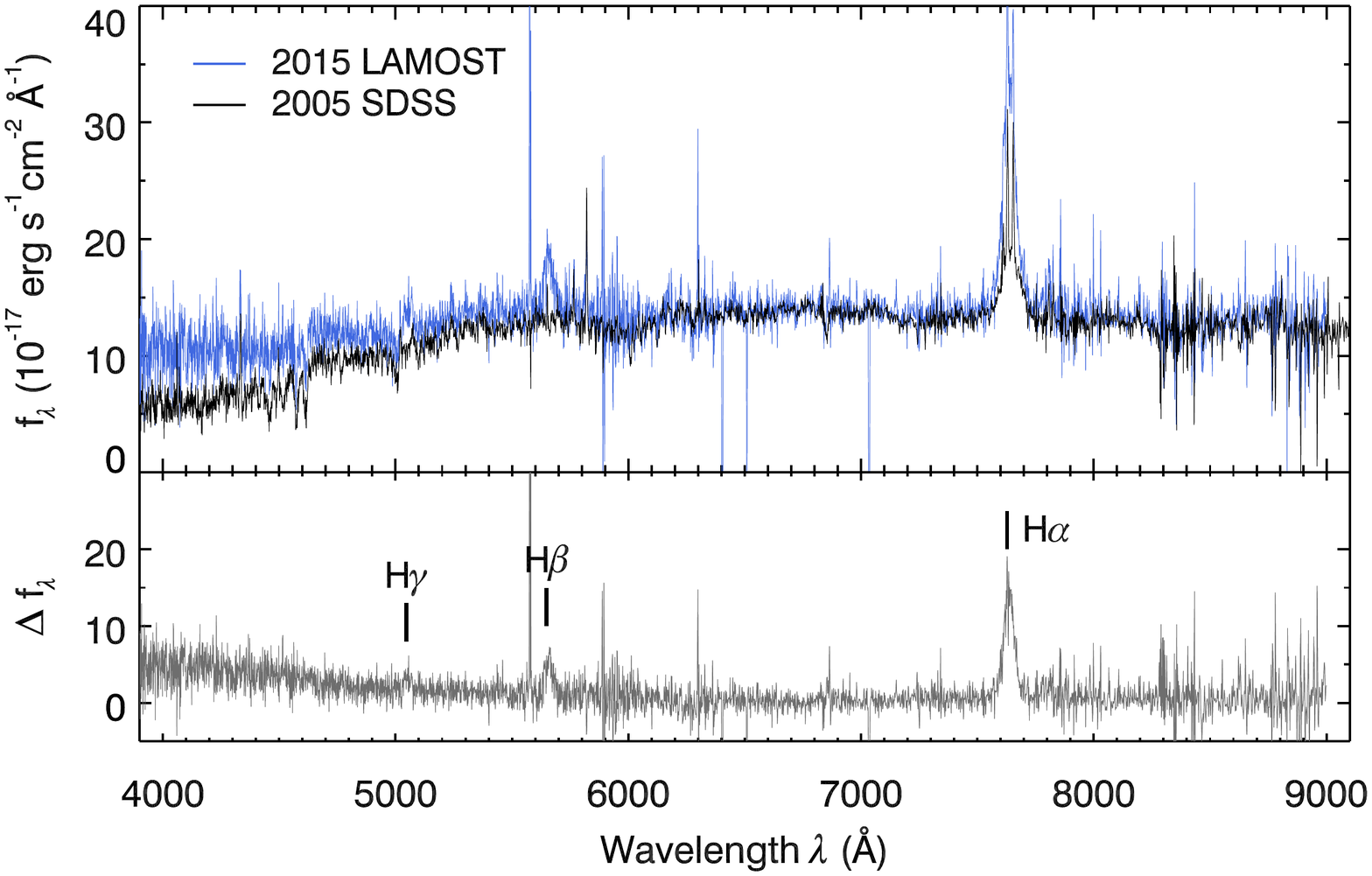}}\\
  \vspace{-2.4cm}

 \centering
 \hspace{0cm}
 \subfigure{
  \hspace{-1.0cm}
  \includegraphics[width=3.9in]{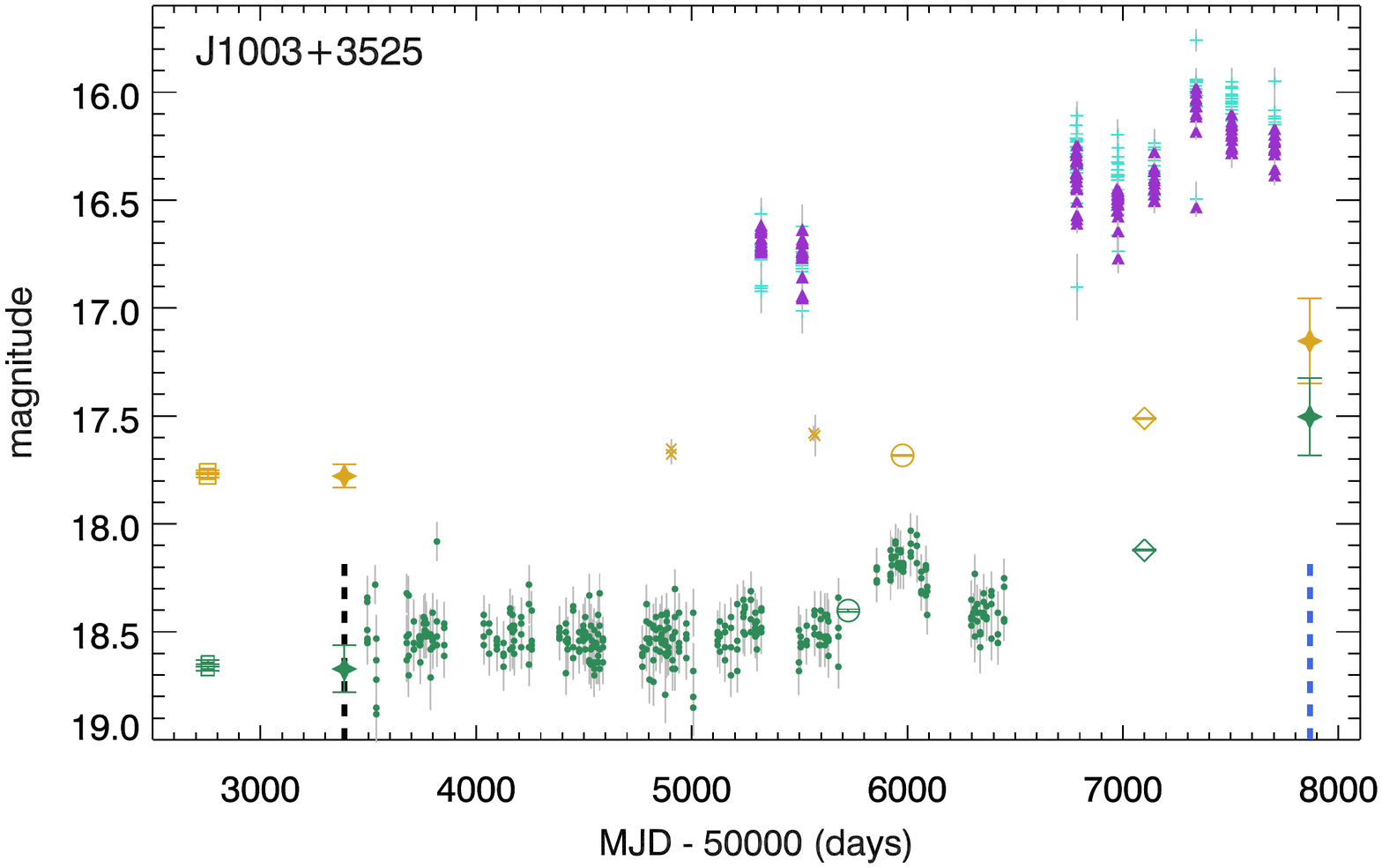}}
 \hspace{-1.4cm}
 \subfigure{
  \includegraphics[width=3.9in]{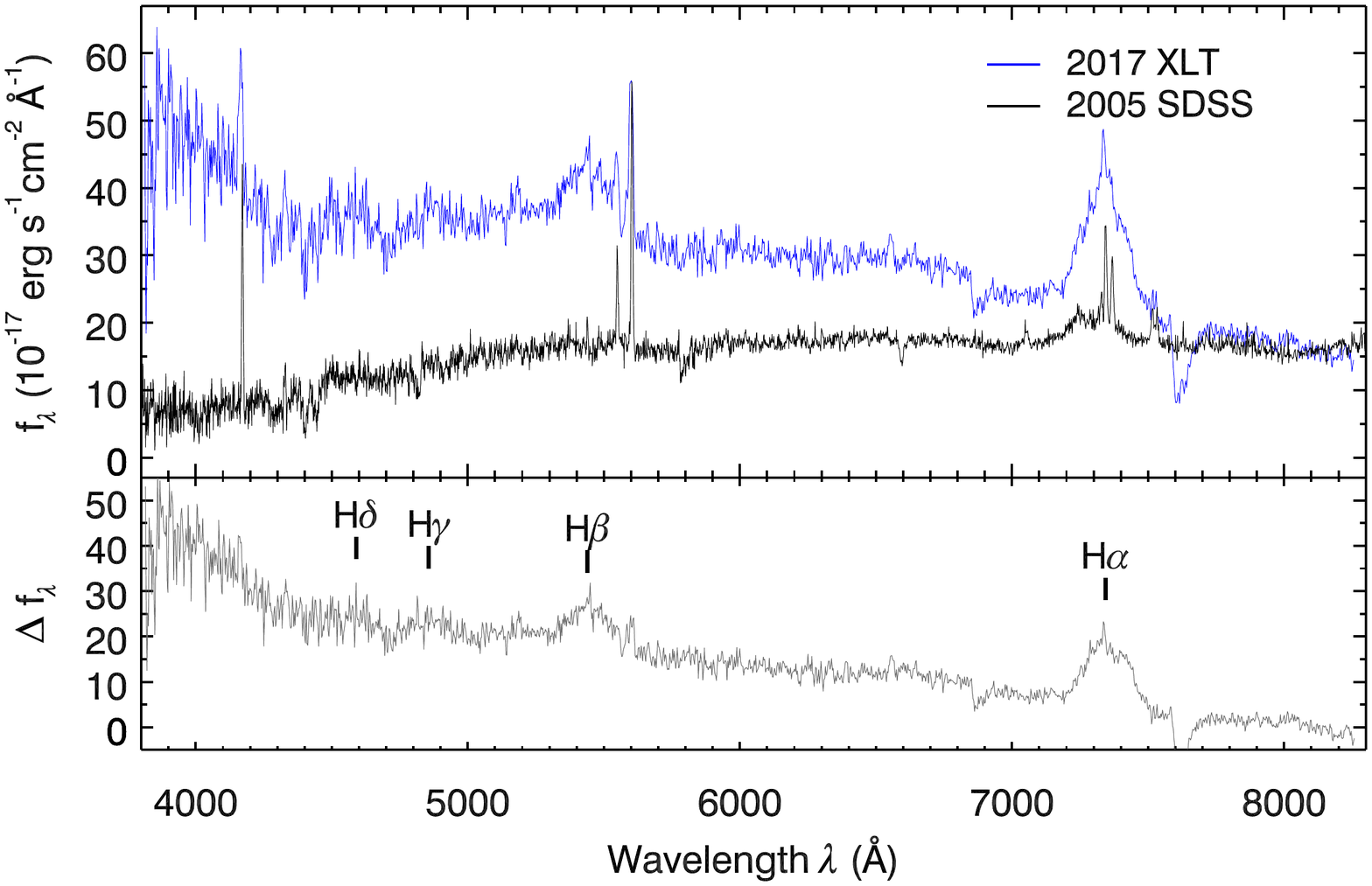}}\\
 \vspace{-1cm}
  \setcounter{figure}{0}
  \caption{\label{fig:cl_others} The same with Figure \ref{fig:cl_example}.}

\end{figure*}

\renewcommand{\thefigure}{A.\arabic{figure}}
\addtocounter{figure}{-1}

\begin{figure*}

 \centering
 \hspace{0cm}
  \subfigure{
   \hspace{-1.0cm}
  \includegraphics[width=3.9in]{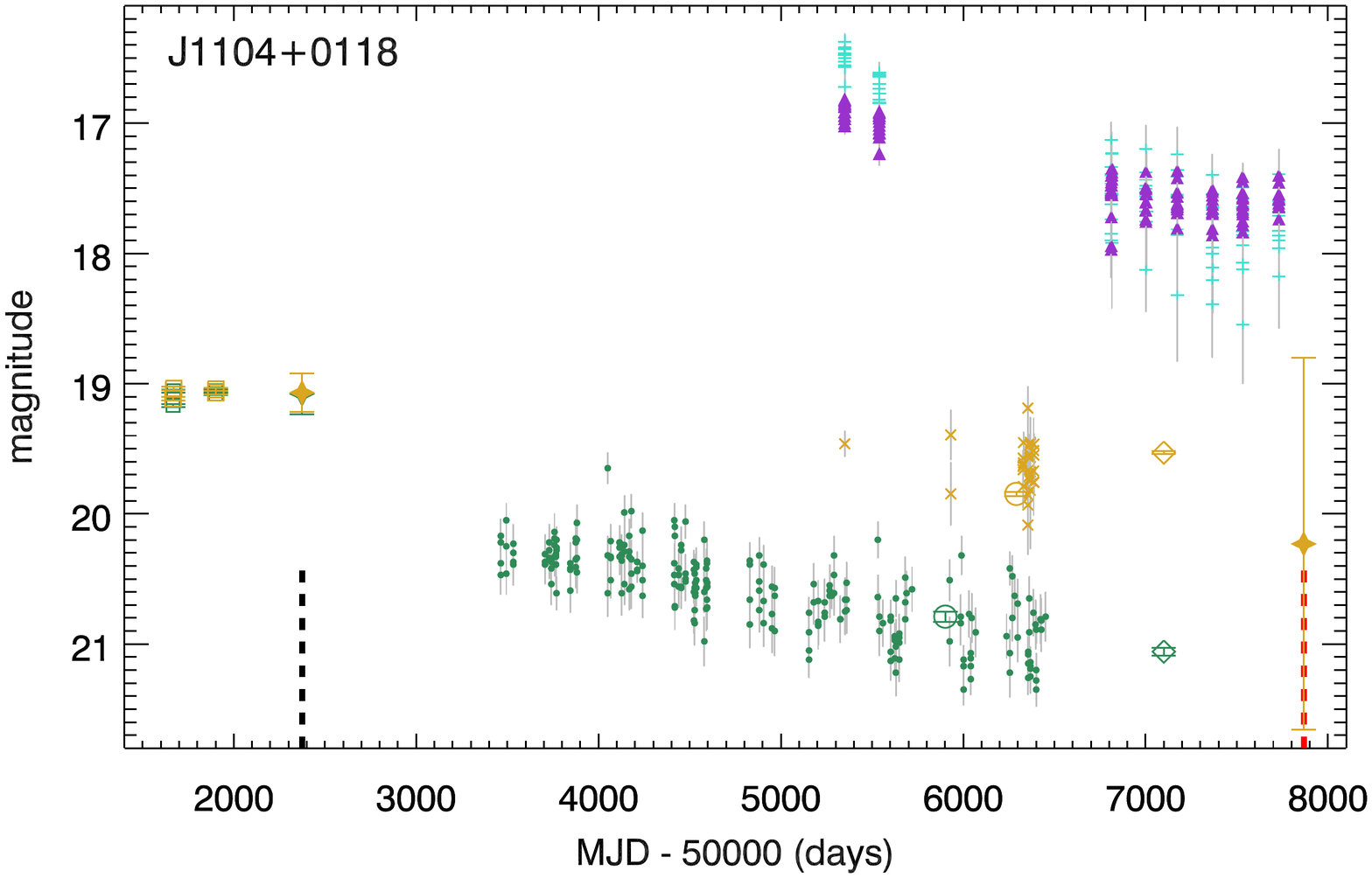}}
 \hspace{-1.4cm}
 \subfigure{
  \includegraphics[width=3.9in]{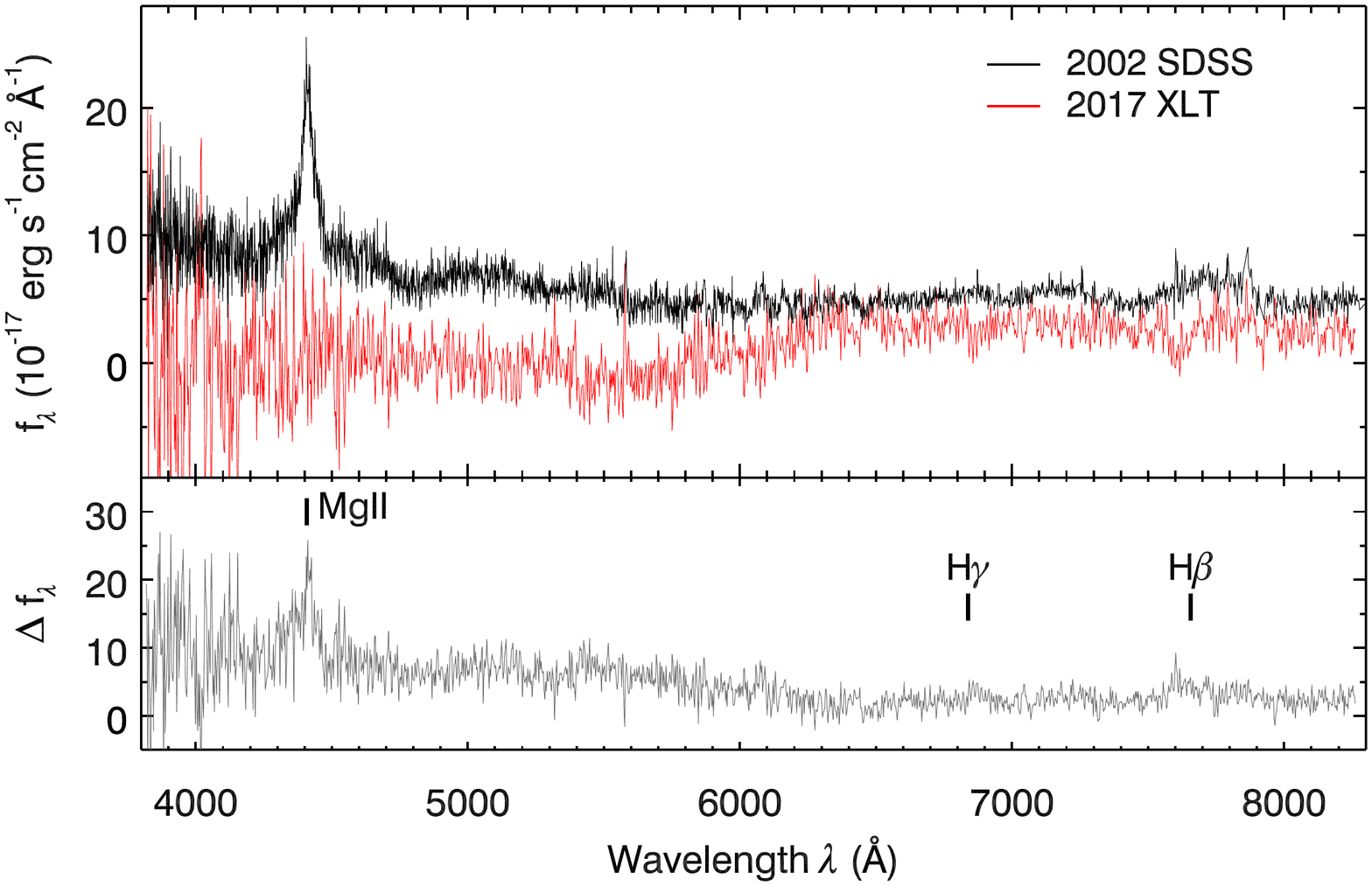}}\\
 \vspace{-2.4cm}

 \centering
  \hspace{0cm}
 \subfigure{
  \hspace{-1.0cm}
  \includegraphics[width=3.9in]{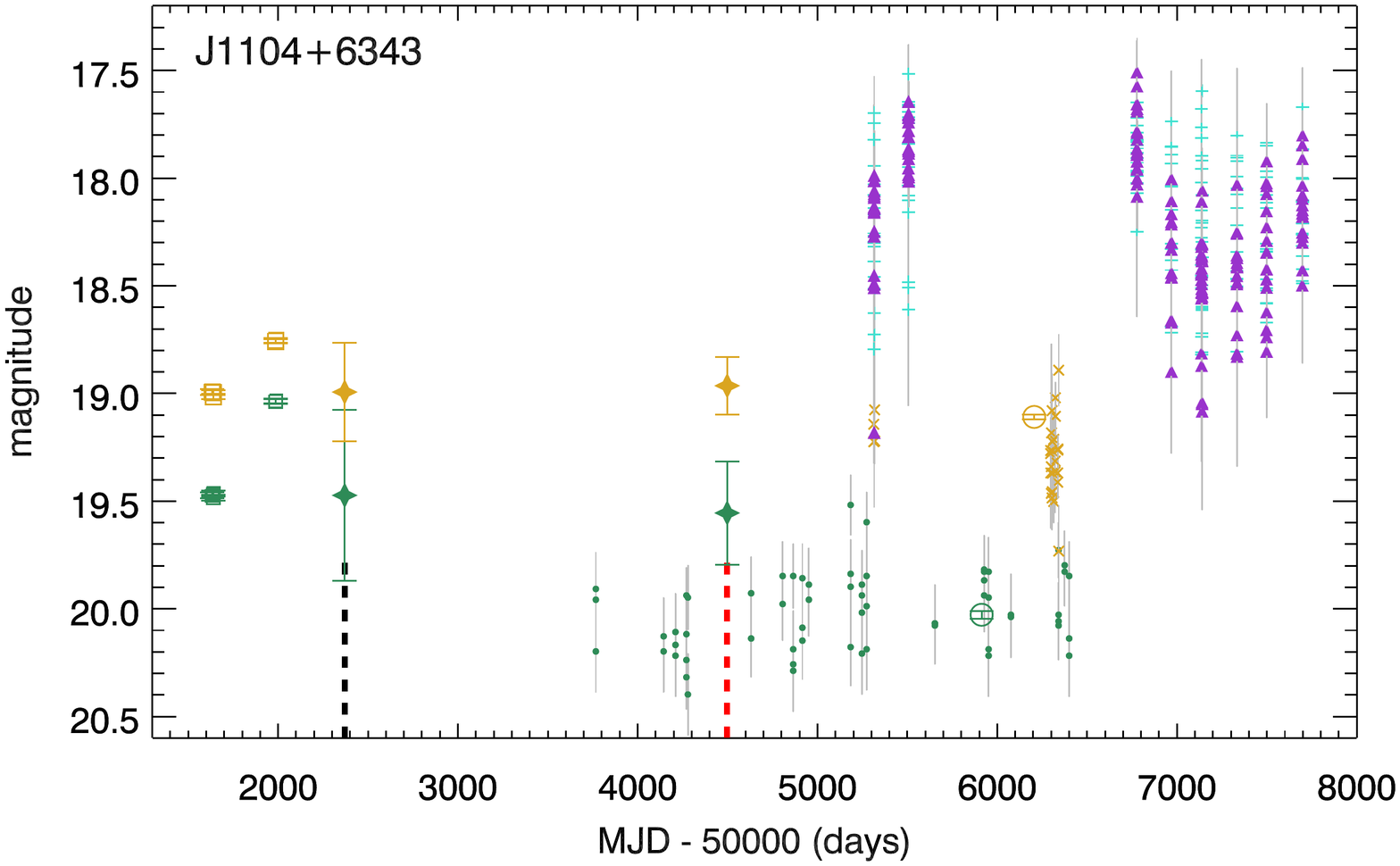}}
 \hspace{-1.4cm}
 \subfigure{
  \includegraphics[width=3.9in]{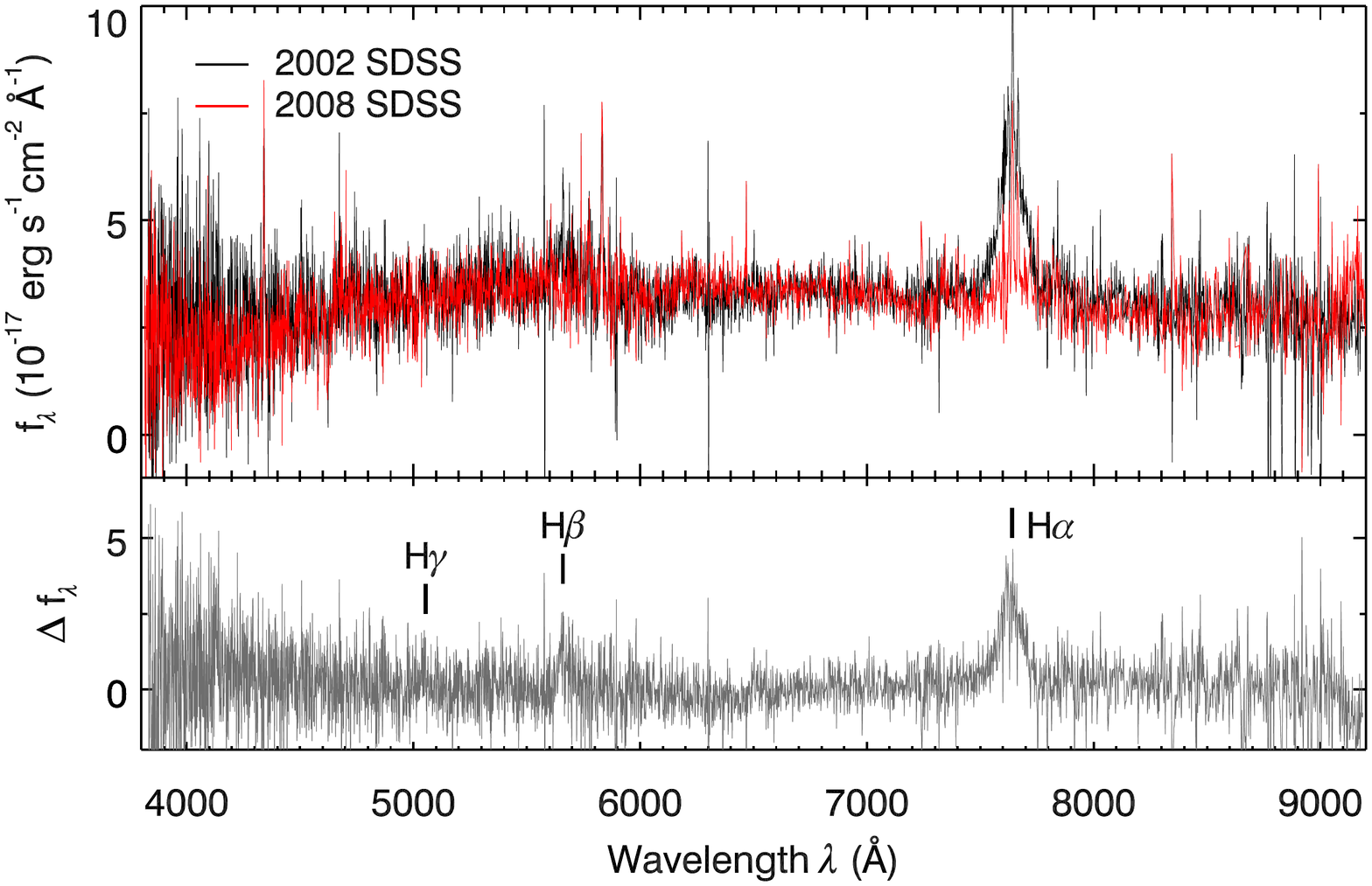}}\\
   \vspace{-2.4cm}

  \centering
 \hspace{0cm}
 \subfigure{
  \hspace{-1.0cm}
  \includegraphics[width=3.9in]{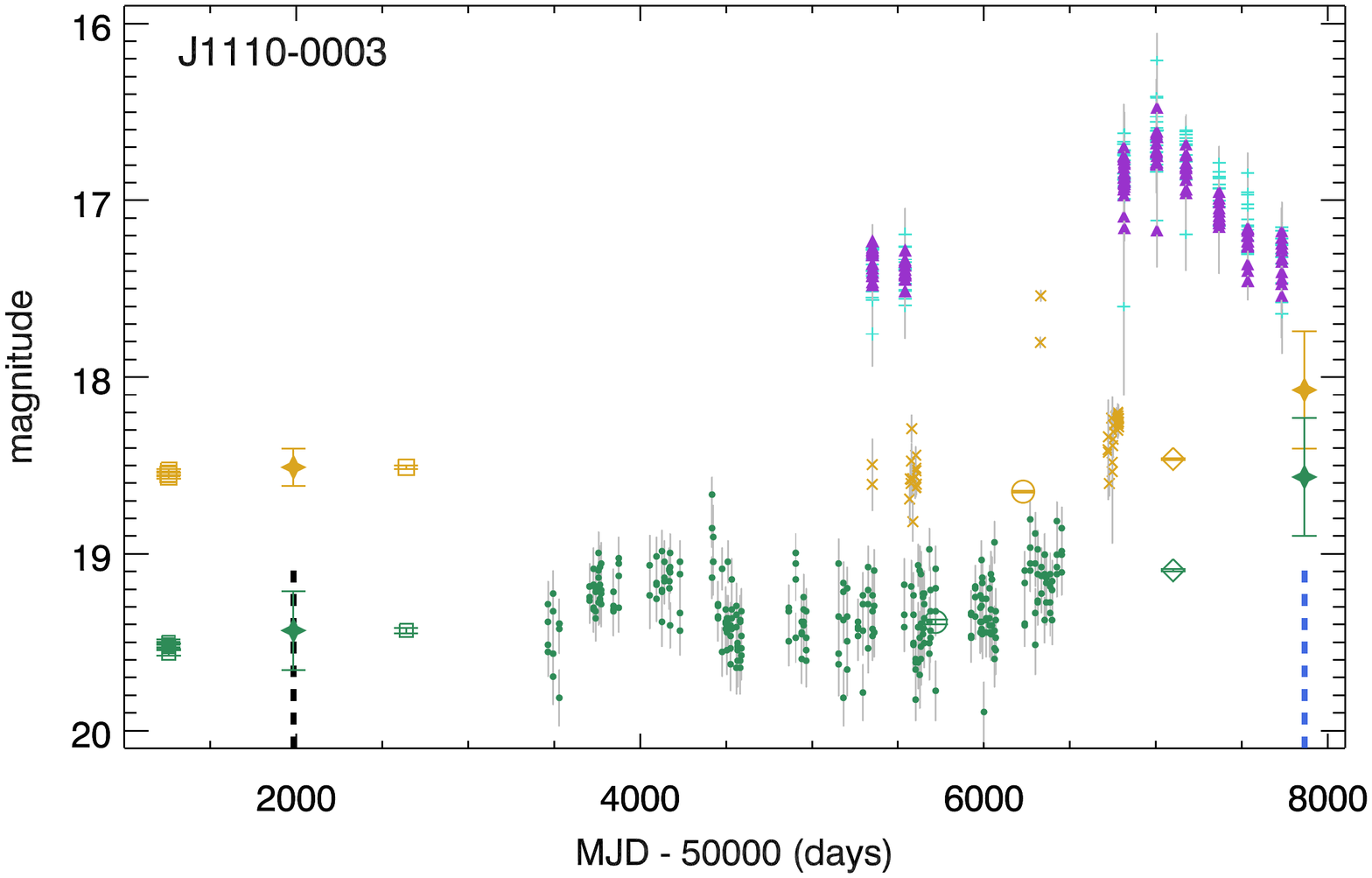}}
 \hspace{-1.4cm}
 \subfigure{
  \includegraphics[width=3.9in]{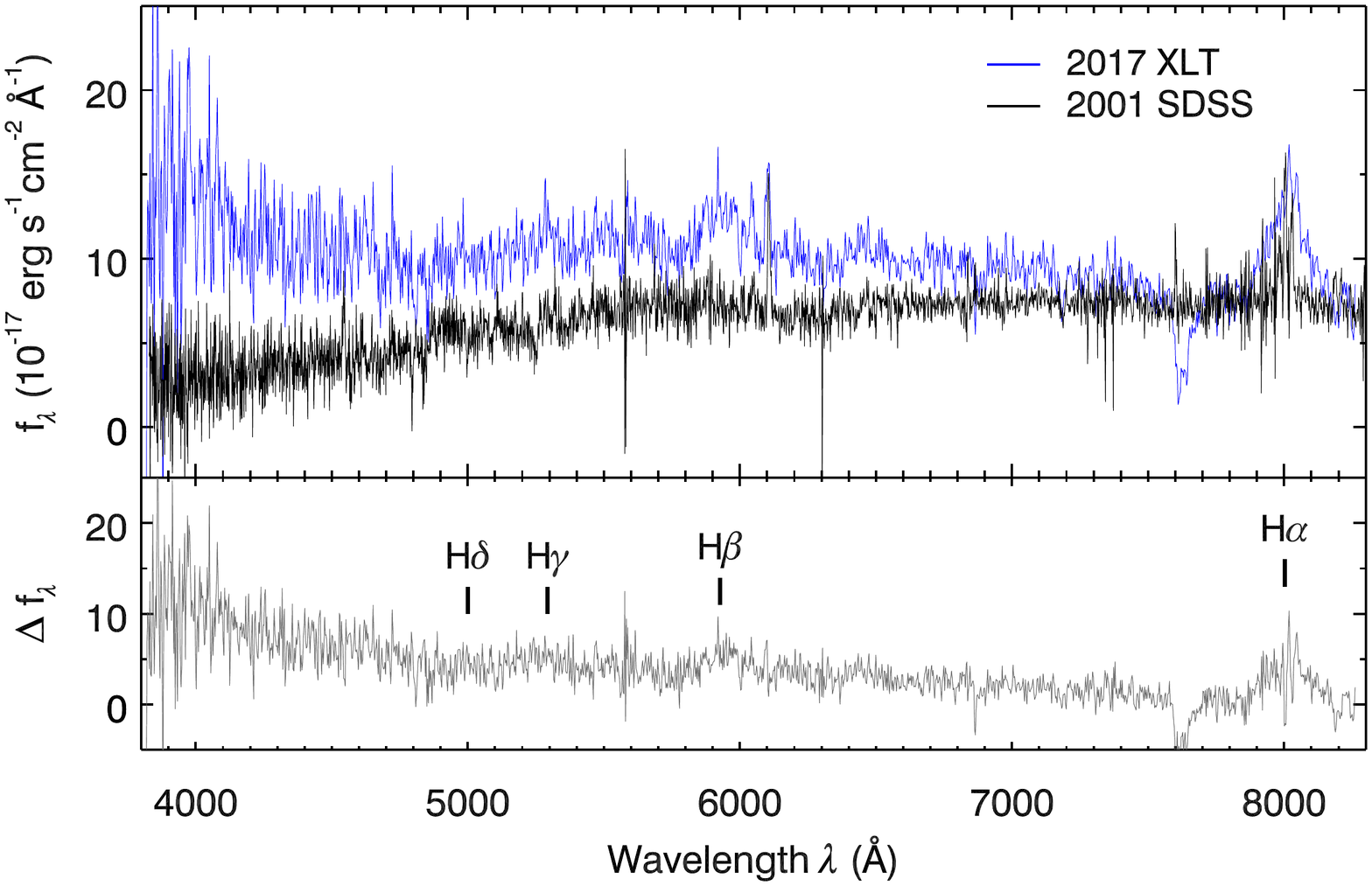}}\\
 \vspace{-2.4cm}

 \centering
 \hspace{0cm}
  \subfigure{
   \hspace{-1.0cm}
  \includegraphics[width=3.9in]{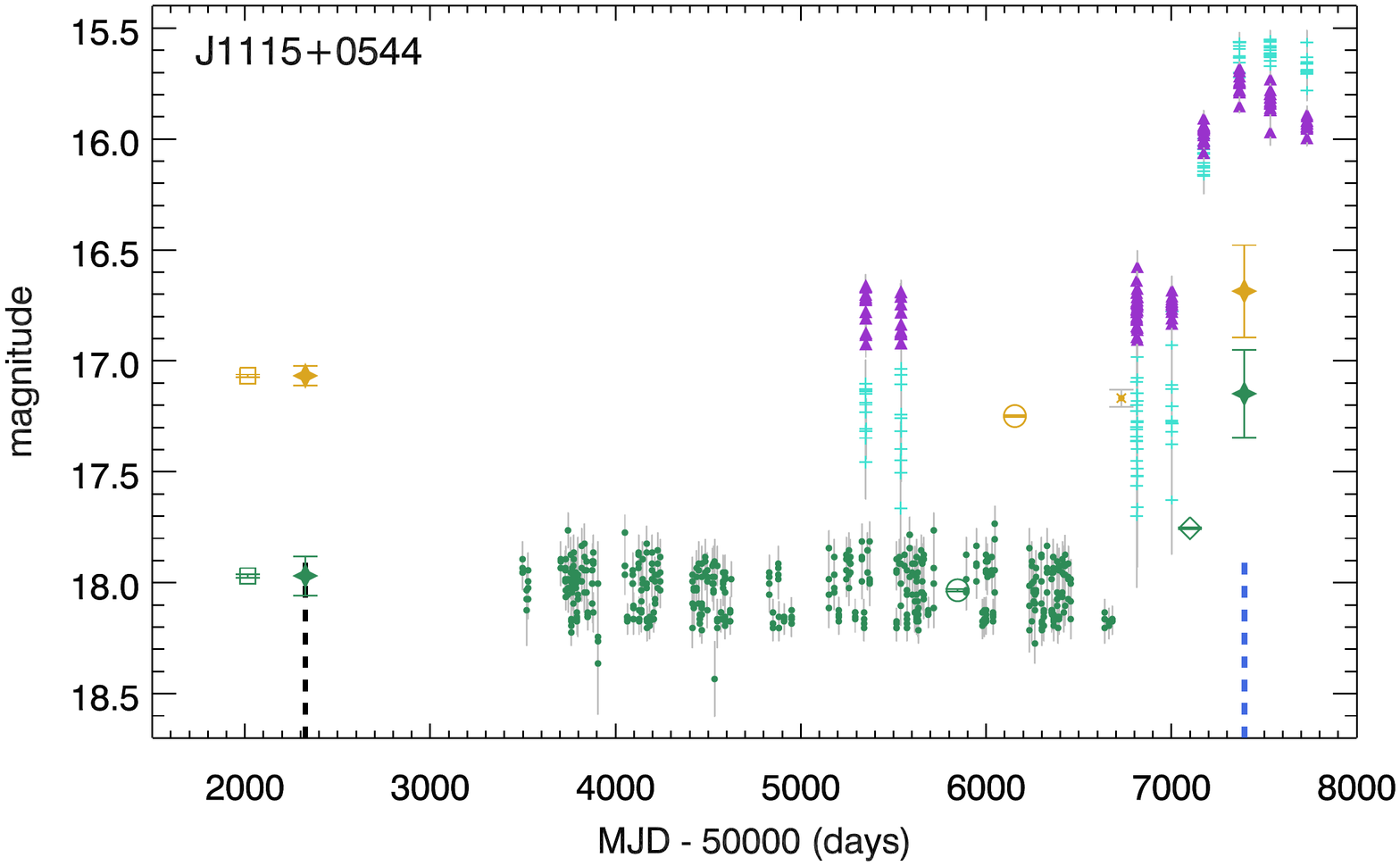}}
 \hspace{-1.4cm}
 \subfigure{
  \includegraphics[width=3.9in]{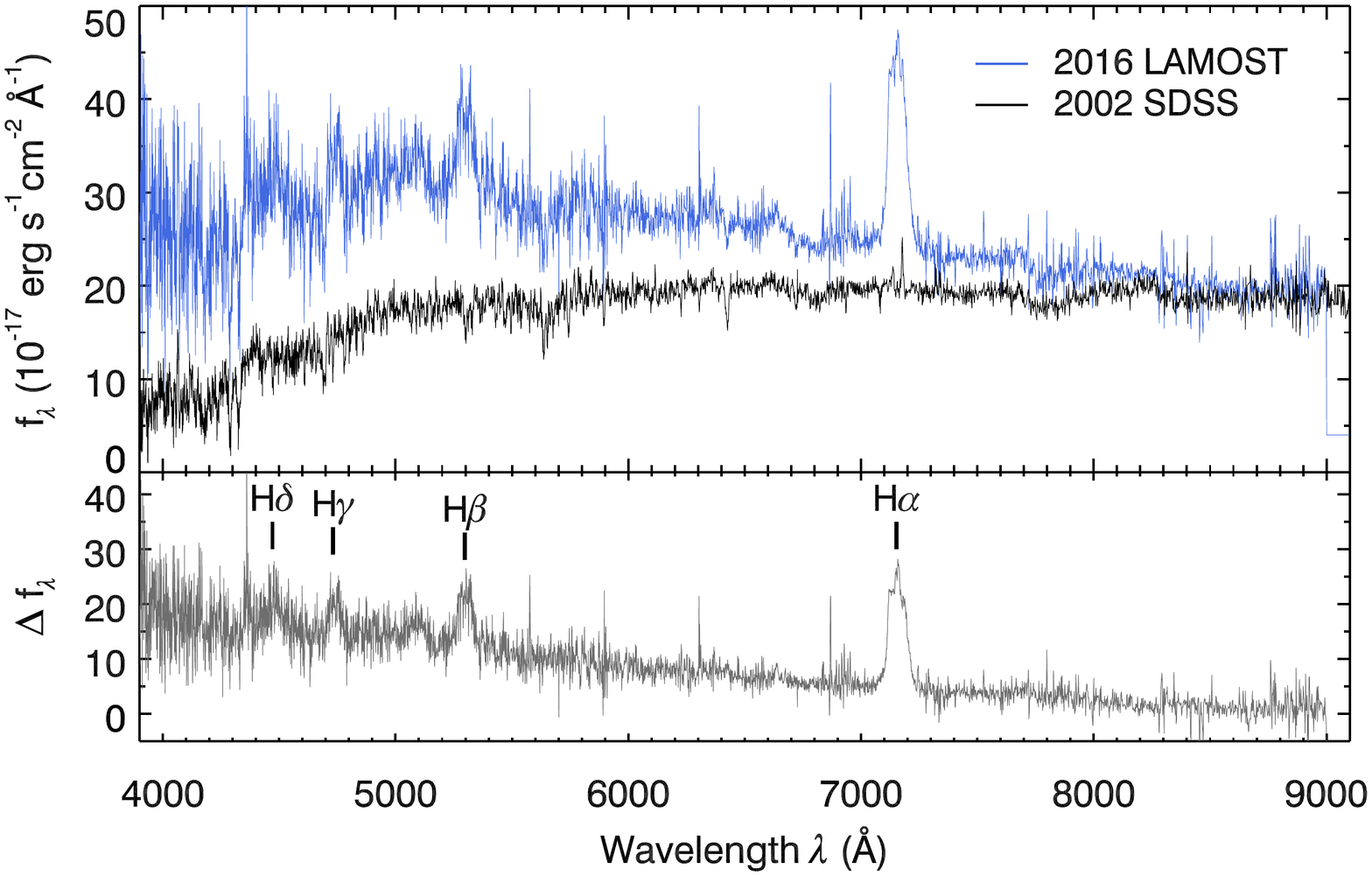}}\\
 \vspace{-1cm}
  \caption{ (Continued.)}
\end{figure*}

\renewcommand{\thefigure}{A.\arabic{figure}}
\addtocounter{figure}{-1}

\begin{figure*}

  \centering
 \hspace{0cm}
  \subfigure{
   \hspace{-1.0cm}
  \includegraphics[width=3.9in]{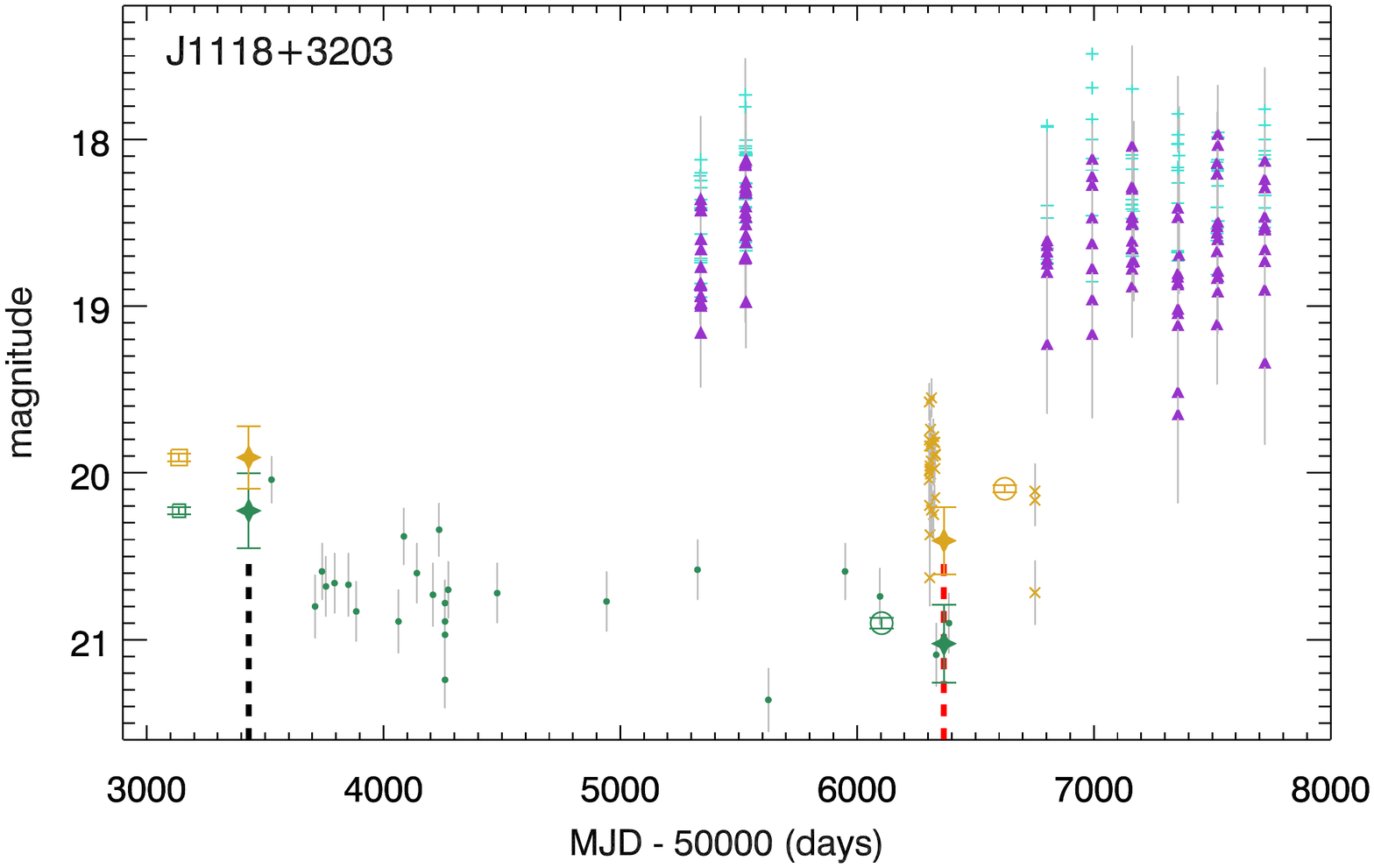}}
 \hspace{-1.4cm}
 \subfigure{
  \includegraphics[width=3.9in]{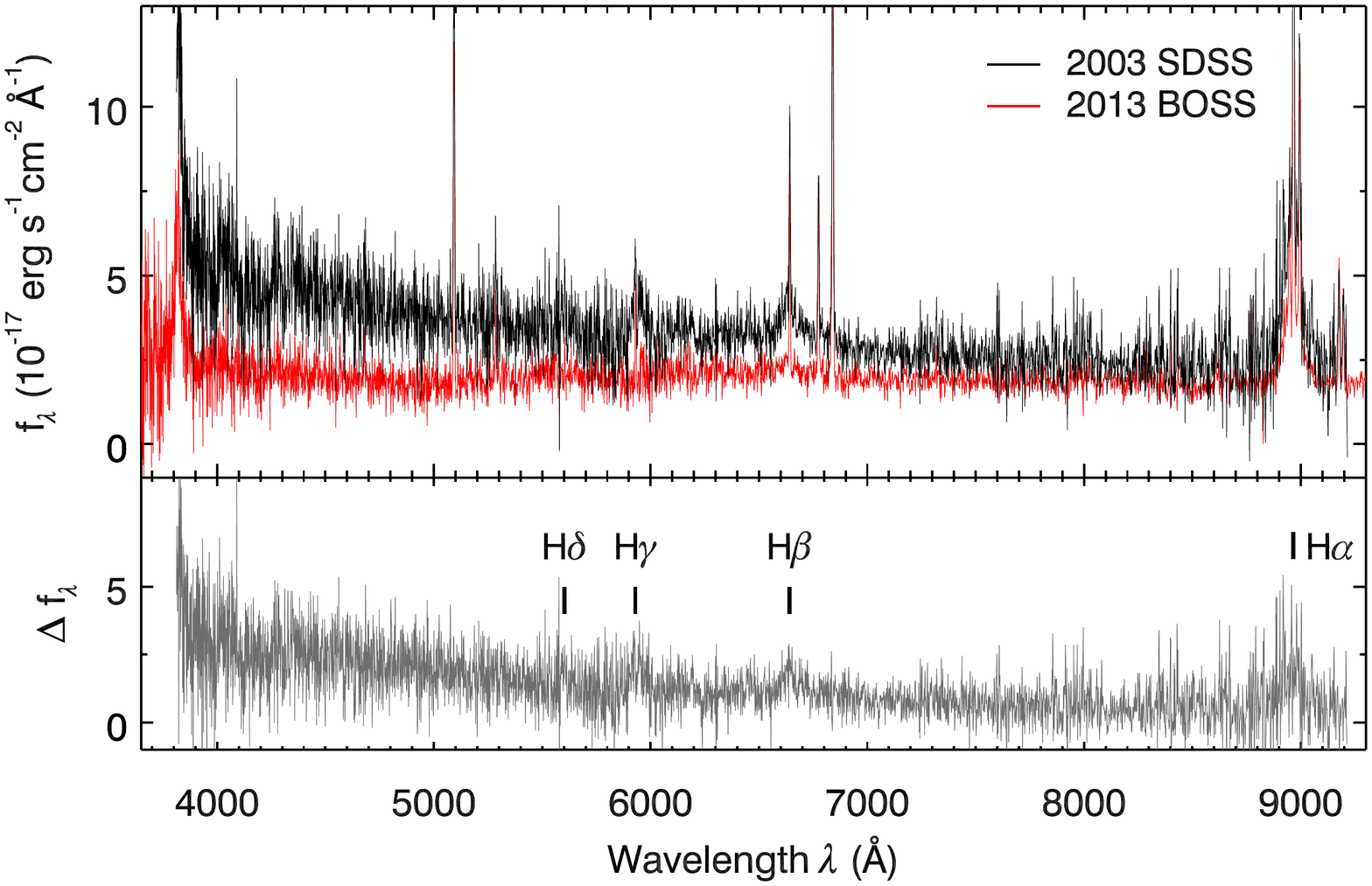}}\\
 \vspace{-2.4cm}

 \hspace{0cm}
  \subfigure{
   \hspace{-1.0cm}
  \includegraphics[width=3.9in]{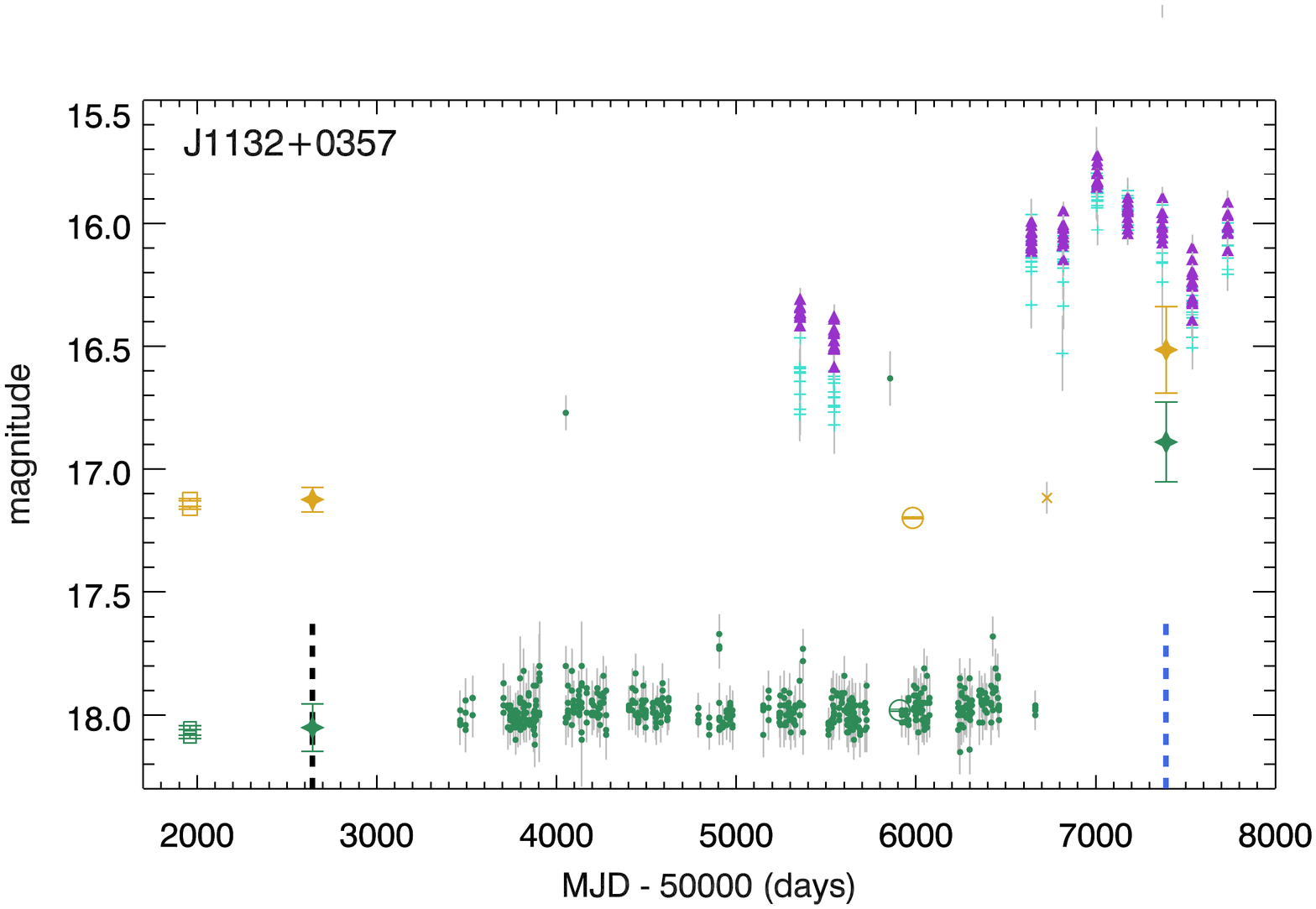}}
 \hspace{-1.4cm}
 \subfigure{
  \includegraphics[width=3.9in]{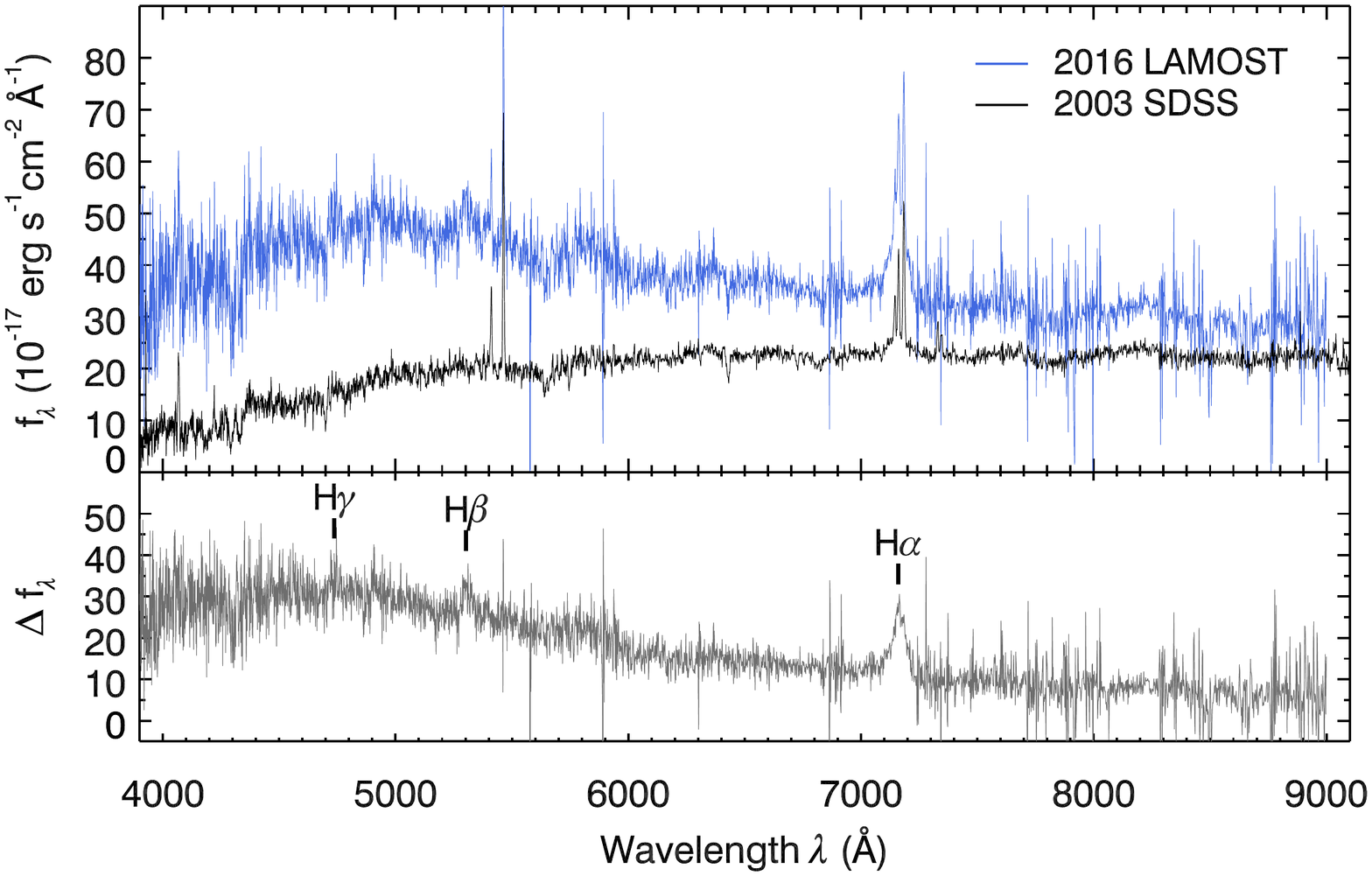}}\\
 \vspace{-2.4cm}

 \centering
 \hspace{0cm}
  \subfigure{
   \hspace{-1.0cm}
  \includegraphics[width=3.9in]{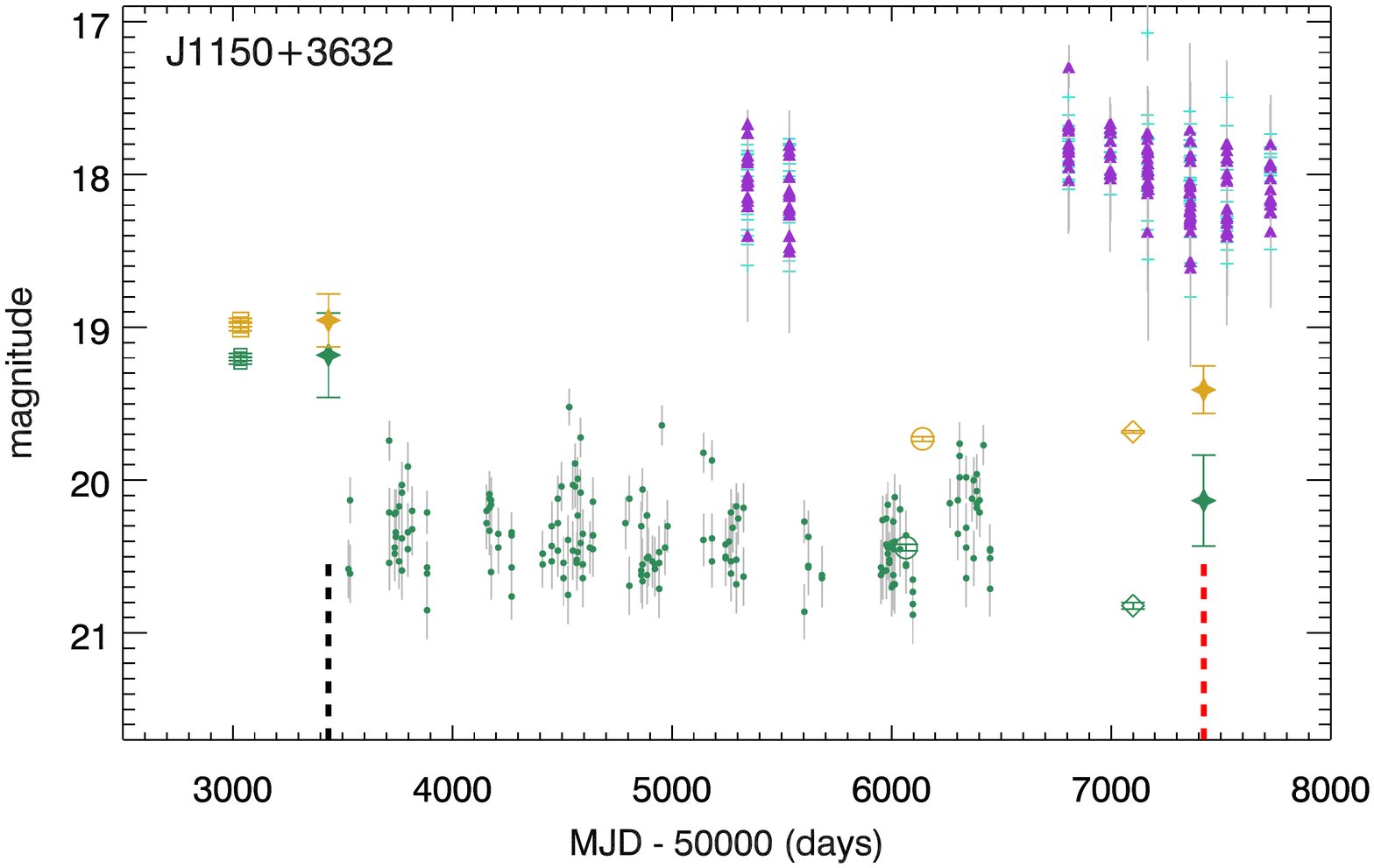}}
 \hspace{-1.4cm}
 \subfigure{
  \includegraphics[width=3.9in]{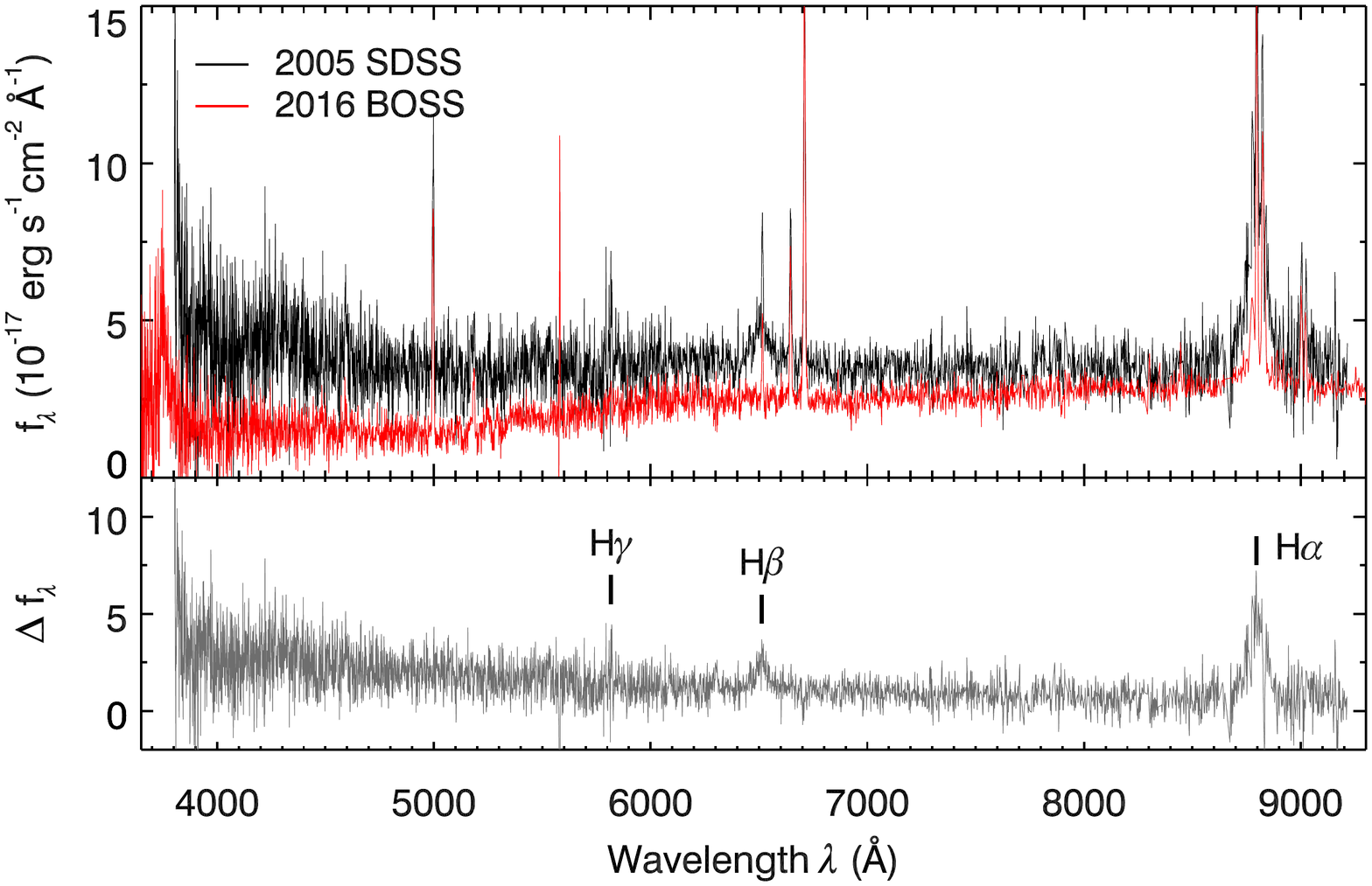}}\\
 \vspace{-2.4cm}

  \centering
 \hspace{0cm}
  \subfigure{
   \hspace{-1.0cm}
  \includegraphics[width=3.9in]{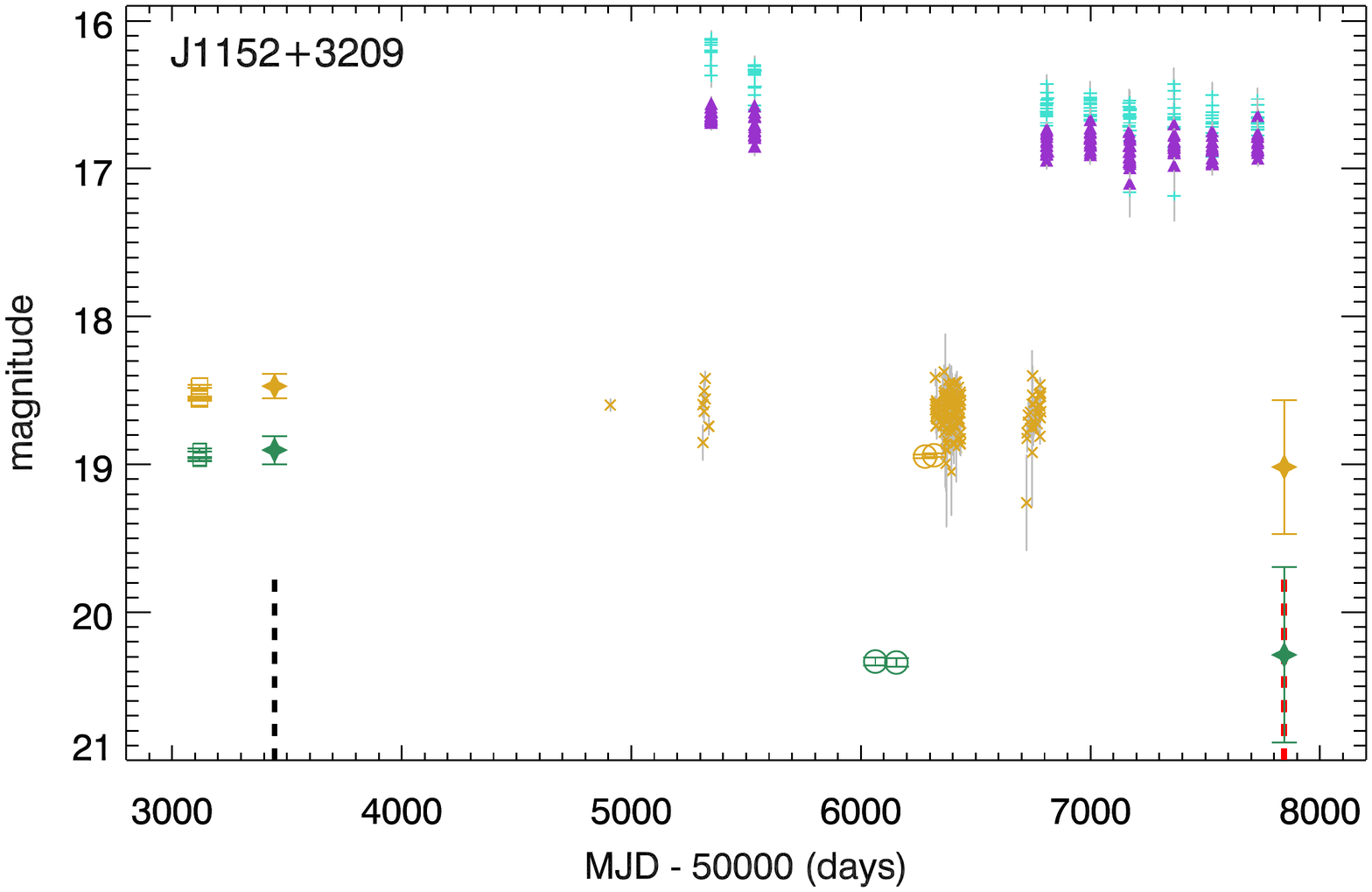}}
 \hspace{-1.4cm}
 \subfigure{
  \includegraphics[width=3.9in]{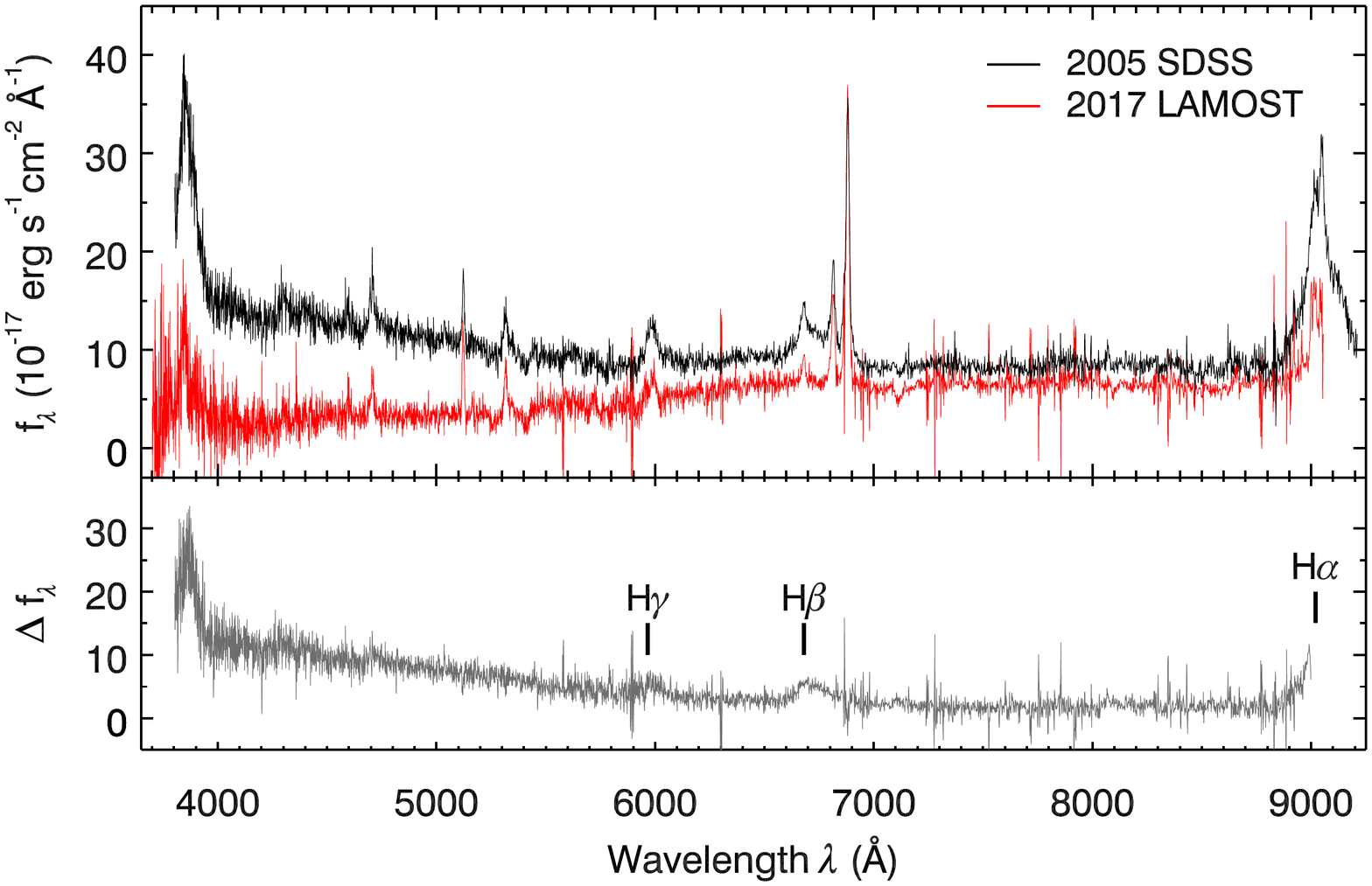}}\\
 \vspace{-1cm}
  \caption{ (Continued.)}
\end{figure*}

\renewcommand{\thefigure}{A.\arabic{figure}}
\addtocounter{figure}{-1}

\begin{figure*}

 \centering
 \hspace{0cm}
 \subfigure{
  \hspace{-1.0cm}
  \includegraphics[width=3.9in]{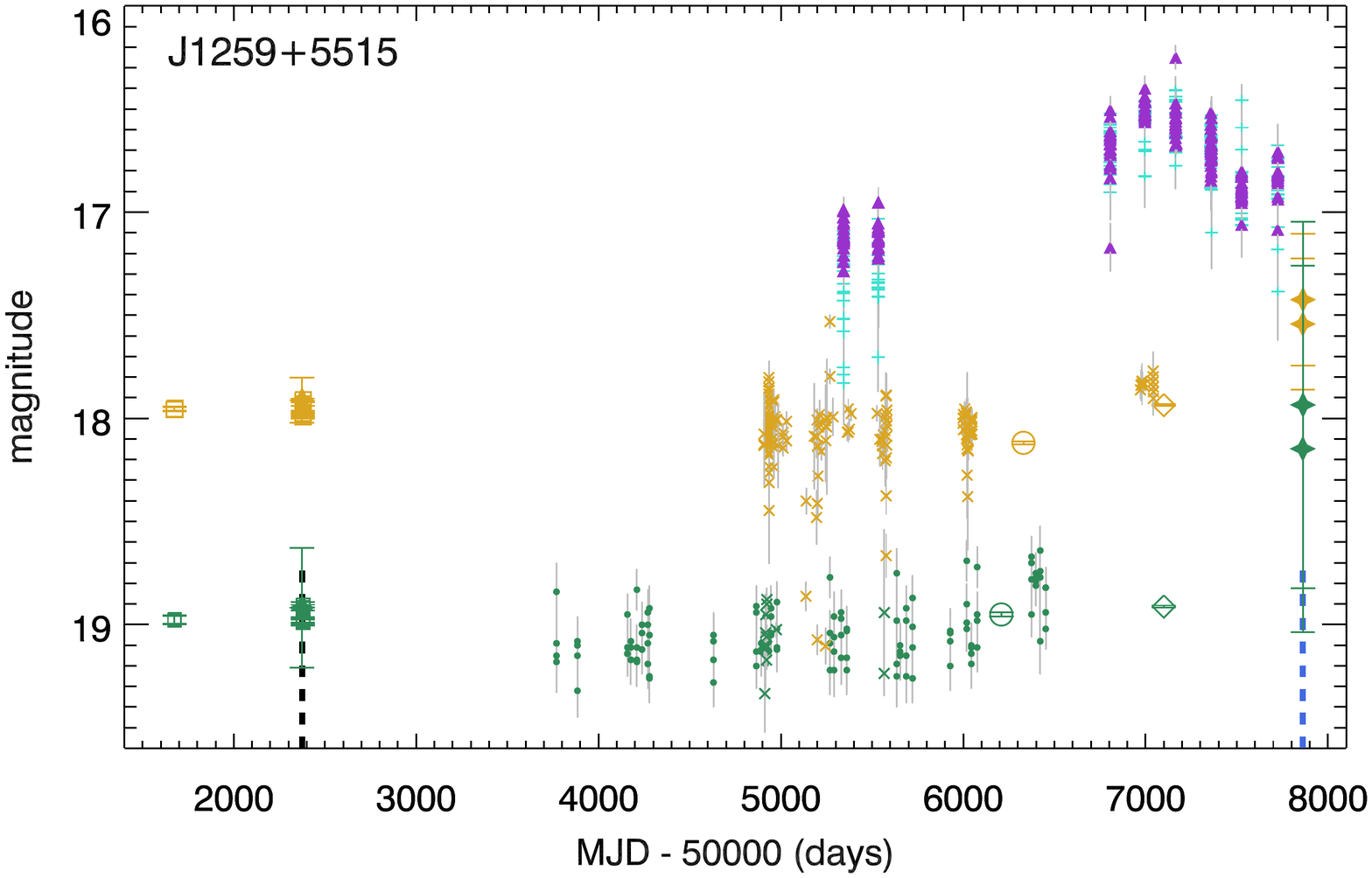}}
 \hspace{-1.4cm}
 \subfigure{
  \includegraphics[width=3.9in]{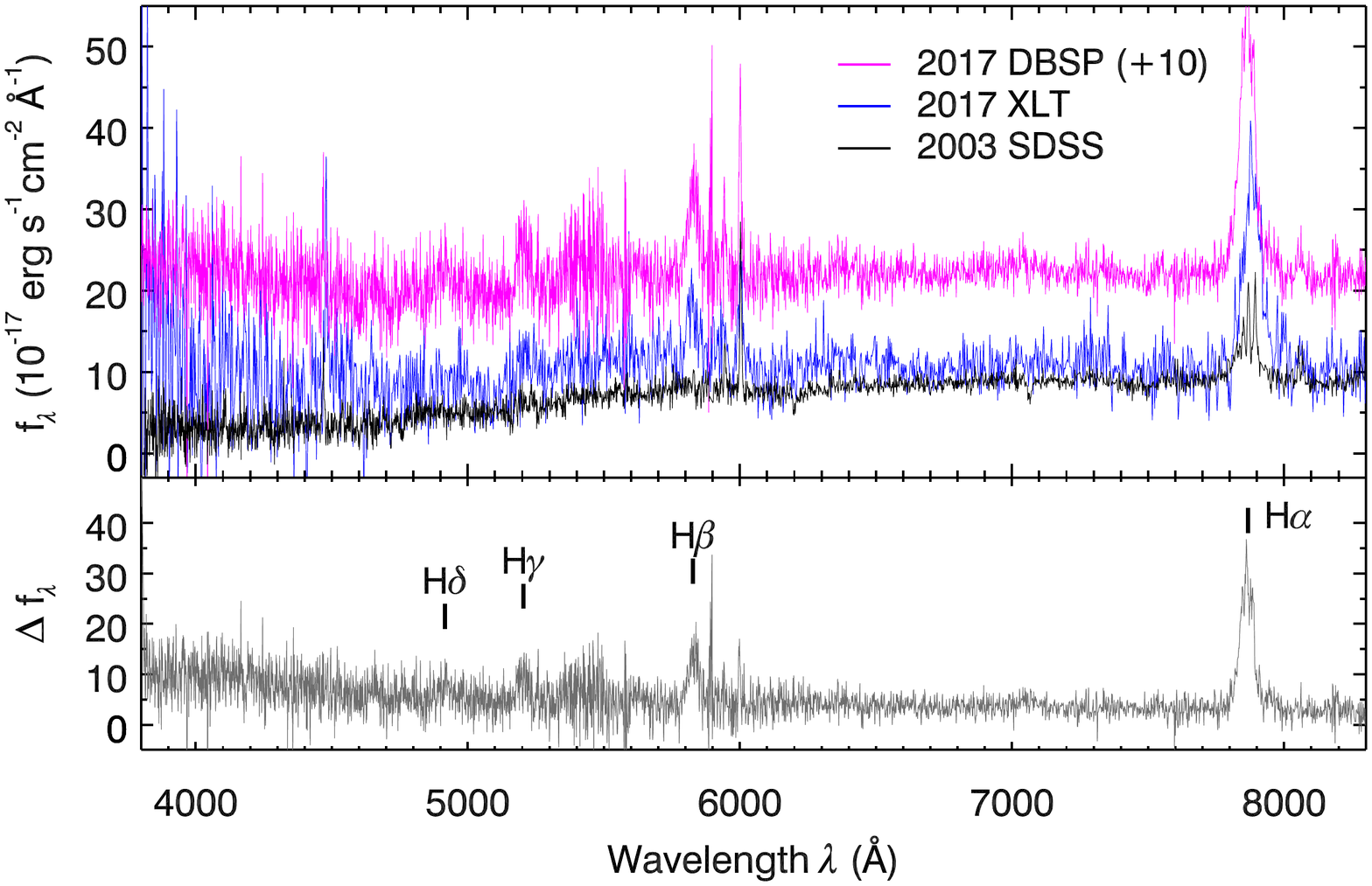}}\\
  \vspace{-2.4cm}

 \centering
 \hspace{0cm}
 \subfigure{
  \hspace{-1.0cm}
  \includegraphics[width=3.9in]{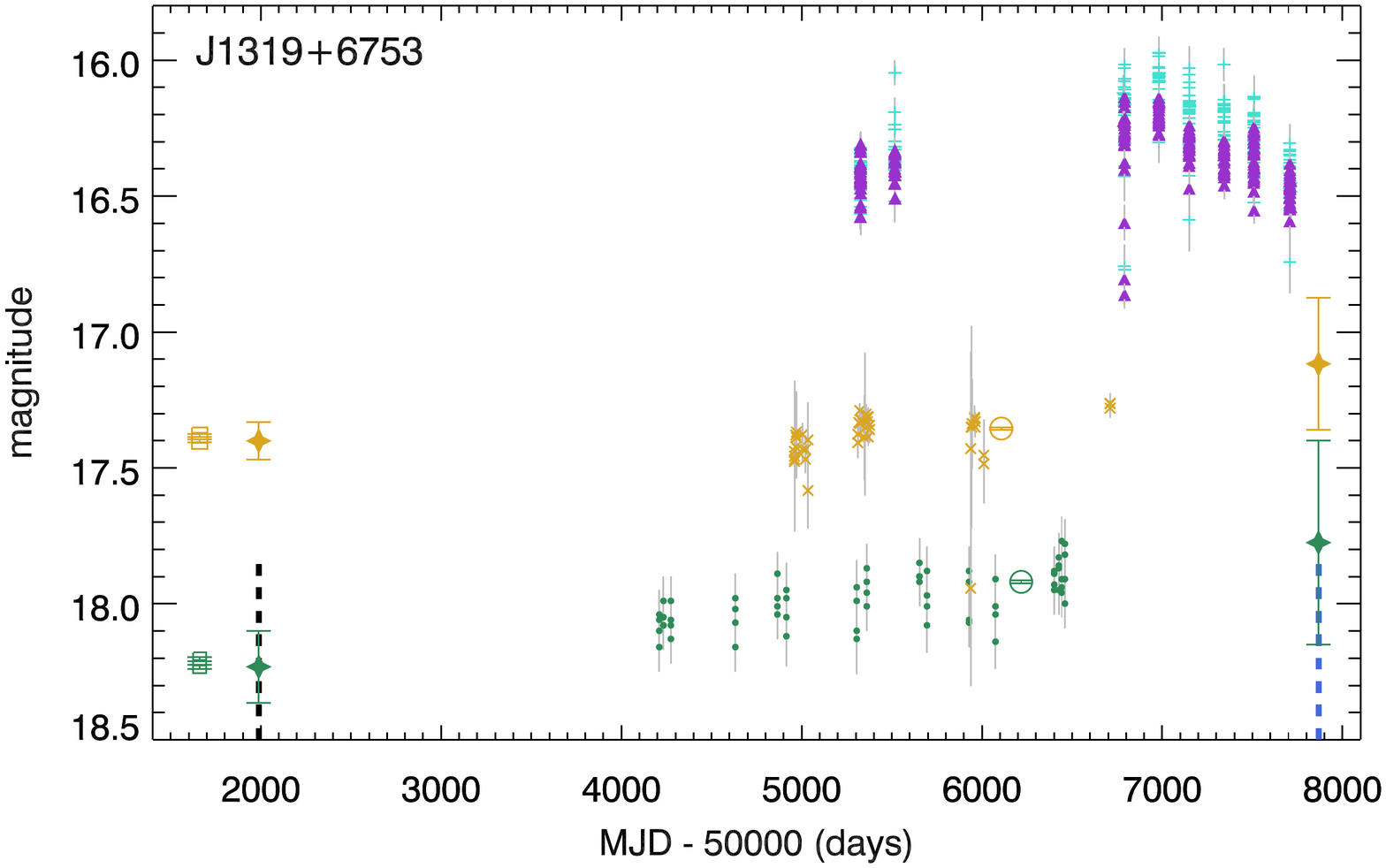}}
 \hspace{-1.4cm}
 \subfigure{
  \includegraphics[width=3.9in]{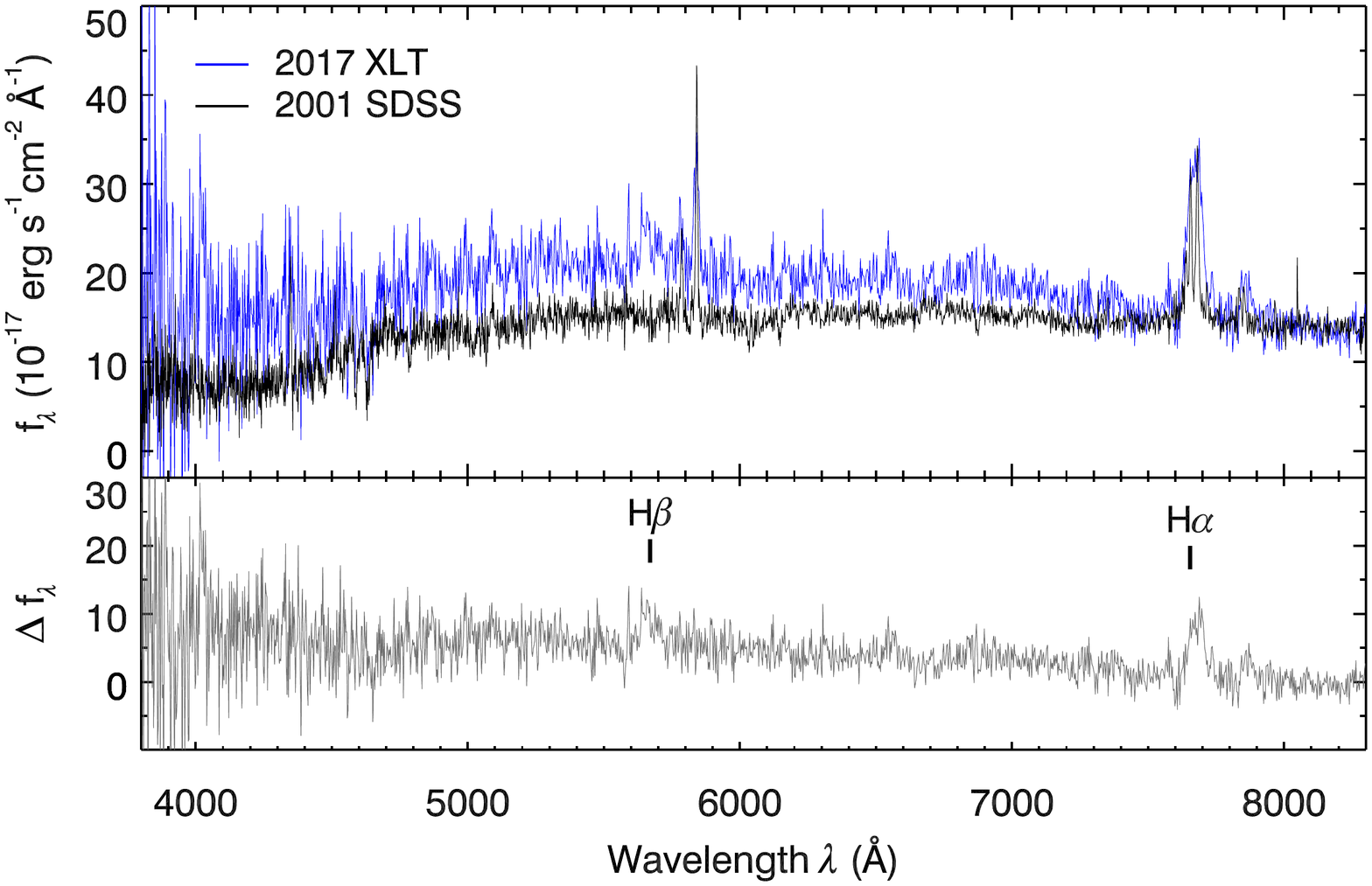}}\\
  \vspace{-2.4cm}

 \centering
 \hspace{0cm}
  \subfigure{
   \hspace{-1.0cm}
  \includegraphics[width=3.9in]{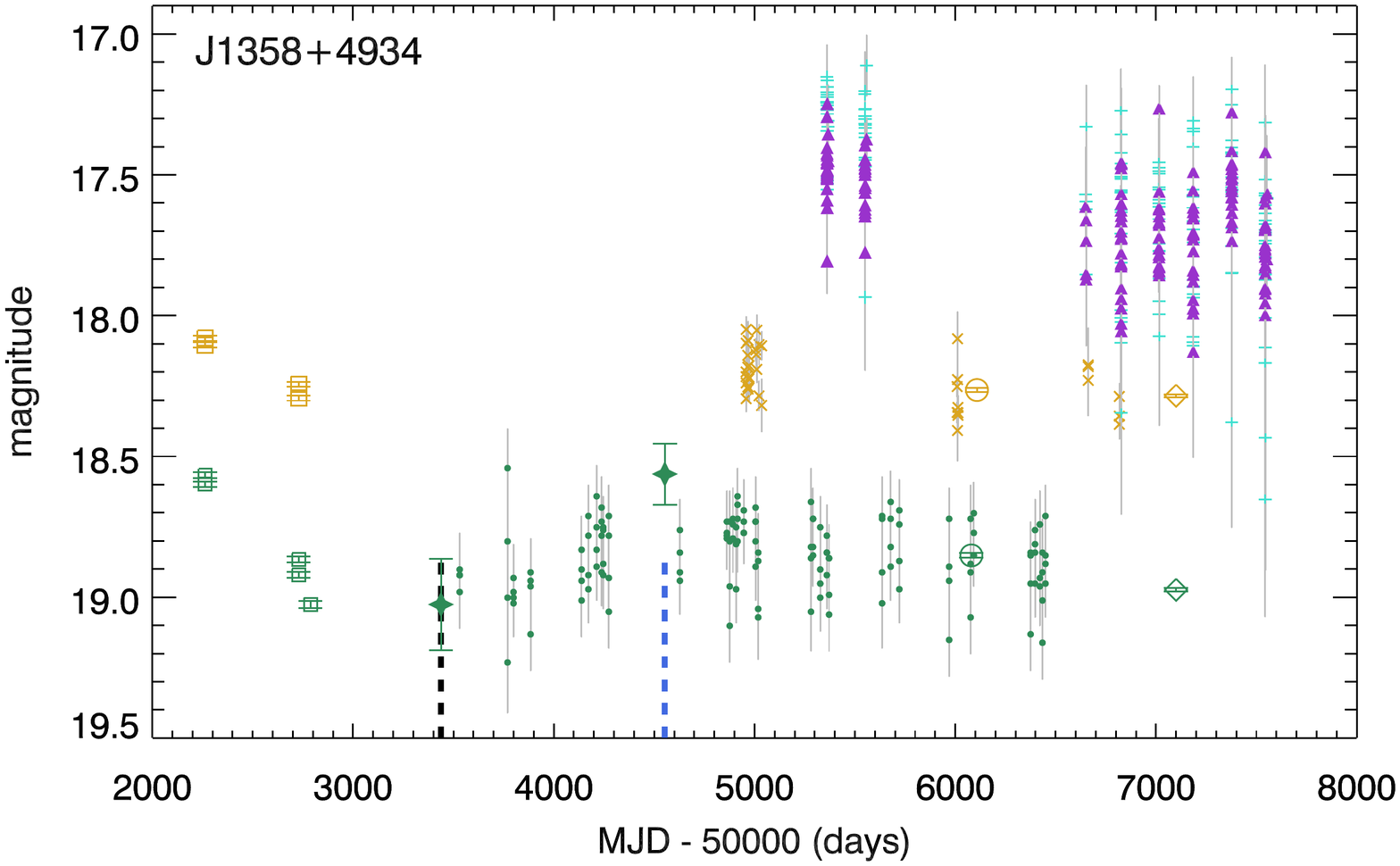}}
 \hspace{-1.4cm}
 \subfigure{
  \includegraphics[width=3.9in]{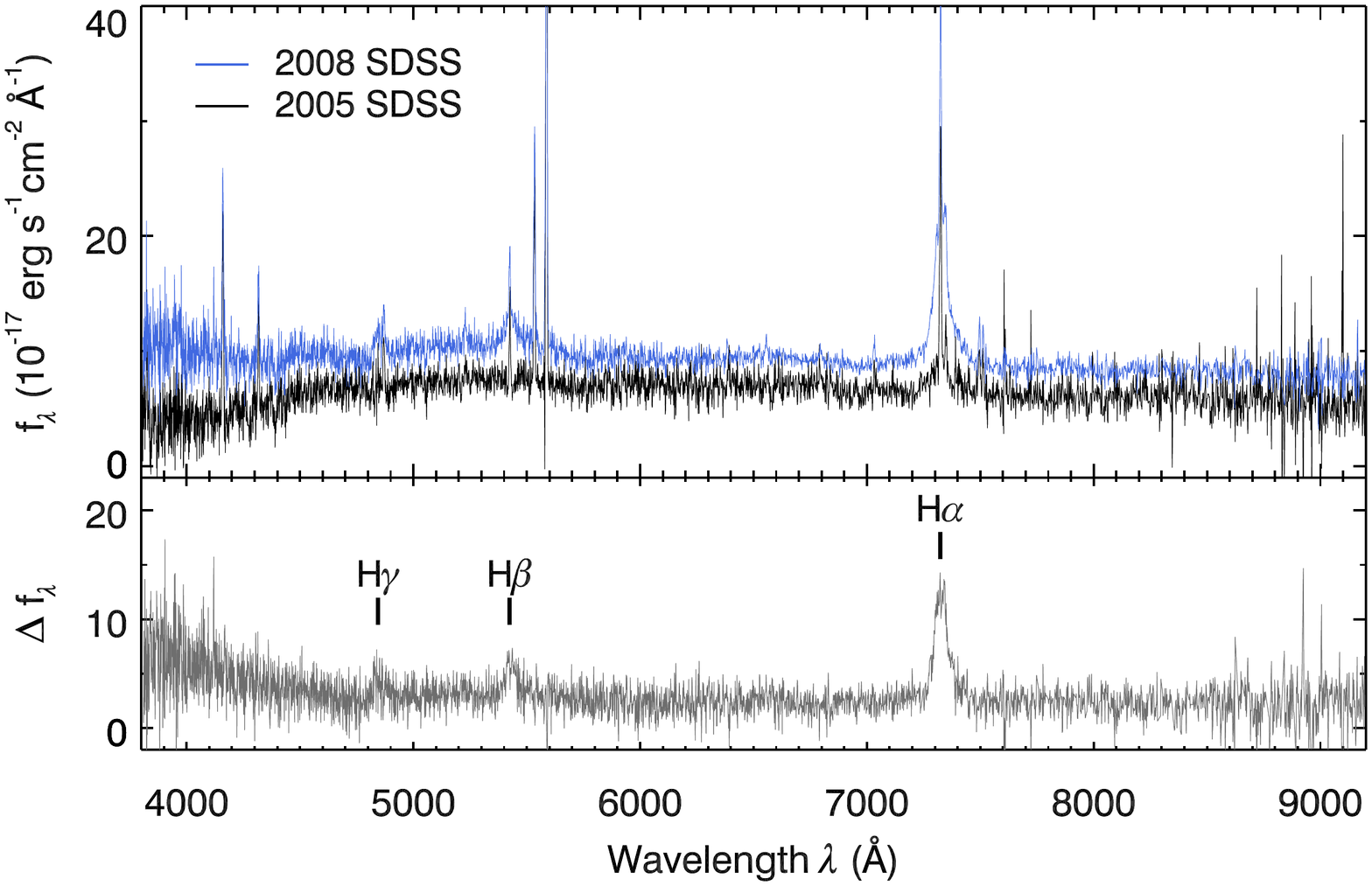}}\\
 \vspace{-2.4cm}

 \centering
  \hspace{0cm}
  \subfigure{
   \hspace{-1.0cm}
  \includegraphics[width=3.9in]{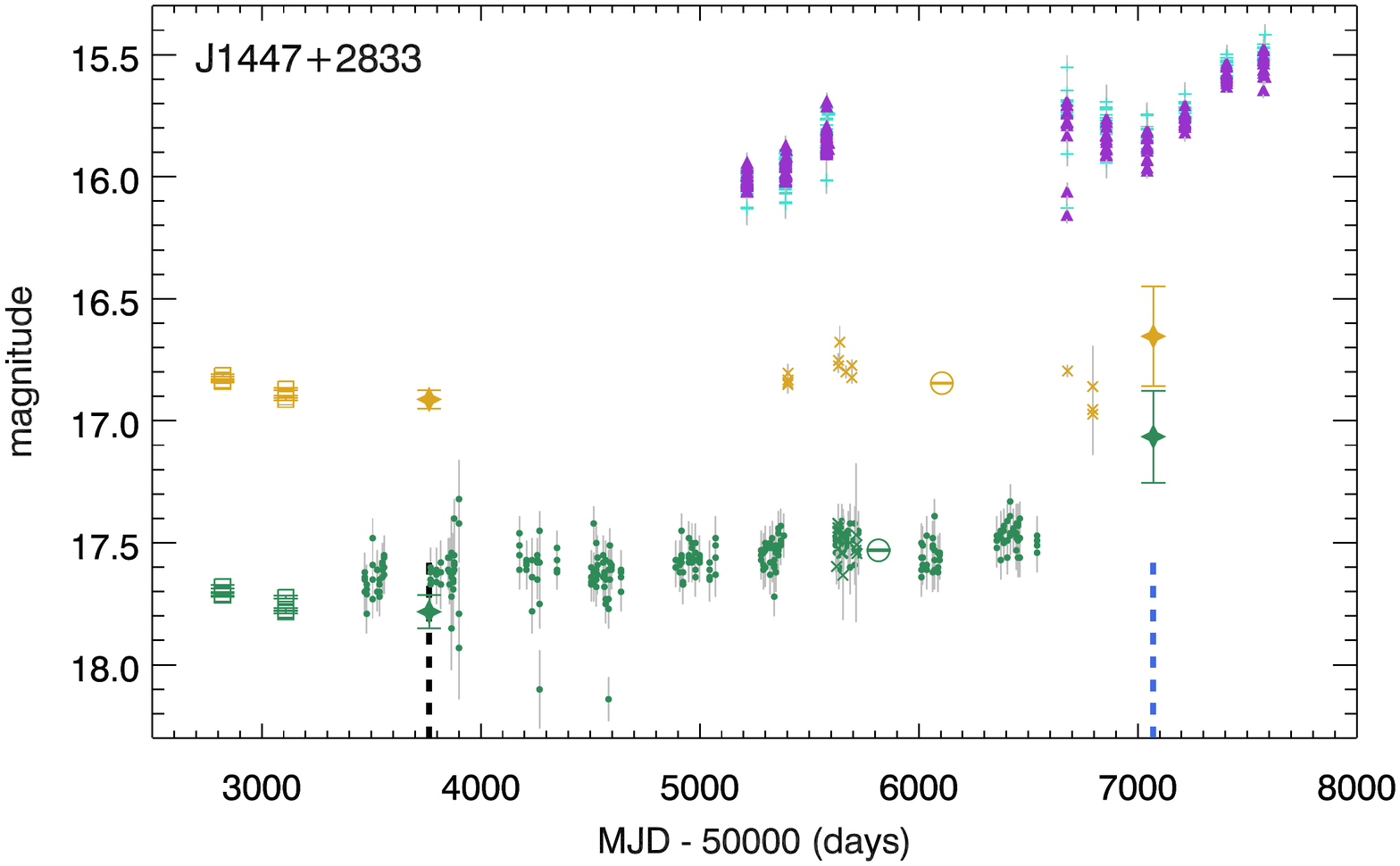}}
 \hspace{-1.4cm}
 \subfigure{
  \includegraphics[width=3.9in]{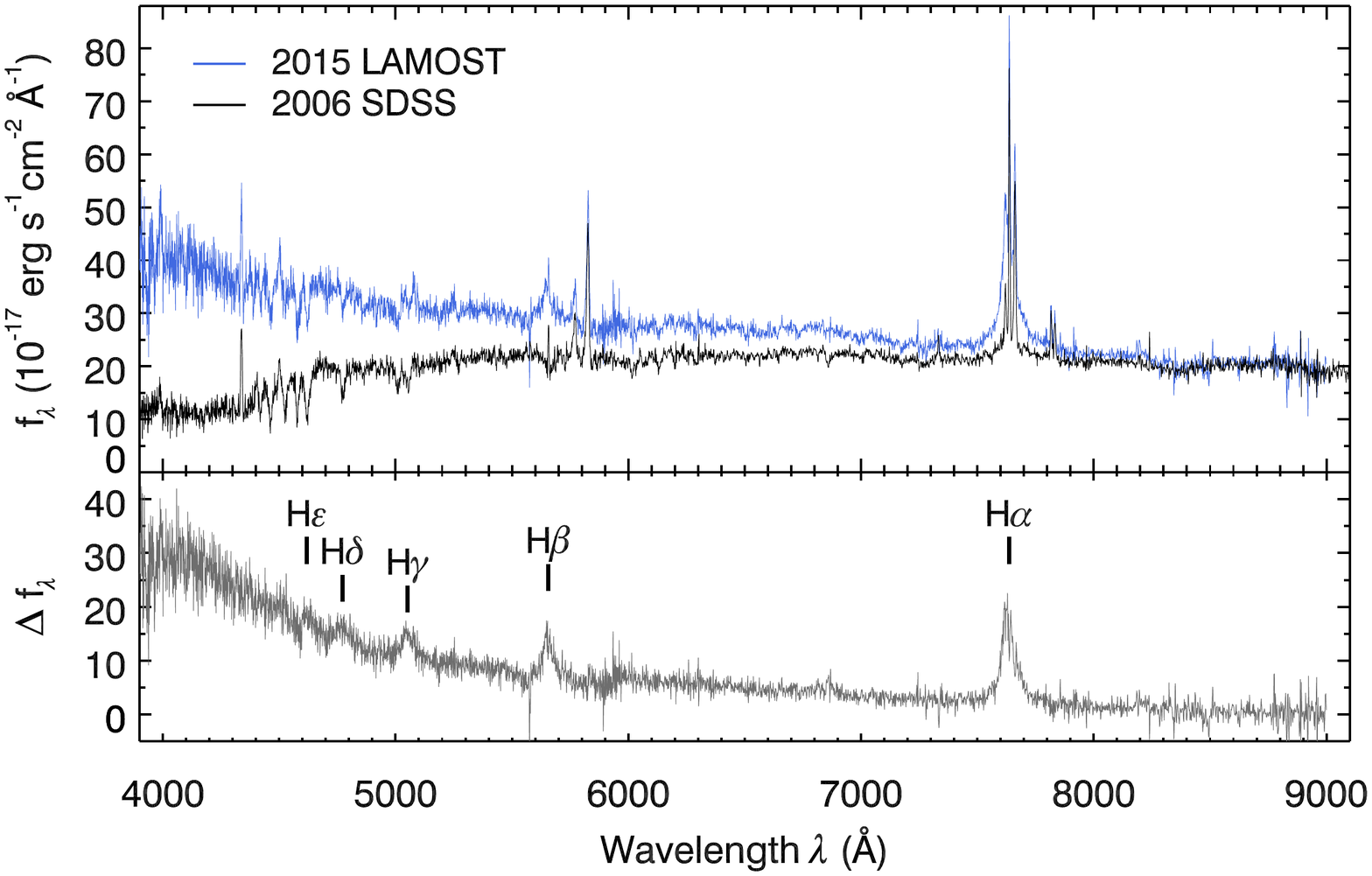}}\\
 \vspace{-1cm}
  \caption{ (Continued.)}
\end{figure*}

\renewcommand{\thefigure}{A.\arabic{figure}}
\addtocounter{figure}{-1}

\begin{figure*}

 \centering
 \hspace{0cm}
  \subfigure{
   \hspace{-1.0cm}
  \includegraphics[width=3.9in]{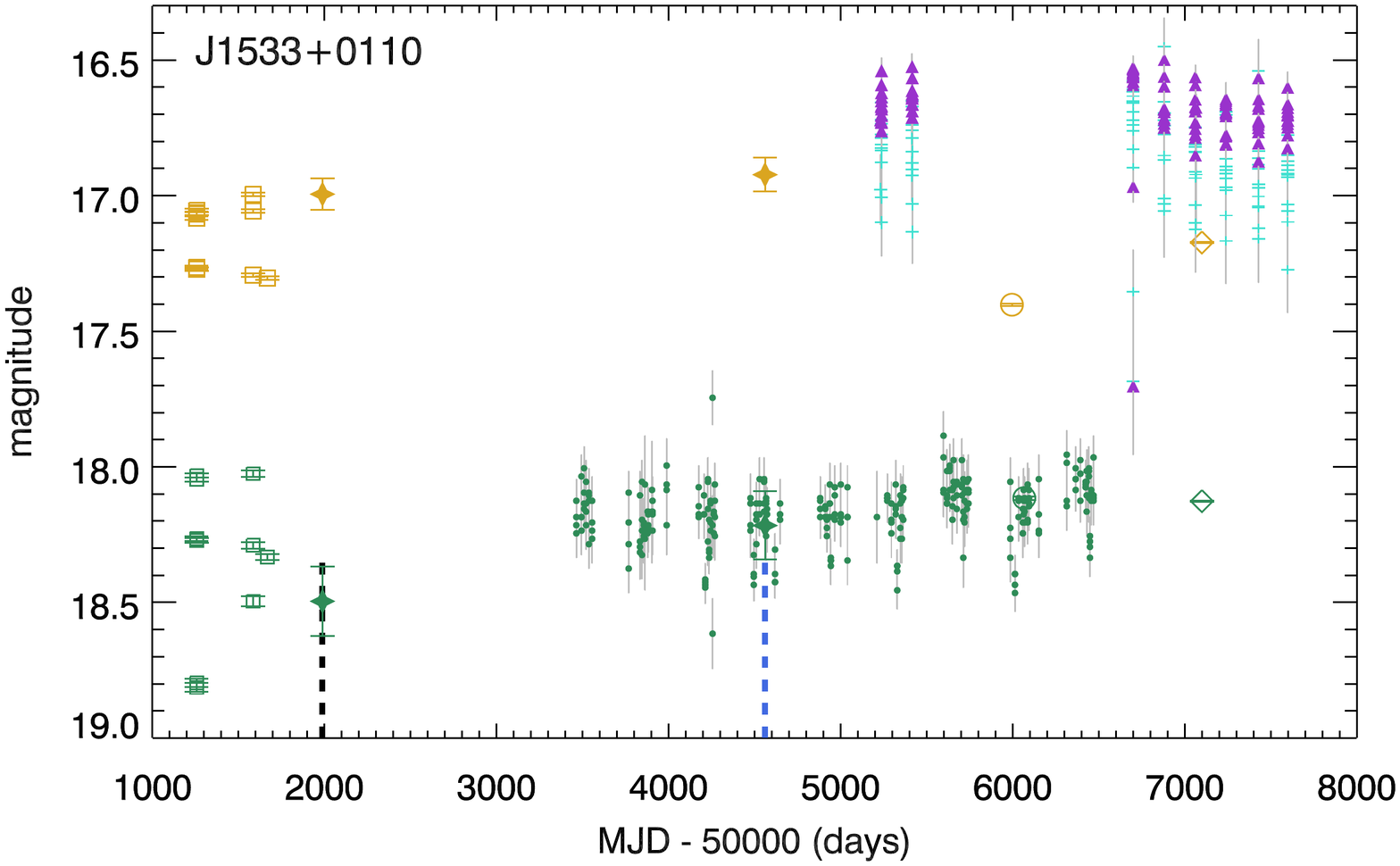}}
 \hspace{-1.4cm}
 \subfigure{
  \includegraphics[width=3.9in]{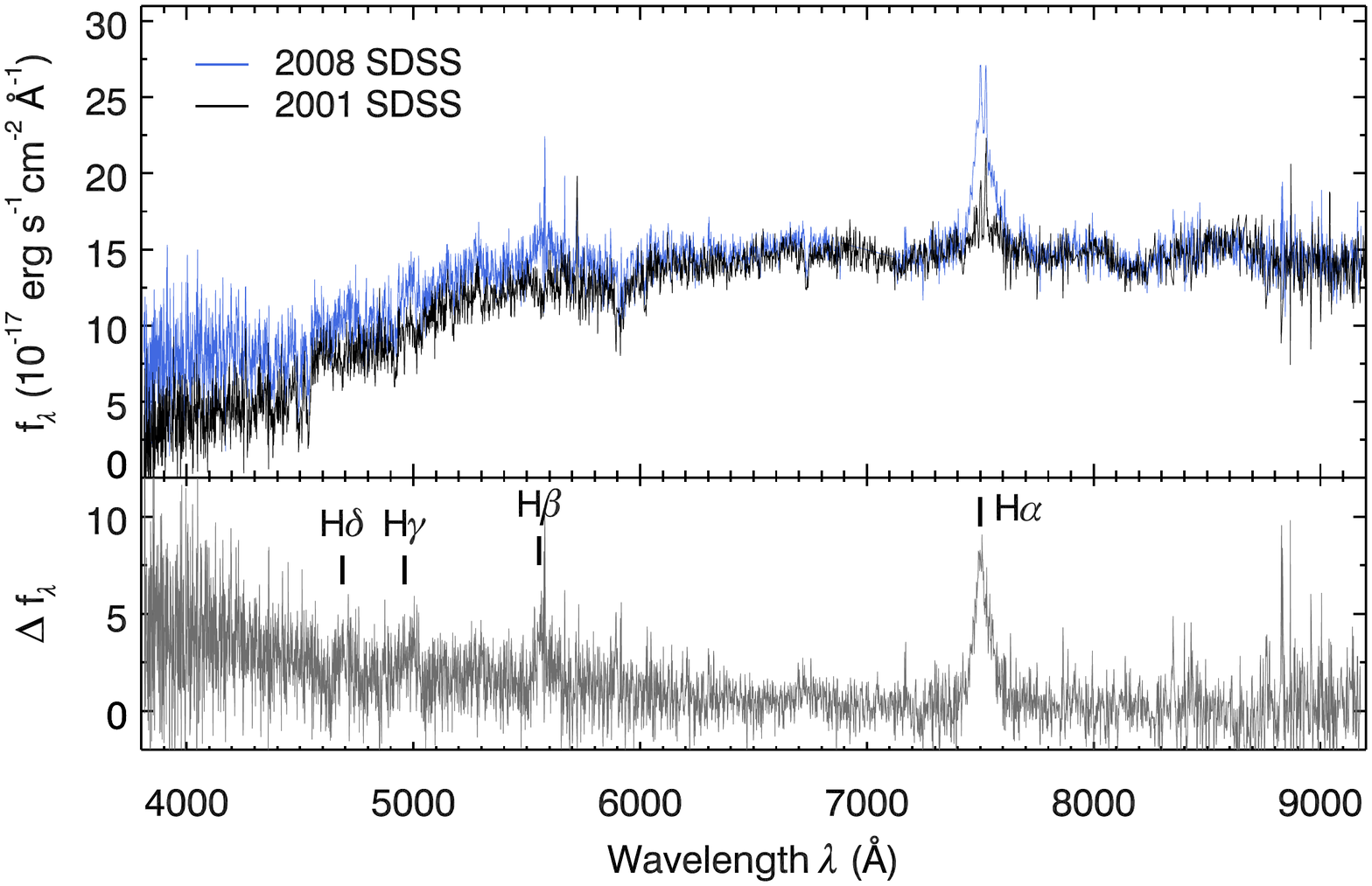}}\\
 \vspace{-2.4cm}

  \centering
  \hspace{0cm}
  \subfigure{
   \hspace{-1.0cm}
  \includegraphics[width=3.9in]{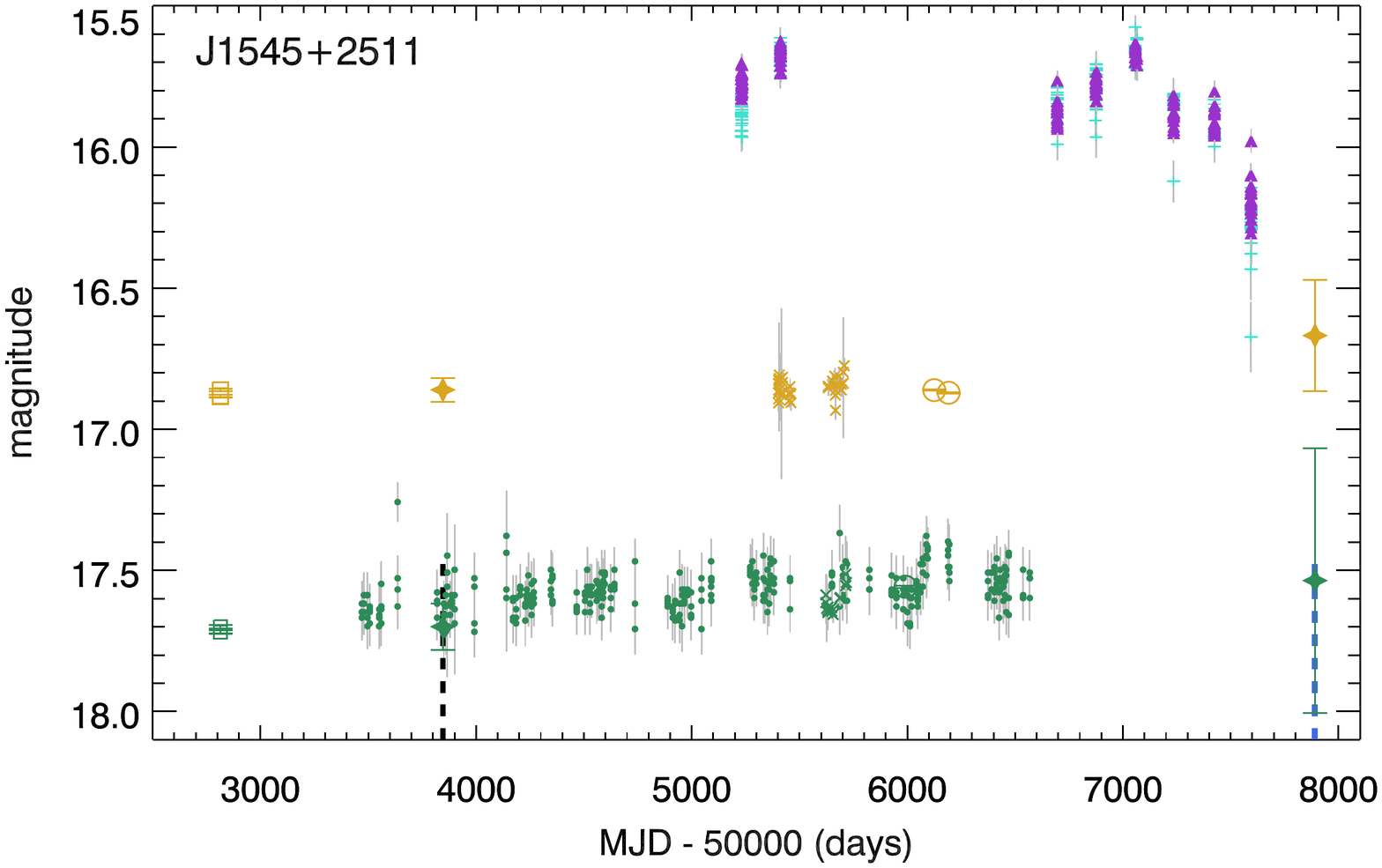}}
 \hspace{-1.4cm}
 \subfigure{
  \includegraphics[width=3.9in]{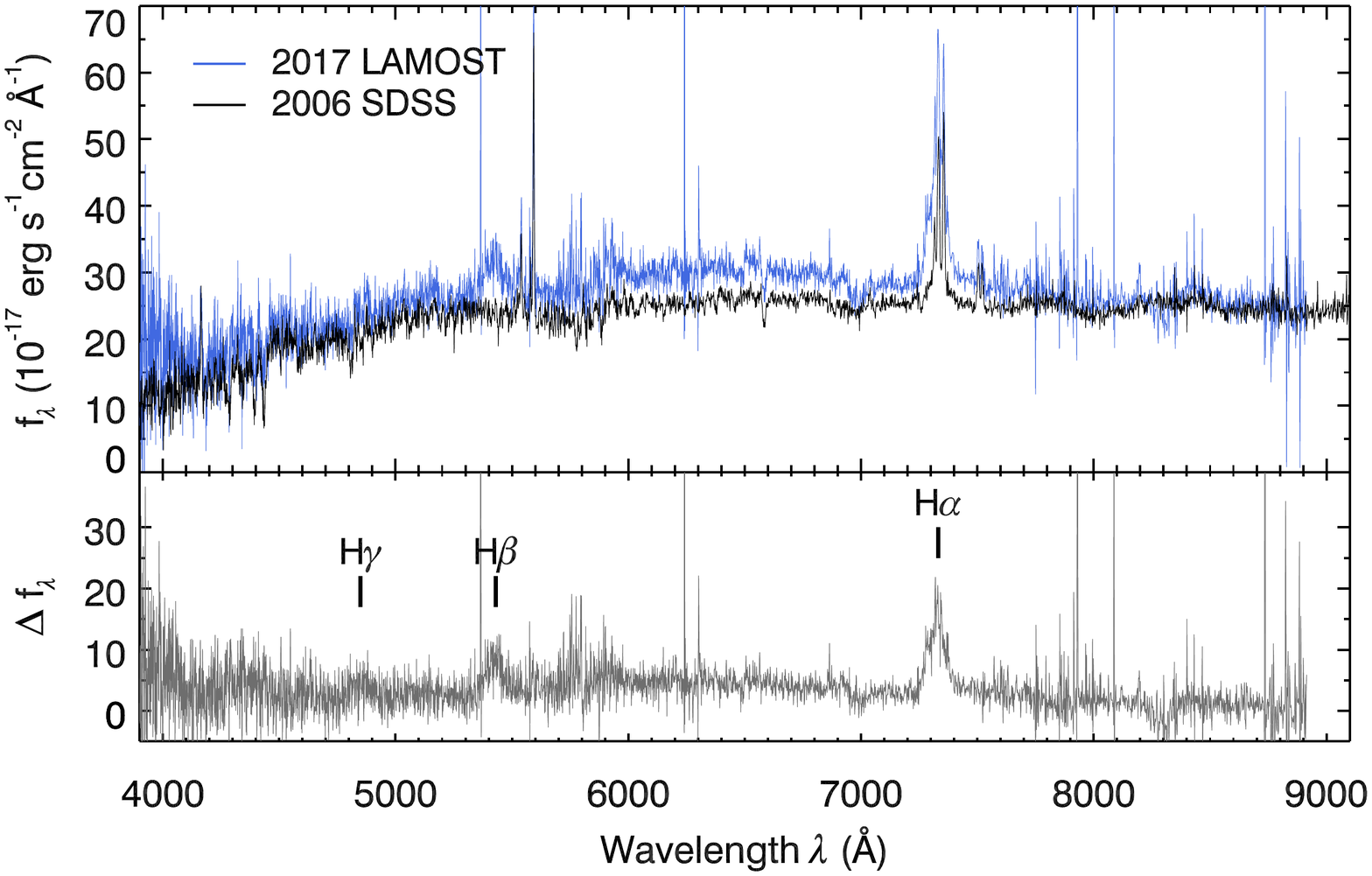}}\\
 \vspace{-2.4cm}

 \centering
 \hspace{0cm}
 \subfigure{
  \hspace{-1.0cm}
  \includegraphics[width=3.9in]{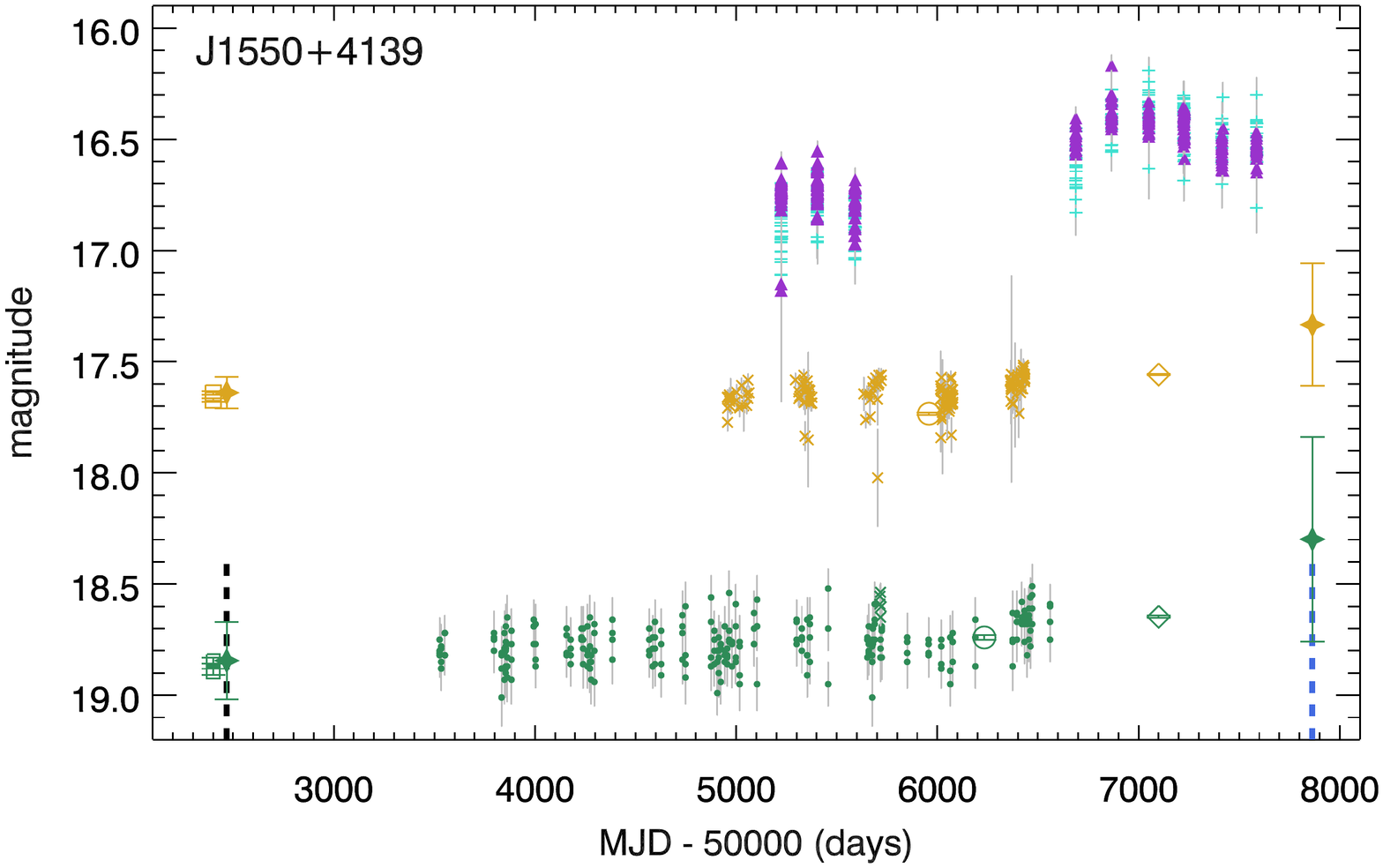}}
 \hspace{-1.4cm}
 \subfigure{
  \includegraphics[width=3.9in]{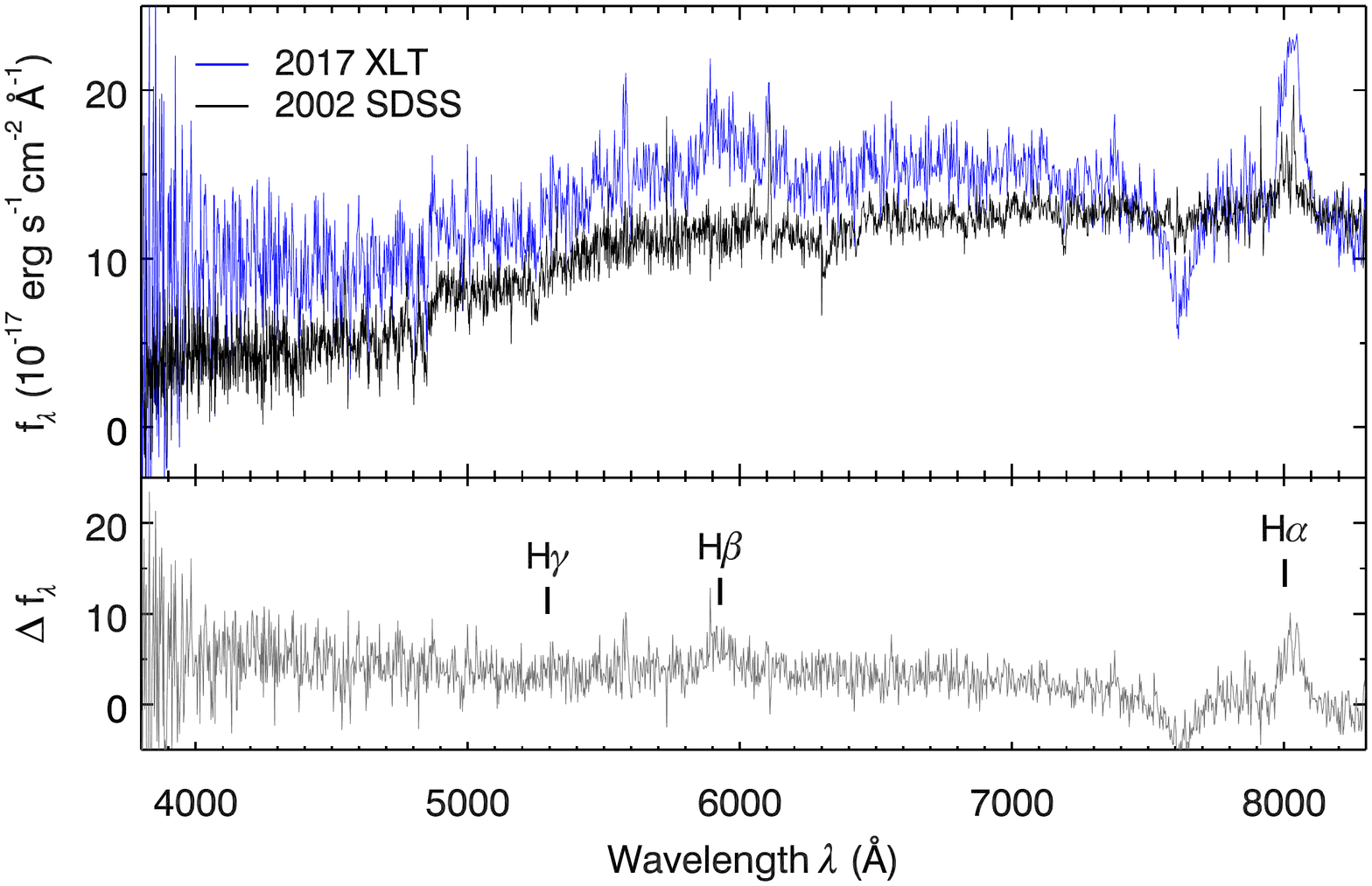}}\\
 \vspace{-2.4cm}

  \centering
  \hspace{0cm}
  \subfigure{
   \hspace{-1.0cm}
  \includegraphics[width=3.9in]{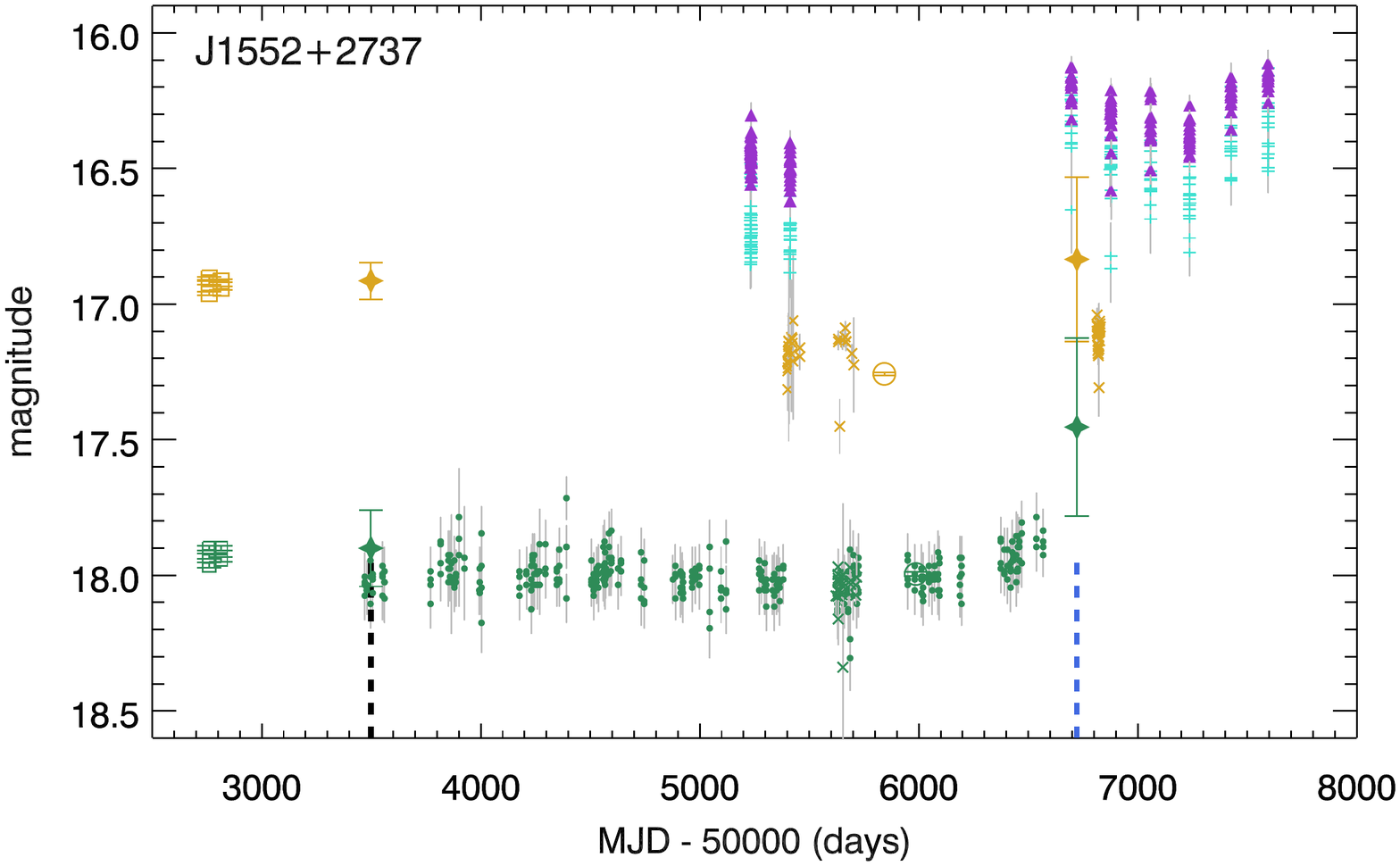}}
 \hspace{-1.4cm}
 \subfigure{
  \includegraphics[width=3.9in]{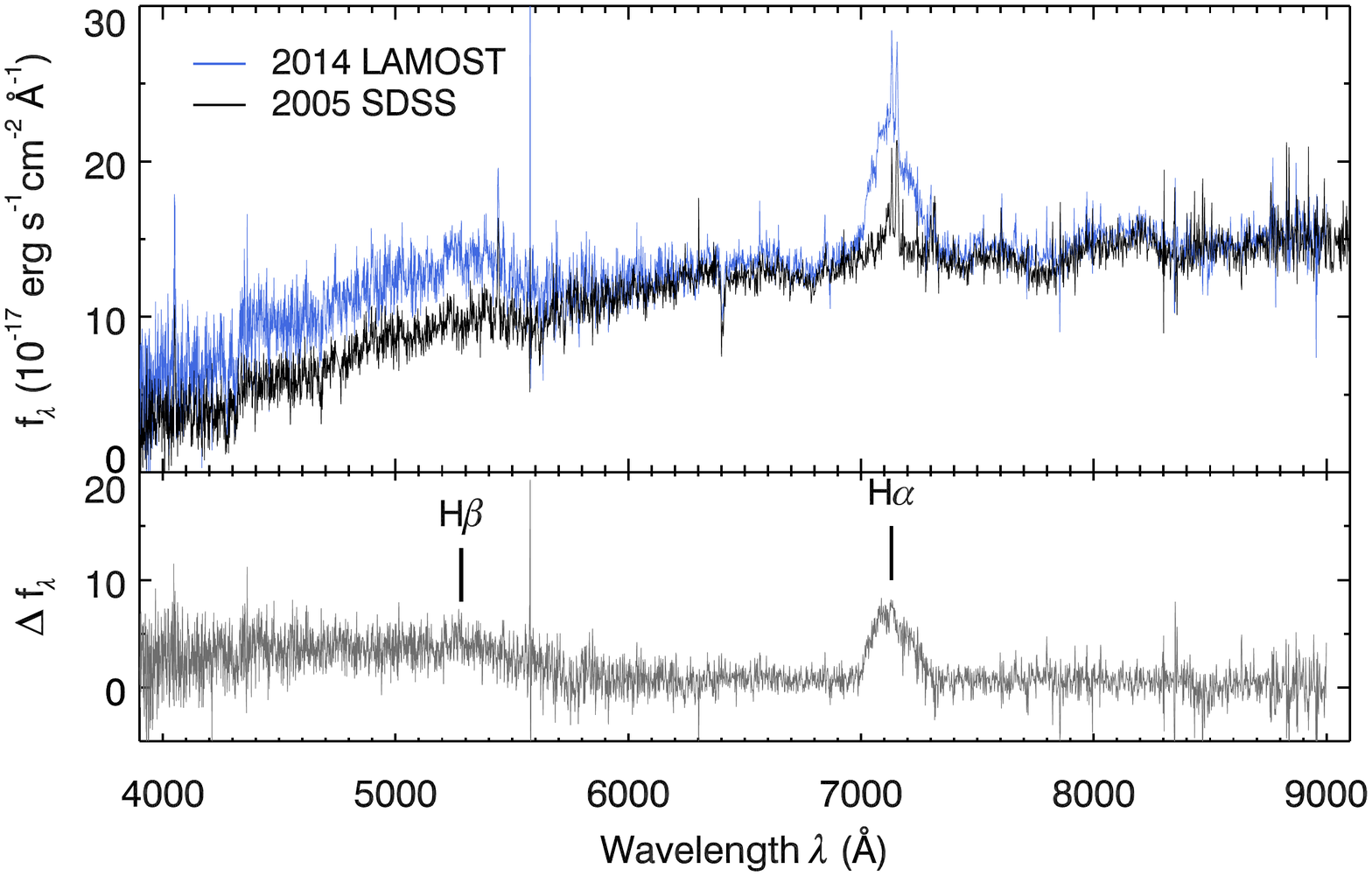}}\\
 \vspace{-1cm}
  \caption{ (Continued.)}
\end{figure*}

\clearpage
\renewcommand{\thefigure}{A.\arabic{figure}}

\begin{figure*}

 \centering
 \hspace{0cm}
  \subfigure{
   \hspace{-1.0cm}
  \includegraphics[width=3.9in]{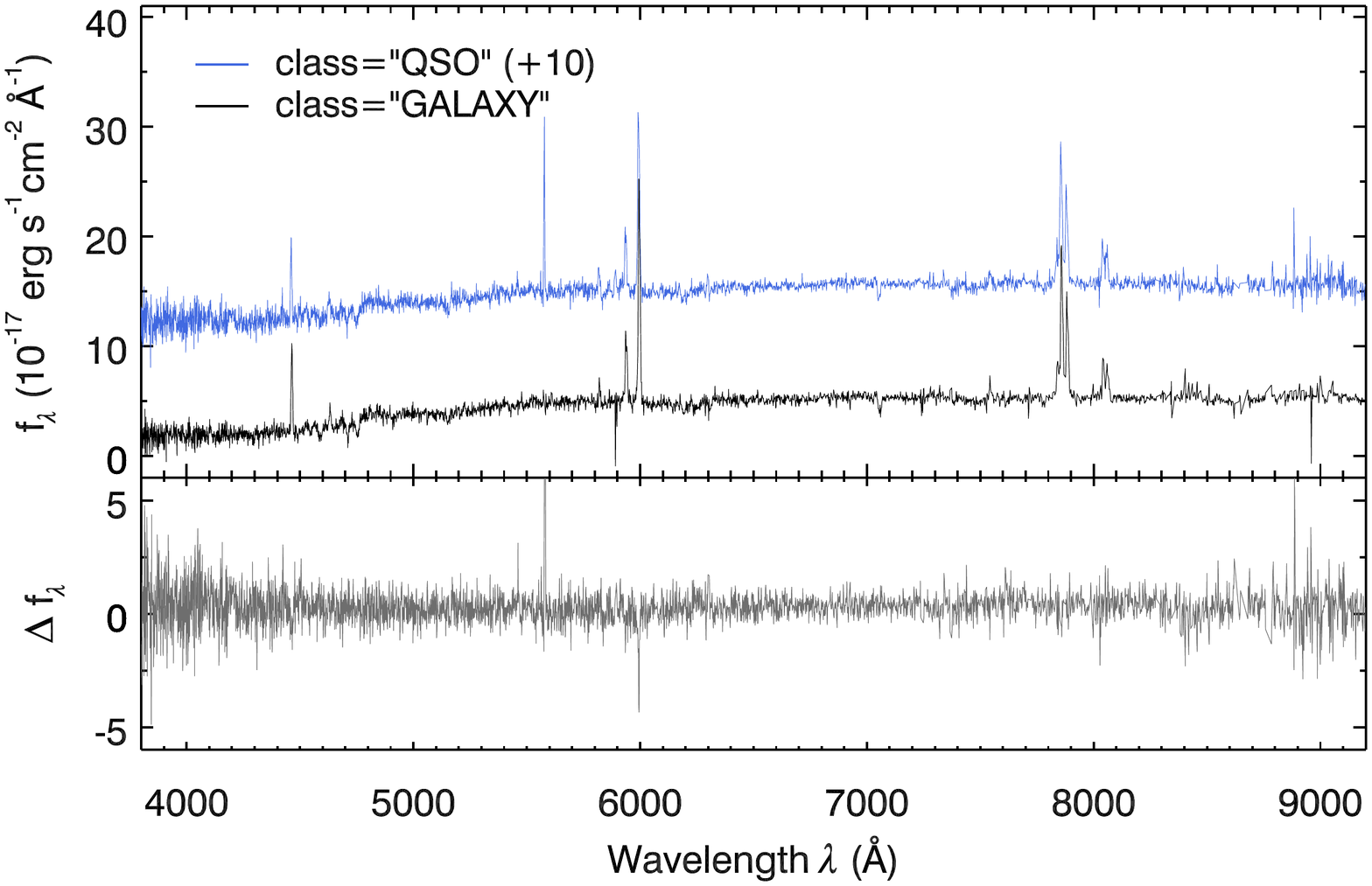}}
 \hspace{-1.4cm}
 \subfigure{
  \includegraphics[width=3.9in]{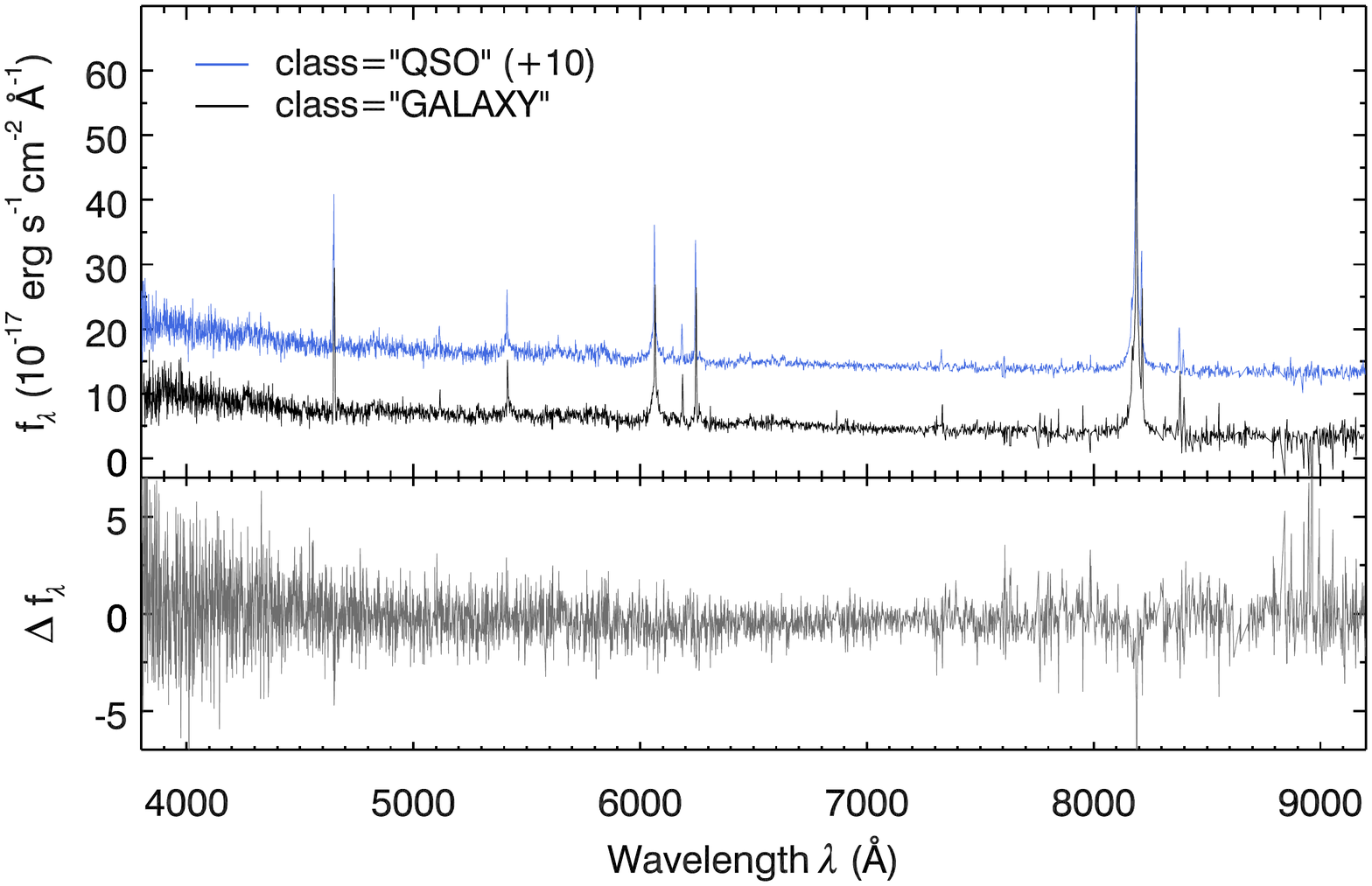}}\\
 \vspace{-2.4cm}

  \centering
  \hspace{0cm}
  \subfigure{
   \hspace{-1.0cm}
  \includegraphics[width=3.9in]{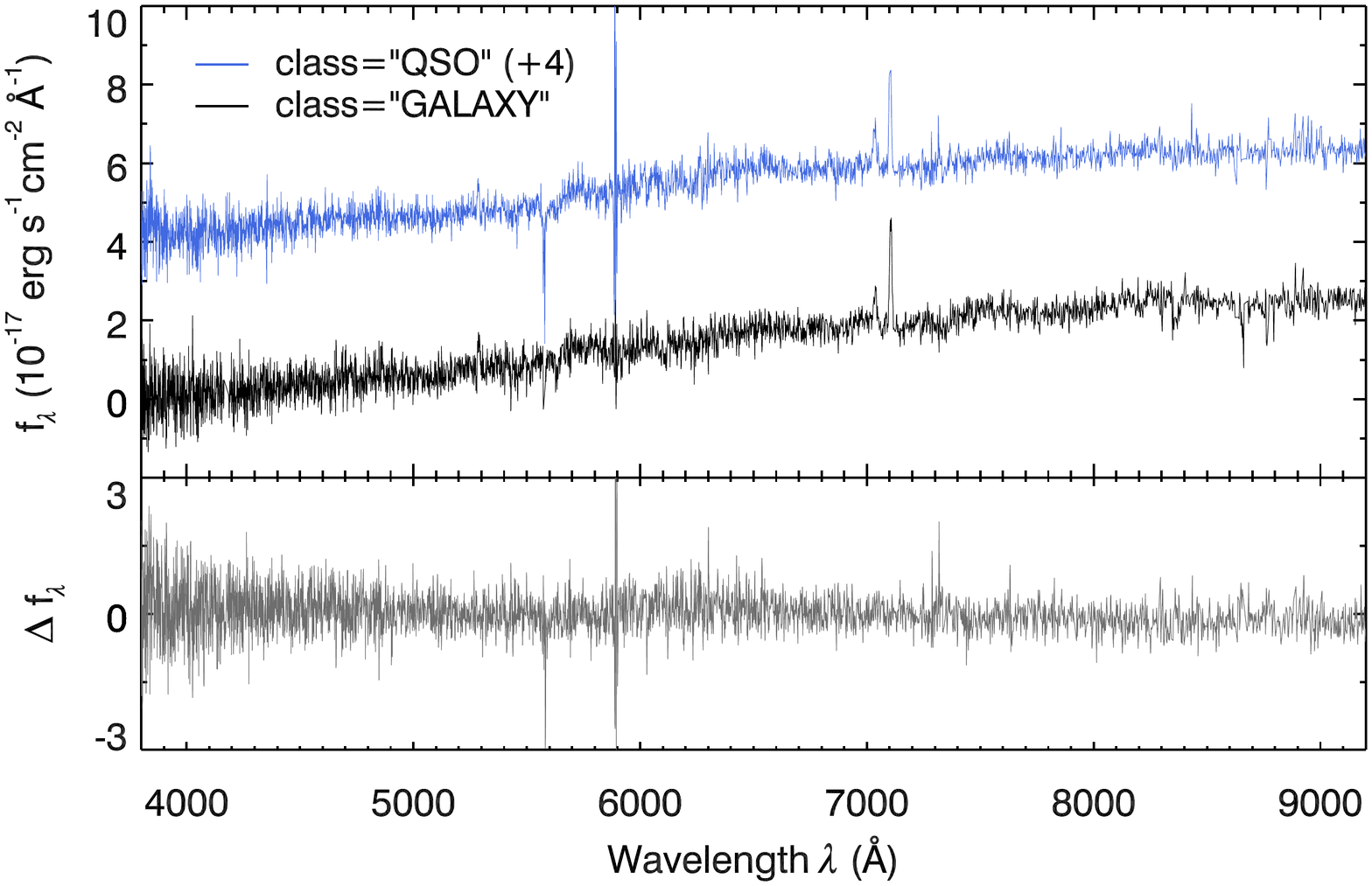}}
 \hspace{-1.4cm}
 \subfigure{
  \includegraphics[width=3.9in]{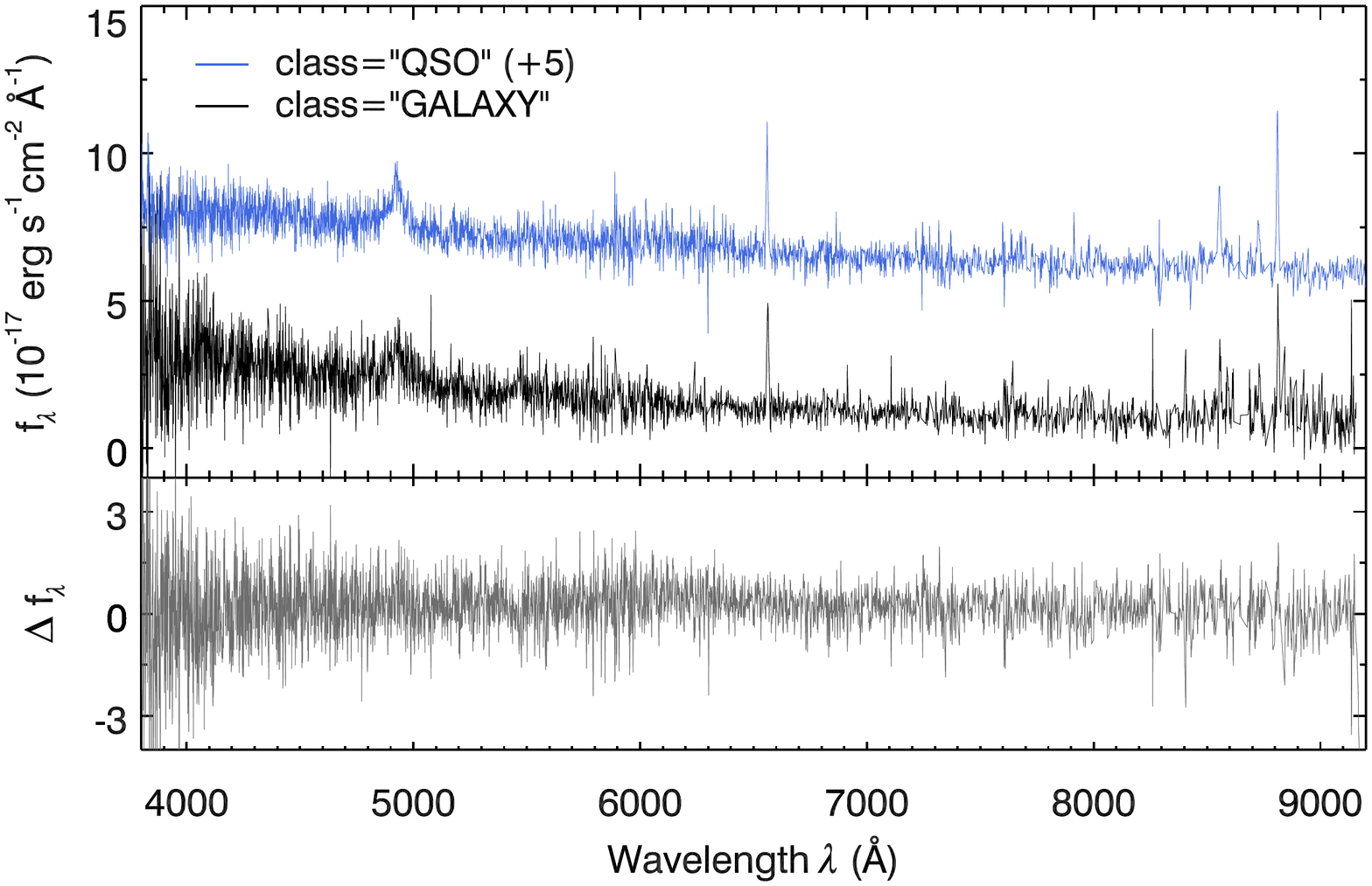}}\\
 \vspace{-1cm}

  \caption{\label{fig:not_cl} Four CL candidates in the SDSS that were rejected by visual inspection. From low to high redshift, the four objects are J0001$-$0005 (left upper panel), J1356$-$0115 (right upper panel), J0803+4258 (left bottom panel), and J0937+3232 (right bottom panel). To show the spectra at different epochs, a constant is added to the spectra that were classified as ``QSO".}
\end{figure*}

Figure \ref{fig:not_cl} shows some examples of CL candidates in the SDSS rejected by our visual inspection. Table \ref{tab:cl_not} summarize some details about the four CL candidates. They were were classified as ``QSO" and ``GALAXY" at different epochs in the SDSS. As shown in the residual spectra, there is no dramatic change. As the SDSS spectroscopic pipelines classify the objects through the comparison of individual spectrum with galaxy, QSO, and stellar templates, these objects were classified as different types possibly due to marginally different spectral S/N at different epochs.

\begin{deluxetable*}{ccclcclc}[htbp]
\tablecaption{\ Examples of CL candidates rejected by visual inspection in the SDSS \label{tab:cl_not}}
\tablewidth{0pt}
\tablehead{
\colhead{Name} & 
\colhead{R.A.} &
\colhead{Decl.} & 
\colhead{Redshift} &
\colhead{Epoch${\rm (GALAXY)}$} &
\colhead{Epoch${\rm (QSO)}$}
}
\startdata
J0001$-$0005 & 00:01:07.52 & $-$00:05:52.3 & 0.19754 & 55477 & 52943 \\
J1356$-$0115 & 13:56:18.49 & $-$01:15:14.0 & 0.24726 & 51942 & 52721 \\
J0803+4258 & 08:03:47.55 & +42:58:38.9 & 0.41881 & 55245 & 55178 \\
J0937+3232 & 09:37:35.46 & +32:32:49.9 & 0.76017 & 54807 & 56310 \\
\enddata
\tablecomments{Epoch${\rm (GALAXY)}$ and Epoch${\rm (QSO)}$ describe the MJDs of the spectra that were classified as ``GALAXY" and ``QSO", respectively.}
\end{deluxetable*}

\renewcommand{\thefigure}{A.\arabic{figure}}

\begin{figure}[htbp]
\hspace*{-0.5cm}
\epsscale{0.7}
\plotone{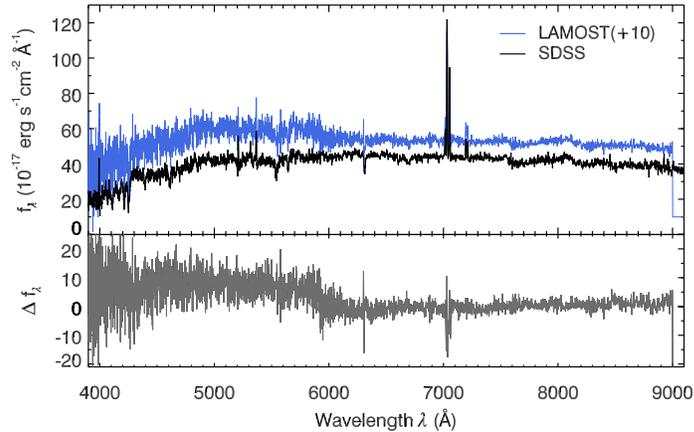}
\vspace{-0.0cm}
\caption{\label{fig:not_lamost} An example of CL candidates in the SDSS and LAMOST that were rejected by visual inspection. }
\end{figure}

The spectra were classified by the LAMOST 1D pipeline into four primary classifications, namely ``STAR", ``GALAXY", ``QSO", and ``UNKNOWN", through matching individual spectrum with templates \citep{Luo2015}. Among the 8171 objects, approximately 36\% of them are classified as ``UNKNOWN" and 7\% are classified as ``STAR". The S/N of objects classified as ``UNKNOWN" is low due to unstable fiber efficiency, non-photometric observational conditions, or they are too faint for the LAMOST survey \citep{Ai2016}. None of the 10 CL AGNs from the LAMOST were classified as ``UNKNOW" or ``STAT". Figure \ref{fig:not_lamost} shows an example of CL candidates in the SDSS and LAMOST that were rejected by visual inspection. There is a break around 5700-5900 ${\rm \AA}$ in the LAMOST spectrum due to inappropriate combination of the spectra in the blue and red arms, resulting in flux intensity variation in the automatic program check process.


\end{document}